\pdfoutput=1

\documentclass[12pt,preprint]{aastex}

\newif\ifpdf\ifx\pdfoutput\undefined\pdffalse\else\pdfoutput=1\pdftrue\fi

\slugcomment{Submitted to Icarus.}

\shorttitle{Two-Dimensional Transport in Protoplanetary Disks}
\shortauthors{F. J. Ciesla}

\begin{document}


\title{Two-Dimensional Transport of Solids in Viscous Protoplanetary Disks}

\author{F. J. Ciesla\altaffilmark{1}}
\affil{Department of Geophysical Sciences, The University of Chicago, 5734 South Ellis Avenue, Chicago, IL 60637}

\begin{abstract}

Large-scale radial transport of solids appears to be a fundamental consequence of protoplanetary disk evolution based on the presence of high temperature minerals in comets and the outer regions of protoplanetary disks around other stars.  Further, inward transport of solids from the outer regions of the solar nebula has been postulated to be the manner in which short-lived radionuclides were introduced to the terrestrial planet region and the cause of the variations in oxygen isotope ratios seen in primitive materials.  Here, both outward and inward transport of solids are investigated in the context of a two-dimensional, viscously evolving protoplanetary disk.  The dynamics of solids are investigated to determine how they depend on particle size and the particular stage of protoplanetary disk evolution, corresponding to different rates of mass transport.  It is found that the outward flows that arise around the disk midplane of a protoplanetary disk aid in the outward transport of solids up to the size of CAIs and can increase the crystallinity fraction of silicate dust at 10 AU around a solar mass star to as much as $\sim$40\% in the case of rapidly evolving disks, decreasing as the accretion rate onto the star slows.
High velocity, inward flows along the disk surface aid in the rapid transport of solids from the outer disk to the inner disk, particularly for small dust.  
Despite the diffusion that occurs throughout the disk, the large-scale, meridonal flows associated with mass transport prevent complete homogenization of the disk, allowing compositional gradients to develop that vary in intensity for a timescale of one million years.  The variations in the rates and the preferred direction of radial transport with height above the disk midplane thus have important implications for the dynamics and chemical evolution of primitive materials.
\end{abstract}


\section{Introduction}

The results of the Stardust mission, which collected roughly 0.001 g of material ejected from Comet Wild 2 \citep{brownlee06}, suggests that large-scale transport in the solar nebula was a fundamental process in shaping the properties of the planetesimals that formed within it, and by extension, the planets that accreted from these building blocks.  Among the materials returned to Earth were crystalline Mg-rich silicate grains which require temperatures in excess of 1000 K to form from the amorphous precursors expected to have been inherited from the natal molecular cloud \citep{hallenbeck00}, or even higher temperatures to condense directly from a gas phase \citep[][and references therein]{wooden07}.  Further, a refractory grain called Inti, whose mineralogy and oxygen isotopic compositions is reminiscent of the Calcium, Aluminum-rich Inclusions (CAIs) found within chondritic meteorites was also found within the Stardust aerogel \citep{brownlee06,zolensky06,mckeegan06}.  These findings have led many to conclude that large amounts of materials that formed or were processed in the hot, inner regions of the solar nebula migrated outwards tens of AU to be accreted by comets. The exact distance that materials were transported is still the matter of debate as comets likely originate from beyond 20 AU \citep{charnoz07}, but the cometesimals from which these objects formed likely originated anywhere from $\sim$5 AU \citep{supulverlin00} to $\sim$100 AU \citep{weidenschilling97,weidenschilling04}.

That dust was mobile in the solar nebula is not surprising, as we have long recognized that transport processes must have operated in our solar nebula.  For example, the chondritic meteorites, which are relatively unaltered products of our solar nebula, come in a variety of types, each characterized by their bulk chemistry, oxygen isotopic ratios, as well as chondrule sizes and abundances
\citep[see review by][]{weisberg06}.  That each of these bodies record such different formation environments despite forming in close proximity to one another (a scale of 1-3 AU in the solar nebula, in all likelihood) and over a time period of 1-3 million years \citep[e.g.][]{kita05} suggests that the solar nebula was either spatially heterogeneous or that its chemical and isotopic properties changed with time due to the transport of materials within \citep{stevenson88,cyr98,cuzzi05,cieslacuz06}.

Further, the components of chondritic meteorites--the refractory inclusions, chondrules, and matrix which contains pre-solar grains--all record very different chemical and isotopic environments as well as different thermal histories prior to their incorporation into common meteorite parent bodies.  It is difficult to imagine how such diverse materials could be brought together without invoking solar nebular transport of some kind.

Fortunately, we can gain insight into the details and driving mechanisms for transport in our solar nebula by examining other protoplanetary disks.  Crystalline silicates are abundant in a number of disks around other stars at levels greater than those seen in molecular clouds, particularly in the hot inner regions, indicating that their formation occurs at high temperatures, while their presence in the cool outer regions of disks is taken as evidence that large-scale radial transport of these materials occurred
\citep{vanboekel04,apai05,watson07,wooden07,wooden08}.  That evidence for this processing and transport is seen in disks  around stars of all masses suggests that this is a fundamental consequence of disk evolution.  Thus by understanding the processes that control or operate during protoplanetary disk evolution we may identify what processes controlled how solids were transported in our own solar nebula.

Protoplanetary disks are known to be dynamic objects that transport mass inward to be accreted by their central stars as part of their stars' last stages of pre-main sequence evolution \citep[see review by][and references therein]{dullemond07}.  Small dust particles that are entrained in the gas would also be subjected to these motions, thus providing a means of large-scale movement.  The exact driving force of this evolution remains the subject of ongoing work, but likely involves instabilities of some kind, such as the magnetorotational instability (MRI) of \citet{bh91} or gravitational torques arising from massive clumps in marginally gravitationally unstable disks 
\citep[e.g.][]{boss02,boley06}.

\citet{hartmann98} demonstrated that the observed accretion rates of T-Tauri stars can be matched by adopting the classical $\alpha$-viscosity model in which the evolution of the disk is treated as arising due to viscous stresses in the differentially rotating gas \citep{ss73,lbp74}.  These stresses serve to drive mass inward with time and allow the disk to spread in the radial direction to conserve angular momentum.  The $\alpha$-viscosity model is a turbulent viscosity model as it assumes that turbulence within the disk acts as the source of the viscosity, because the molecular viscosity of the gas is orders of magnitude too small to explain the observed mass accretion rates.  A similar model, the $\beta$-viscosity model, has also been proposed in which the turbulence specifically arises due to hydrodynamic instabilities in the disk \citep{richard99}.  This latter model assigns a viscosity to the disk whose functional dependence on disk properties differs slightly from the  $\alpha$-viscosity model.

Because of the turbulent nature of the gas, the random motions associated with its evolution lead to a diffusivity of material within it, allowing dust particles to be redistributed in a manner to smooth out concentration gradients.  This diffusion has previously been investigated  to evaluate whether particles that formed or were processed in the hot, inner regions of the solar nebula could have been carried outwards against the inward flows associated with mass transport to be delivered to where comets formed \citep{gail01,gail04,bockelee02}.  In the case of \citet{bockelee02}, they found that such a scenario was possible, however, in order to get a large fraction of the silicate grains in the outer nebula to be crystalline, in agreement with the observations of comets Halley and Hale-Bopp, the nebula must have started out ``warm'' and compact.  That is, the entire nebula mass was initially distributed within 10 AU of the Sun, resulting in mass accretion rates in excess of 10$^{-5}$ M$_{\odot}$/year.  Such a mass accretion rate is greater than observed for disks around roughly solar mass stars \citep[e.g.][]{calvet05}, though the data is lacking for ages less than 10$^{5}$ years old.  \citet{gail01,gail04} also found that high mass accretion rates favored the outward transport of high temperature materials, though the fractions of crystalline forsterite in these models were small ($\sim$1\%).

Previous studies of dust transport in protoplanetary disks, such as those described above, have largely relied on one-dimensional models where the changes in the surface (or column) density of the dust are calculated as a function of time and location as the disks evolved.  Thus dust at a given location of a disk is expected to follow the same dynamical behavior wherever it is located in that column. However, as discussed in detail in the next section, the dynamics of dust particles depend on the local gas density, pressure, and their respective gradients, all of which vary with height in a protoplanetary disk.  The importance of such considerations for the chemical evolution of protoplanetary disks has been recognized in recent work by \citet{tschar07}, \citet{cieslasci07}, and \citet{wehrstedt08} in which the rates as well as the preferred direction of radial transport were found to vary with height above the disk midplane.   Here, the model of \citet{cieslasci07} is extended to examine the roles that these variations played in the transport of primitive materials in our solar nebula as well as other protoplanetary disks.  The model of \citet{cieslasci07} is used as it accounts for the motions that develop as a result of gas drag and vertical settling, which are not included in \citet{tschar07} and \citet{wehrstedt08}, and therefore can be applied to cases beyond vapor species and sub-micron dust particles.

In the next section, the two-dimensional (radial and vertical) model of particle transport in protoplanetary disks is described, with the key differences with one-dimensional (radial) models highlighted.  Section 3 describes the protoplanetary disks to which this two-dimensional model is applied, with the specific numeric approaches described in Section 4.  Section 5 presents the results for transport calculations for disks of the various structures described.   The implications that these results have for the evolution of primitive materials in our solar nebula and other protoplanetary disks are then discussed.

\section{Model Description}
A key parameter in quantifying how a solid particle responds to the motions of the gas in a protoplanetary disk is given by the particle stopping time:
\begin{equation}
t_{s} = \frac{\rho c}{\rho_{g} a}
\end{equation}
where $\rho$ is the material density of the solid particle,  $c$ is the local speed of sound, $\rho_{g}$ is the local gas density and $a$ is the particle radius.  This is approximately how long it takes for a particle to lose its relative velocity with respect to the gas in which it is embedded.  Particles with short stopping times are very well coupled to the gas, moving almost as gas molecules themselves, whereas particles with long stopping times decouple from the gas and begin to follow a separate evolutionary path.  Strictly speaking, this particular formulation is only true for particles whose radii are less than the mean free path of the gas.  As the focus here is on the primitive materials found in chondritic meteorites, comets, or observed in disks around other stars, all of which are no more than 1 cm in size,   this particular formula can safely be used for all cases considered here. 

Given that the stopping time is inversely proportional to the gas density, it is not surprising that the particular dynamics of the particles will vary with location within a protoplanetary disk. Particles will generally have shorter stopping times and be trapped to the gas in the inner disk, whereas particles in the outer disk respond more slowly, and decouple to greater degrees.  Similar variations in behavior are expected as one considers the gas density profile above the disk midplane which is given by:
\begin{equation}
\rho_{g} \left( z \right) = \rho_{g,0} \mathrm{exp}\left(-z^{2}\mathrm{/}H^{2} \right)
\end{equation}
where $\rho_{g,0}$ is the density of the gas at the disk midplane, $z$ is the height above the midplane, and $H$ is the local gas scale height, given by $c$/$\Omega_{K}$ where $\Omega_{K}$ is the local Keplerian frequency.  As the stopping time simply scales as the inverse of the gas density, this means that the variations with height will go as:
\begin{equation}
t_{s} \left( z \right) = t_{s,0}  \mathrm{exp} \left( z^{2} \mathrm{/} H^{2} \right)
\end{equation}
In other words, the stopping time of a particle at $z$=$H$ will be 2.7 times what it is at the disk midplane, at $z$=2$H$, it will be 55 times, and at $z$=3$H$ it will be $\sim$8000 times.  Thus particles will respond differently to gas motions depending on where they are in the column of gas at a given radial location in a protoplanetary disk.

Not only does the response of a particle vary with height, but the particular forces that it will be subjected to will vary with height as well.  For example, gas drag migration of particles arises due to the differential rotation of the gas and solids in a protoplanetary disk.  This differential rotation is due to the fact that the solids feel a central force due to the gravity of the central star whereas the central force that the gas experiences is modified by the radial pressure gradient \citep{adachi76,weidgd77}.  
Around the midplane, the radial pressure gradient is largely negative as disks are expected to have hot, dense gas near the star and cool, sparse gas further out.  However, at higher altitudes, the radial pressure gradient will actually become positive.  This is due to the fact that while the surface density of the disk generally decreases with
distance from the star, its vertical extent grows with distance, except under the most extreme conditions (that is since $H=c$/$\Omega_{K}$, as long as $c$ does not fall off more steeply than $r^{-\frac{3}{2}}$, or $T$ as $r^{-3}$, the scale-height will increase with distance from the star).  \citet{tl02} demonstrated
that for a vertically isothermal disk,  whose surface density and temperature can be described by the
power laws:
\begin{equation}
\Sigma \left(r \right) = \Sigma_{\mathrm{AU}} \left( \frac{r}{\mathrm{AU}} \right)^{p}
\end{equation}
\begin{equation}
T \left(  r \right) = T_{\mathrm{AU}} \left( \frac{r}{\mathrm{AU}} \right)^{q}
\end{equation}
where $r$ is the radial coordinate, and the AU subscripts denote the values of the surface density and
temperature at 1 AU, the radial pressure gradient could be expressed by:
\begin{equation}
\frac{\partial P}{\partial r} = r \Omega_{K}^{2} \rho_{g} \left( \frac{H}{r} \right)^{2} 
\left[ p + \left( \frac{q-3}{2} \right) + \left( \frac{q+3}{2} \right)\frac{z^{2}}{H^{2}}\right]
\end{equation}
With the pressure gradient being negative at the midplane, it switches sign at 
$z$= $\pm \left[ \left( 2p+q-3 \right) \mathrm{/} \left(q+3\right)\right]^{1\mathrm{/}2} H$, meaning
that above this height the gas will have an orbital velocity which exceeds the Keplerian rotation rate.  This produces  a tailwind which pushes materials outwards for the same reasons that materials migrate
inwards due to gas drag around the disk midplane.  Thus the rate and preferred direction of transport via gas drag will vary with height above the disk midplane.

The variations in transport become even more pronounced when one considers the motions of
materials in a viscously evolving disk.  In one-dimensional (1D) disk models of viscous
protoplanetary disks, the evolution of the gas is found by vertically integrating the stresses 
exerted on each annuli of gas in the disk and determining the net radial motions 
\citep{ss73,lbp74}.  In reality, however, the stresses exerted within the disk will be functions of the
local gas properties and will vary with height as these properties change, potentially leading to very different behavior than the net radial motions.  These variations were
investigated in previous studies by \citet{urpin84}, \citet{tl02}, and \citet{kg04} who found analytic
expressions for the variations in the gas motions in the disk with height above the disk midplane.  
These authors found that, in agreement with numerical simulations 
\citep{kleylin92,rozyczka94}, meridonal flows develop around the disk midplane.  That is, in
steady-state disks in which the mass accretion rate of material in the disk is constant in time and 
location in the disk, the net-flux of gas continues to be inward.  However, the inward movement of
material is largely relegated to the gas in the upper part of the disk, whereas the gas around the midplane moved outwards.  This outward movement of material is due to angular momentum 
conservation--that is, as mass is moving inward through a disk, some amount is pushed outwards to
conserve angular momentum.  That this happens around the disk midplane is due to how the mass
of the disk is distributed, in that the gas density has its steepest radial gradient along the disk midplane,
resulting in stresses that drive material outward.

The structure of the disk flow can be found, following \citet{urpin84} ,\citet{tl02} and \citet{kg04}, by solving the azimuthal component of the momentum equation for rotating viscous fluid in cylindrical coordinates:
\begin{equation}
2 \pi r \rho_{g} \left( v_{r} \frac{\partial}{\partial r} + v_{z} \frac{\partial}{\partial z} \right) \left(r^{2} \Omega_{g} \right) =
2 \pi \left[ \frac{\partial}{\partial r} \left( r^{3} \rho_{g} \nu \frac{\partial \Omega_{g}}{\partial r} \right) +
\frac{\partial}{\partial z} \left( r^{3} \rho_{g} \nu \frac{\partial \Omega_{g}}{\partial z} \right) \right]
\end{equation}
simultaneously with the equation for mass conservation:
\begin{equation}
\frac{\partial \rho_{g}}{\partial t} + \frac{1}{r}\frac{\partial}{\partial r} \left(r \rho_{g} v_{r} \right) +
\frac{\partial}{\partial z} \left( \rho_{g} v_{z} \right) = 0
\end{equation}
where $v_{r}$ and $v_{z}$ are the radial and vertical velocities of the gas,  
$\Omega_{g}$ is the orbital frequency of the gas, and $\nu$ is the viscosity of the gas.  

Thus, given a steady-state protoplanetary disk with a given structure, it is possible to calculate the
rates of transport of particles at any location due to the large-scale viscous flows, gas drag migration,
and turbulent motions.  Calculating the dynamical evolution of particles in the disk requires solving 
the two-dimensional advection-diffusion equation in the radial and vertical directions:
\begin{equation}
\frac{\partial C}{\partial t} + \frac{1}{\rho_{g} r} \frac{\partial}{\partial r} \left( r V_{r} \rho_{g} C \right) +
\frac{1}{\rho_{g}} \frac{\partial}{\partial z} \left( V_{z} \rho_{g} C \right) =
 \frac{1}{\rho_{g} r} \frac{\partial}{\partial r} \left( r \rho_{g} \nu \frac{\partial C}{\partial r} \right) +
 \frac{1}{\rho_{g}} \frac{\partial}{\partial z} \left(\rho_{g} \nu \frac{\partial C}{\partial z} \right)
\end{equation}
where $V_{r}$ is the total radial velocity of the dust particles at a given location due to both the 
large-scale flows of the disk and gas-drag migration, $V_{z}$ is the total vertical velocity of the
particles due to both the large-scale flows and vertical settling due to gravity, and $C$ is the
concentration of the particles defined as $\rho_{a}$/$\rho_{g}$ where $\rho_{a}$ is the mass
density of the particles in the disk.  Thus the concentration, $C$, purely reflects the concentration of material relative to the ambient gas and not necessarily relative to other dust that would be present in the disk.

For this study, rather than assuming that the protoplanetary disk mass density and thermal structure can be defined by the power laws in Equations (4,5), a self-consistent solution for the disk structure is found given a particular mass accretion rate.  The process by which the surface density and temperature are found at each location is described below.  The vertical structure of the disk is found by assuming hydrostatic equilibrium.  This is a bit of a simplification as the vertical motions of the gas imply that perfect hydrostatic equilibrium is not completely achieved.  However, because of the relatively small values of the gas vertical velocities, both \citet{tl02} and \citet{kg04} showed that this assumption is accurate to order $H^{2}$/$r^{2}$ (0.1 to 1\%).

\section{Disk Structure}

The structure of the disks in the models considered here were found by assuming that the disk was in steady state, that is, that the accretion rate through the disk $\dot M$, is constant with time and location.  
Under such an assumption, the surface density of the disk and the viscosity at a given location, $r$, are related by \citep{lbp74}:
\begin{equation}
\Sigma \nu = \frac{1}{3 \pi} \dot M \left(1-\sqrt{\frac{r_{\mathrm{in}}}{r}}\right)
\end{equation}
where $\Sigma$ is the local surface density of the disk and $r_{in}$ is the inner edge of the disk.  

The midplane temperature of the disk, $T_{m}$, which helps define the viscosity, is determined by balancing the rates of heating with the rate at which the disk cools from radiation.  Two sources of heating were considered here: that due to viscous dissipation and that due to irradiation from the central star.  The rate of viscous heating in the disk is given by \citep{rp91,step98}:
\begin{equation}
Q_{visc} = \frac{9}{4} \Sigma \nu \Omega_{K}^2
\end{equation}
which under steady-state assumptions becomes:
\begin{equation}
Q_{visc} = \frac{3}{4 \pi} \dot M \Omega_{K}^2  \left(1-\sqrt{\frac{r_{\mathrm{in}}}{r}}\right)
\end{equation}
The heating rate due to irradiation from the star is given by:
\begin{equation}
Q_{irr} = 2 \phi \frac{L_{*}}{4 \pi r^{2}}
\end{equation}
where $\phi$ is the angle at which the photons from the star impact the disk surface and $L_{*}$ is the luminosity of the star at the time of interest \citep{chiang97,brauer08}.  

Given that the cooling rate of the disk is given by:
\begin{equation}
Q_{cool} = 2 \sigma T_{e}^{4}
\end{equation}
where $T_{e}$ is the effective temperature of the disk surface, the effective temperature of the disk due to each individual effect can be calculated as:
\begin{equation}
T_{e,visc} =\left[ \frac{3 G M_{\odot} \dot M}{8 \pi \sigma} \left(1-\sqrt{\frac{r_{\mathrm{in}}}{r}}\right)\right]^{\frac{1}{4}} r^{-\frac{3}{4}}
\end{equation}
\begin{equation}
T_{e,irr} = \left( \frac{ \phi L_{*}}{4 \pi \sigma}\right)^{\frac{1}{4}} r^{-\frac{1}{2}}
\end{equation}
respectively.  These expressions give the effective temperatures of the disk if viscous dissipation or stellar irradiation dominated the heating of the disk respectively.
Here, however, the important quantity for determining the viscosity of the disk is the midplane temperature.  This can be estimated using the expression:
\begin{equation}
T_{m}^{4} = \frac{3}{8} \tau  T_{e,visc}^{4} + T_{e,irr}^4
\end{equation}
where $\tau$ is the optical depth from the disk surface to the disk midplane (C. Dullemond, pers. communication).  This formula accounts for the the fact that the energy generated interior to the disk by viscous dissipation must diffuse to the surface before being radiated away. The optical depth is given by $\kappa \Sigma$/2, where $\kappa$ is the opacity of the disk.  

Given values of $\dot M$, the stellar luminosity, and values for $\kappa$ and $\nu$ everywhere, the surface density of the disk can thus be solved using an iterative approach.  Here, values of $\dot M$=10$^{-6}$, 10$^{-7}$, and 10$^{-8}$ $M_{\odot}$/yr were investigated and are discussed below.  In all cases, the stellar luminosity was given by $L_{*}$=4$\pi R_{*}^{2} \sigma T_{*}^{4}$, with $R_{*}$=3$R_{\odot}$ and $T_{*}$=4000 K as in \citet{rp91} and \citet{cieslacuz06}.  The incidence angle was taken as $\phi$=0.05 for all simulations here, a value typical for disks where $H$/$r$ increases with distance from the central star \citep[e.g.][]{brauer08}.  While the actual value of $\phi$ will vary with location and time in a disk, given that the heating rate goes as $\phi^{\frac{1}{4}}$, slight changes in its value will not produce significant shifts in the final temperature.  A simple opacity law was adopted, where $\kappa$= 5 cm$^{2}$/g everywhere  \citep{cassen94}.  Figure 1 shows the surface density and temperature profiles for the disks explored here.

\section{Numerical Approach}
 
In order to solve for the dynamical evolution of particles in the disk, the structure of the disk is defined
on a two-dimensional, cylindrical ($r-z$) grid.  The grid is logarithmically spaced in both the radial and vertical directions with $r_{i+1}$/$r_{i}$=1.05 and $z_{j+1}$/$z_{j}$=1.1, where $r_{0}$=0.5 AU and
$z_{0}$=8$\times 10^{10}$ cm.  Given the surface density and thermal structures of the disk, the density at each location on the grid is defined based on the assumptions described in the previous section.

The large-scale flows of the gas are calculated using an iterative finite-difference scheme.  The radial
velocity of the disk is first calculated using equation (7) and assuming $v_{z}$=0.  These values of the
radial velocity are then used in equation (8) to determine the values of $v_{z}$.  These updated values
are then applied to equation (7) and then the process repeated until the velocities converge to values that do not change by more than 1\% in subsequent iterations.   An example of the calculated flow structure is shown in Figure 2, which is qualitatively representative of the flow fields throughout the disk for each mass accretion rate shown.

As the particles to be considered here
are largely small with $t_{s}$ much less than an orbital period throughout the
protoplanetary disk, their motions due to the large-scale motions of the gas are assumed to be the
same as that of the gas.
The velocities of the particles due to vertical settling and gas drag migration are calculated as in \citet{tanaka05}.  The net velocities of the dust particles at any point in the disk are thus found by summing those due to the large-scale flows of the disk and those due to gravitational settling and gas drag.  Figure 3 shows the outward flow rates of the gas and the inward migration rates of the $a$=0.5 mm particles at the disk midplane for the different disks considered here.  For comparison, the inward drift rates of the $a$=5 $\mu$m dust would be $\sim$100x smaller at the midplane \citep{cuzzweid06}.

All transport calculations are done using  a \citet{vanleer77} interpolation scheme.  This significantly limits the amount of numerical (or artificial) diffusion that would occur as the calculations are performed.  The calculations are explicit in nature, and thus the timestep is limited to the smallest time it takes for materials to advect or diffuse across a grid cell, thus satisfying  the Courant-Friedrichs-Lewis stability criterian \citep{nr}.

In order to ensure stability in the calculations, all transport in the disk is limited to occur within three
gas pressure scale-heights of the disk midplane.  It is within this range that more than 99\% of the 
mass of the disk is contained, allowing such a consideration to be made without significant loss of accuracy.  To impose this condition, all vertical velocities above $z$=$3H$ at all radial locations are set to be equal to -10 km/s, and the diffusivity of all particles are set to zero.  This forces all particles that are transported to these high altitudes, either through vertical diffusion or through radial transport from an adjacent radial bin which is below the 3$H$ computational roof, to rain back down to the computational grid without any loss of mass from the simulation.  

This approach contrasts with the two-dimensional transport models of \citet{kg04} and \citet{wehrstedt08} which used spherical geometries to study disks that extended 10$^{\circ}$ and 4.01$^{\circ}$ above the disk midplane respecitvely.  While a direct translation into the effective scale-heights that these figures correspond to is difficult as it would depend on the temperature and location in the nebula, they would be roughly $\sim$1-2$H$ for typical nebular parameters.  For comparison, the scale-heights at 1 and 20 AU for the different disk models considered here are listed in Table 2.  These heights correspond to 3-8$^{\circ}$ above the midplane at 1 AU and 4-8$^{\circ}$ at 20 AU.  Thus \citet{kg04} and \citet{wehrstedt08} focused on the region of the disk below roughly one scale height, and did not consider the motions that developed above this height \citep[see, for example, Figures 3 and 4 of][]{wehrstedt08}.  Given the flow structure of the disk, this means that the rapid inward transport of the gas at high altitudes (above $\sim$ 0.35-0.5 AU in Figure 2) are not modeled.  While the gas density (and under well-mixed conditions the solid density as well) at these altitudes is much less than around the midplane, the rapid inward velocities at these altitudes means that the mass flux is still significant.  Neglecting these higher altitudes leads to focusing on the region of the disk where outward transport of materials is more likely, and thus would lead to overestimates of the outward fluxes of materials in a protoplanetary disk.

The computational domain was chosen to ensure that its vertical scale exceeded the computational roof throughout the disk in the model.  While this roof makes the  particular choice of boundary conditions on the upper level unimportant, $\partial C$/$\partial z$ = 0 was used in the actual computation.  Reflective boundary conditions
($\partial C$/$\partial z$ = 0) were imposed at the disk midplane due to the assumed symmetry.
At the inner and outer radial boundaries, $C$=0 was imposed throughout the simulations.  At the inner boundary,
this allowed materials that were migrating inwards to fall onto the star, being lost from the computational grid.  This also meant that particles which were transported beyond the outermost radial grid were
lost from the simulation.  As the radial grid extended to $\sim$80 AU this meant only a minimal amount of material was lost during a given simulation, and as the main area of interest here is for $r <$ 30AU, this is expected to have minimal impact on the results presented here.

A value of 
 $\alpha$=10$^{-3}$ was assumed throughout the disk, with the viscosity given by $\nu$=$\alpha c H$, and the diffusivity of solids equal to $D$ = $\nu$/(1+$t_{s}^{2} \Omega_{K}^{2}$) \citep{yl07}.  In adopting this value of $\alpha$, it is explicitly assumed that it is constant with time and location.  However, if the MRI were responsible for driving disk evolution, it would operate most effectively in the regions of the disk in which ions were abundant enough for the disk to couple to magnetic fields.  This would only be achieved in parts of the disk where significant ionization had occurred, meaning  that parts of the disk would be located in the so-called ``Dead Zone''  \citep{gammie96,glassgold97}.  As the MRI would not operate here, it was thought such a region would not be subjected to the stresses that drive mass and angular transport in protoplanetary disks, which would thus result in a different evolution than that described here.
However, \citet{flemstone03} showed that overshoots of turbulent eddies from an MRI-active region could ``stir up'' stresses and turbulence in a Dead Zone, meaning that region would still be dynamically active.  Further, recent work by \citet{turnersano08} showed that viscous stresses could be produce in a neutral region of the disk, resulting in laminar (or less turbulent) accretion flows.  This means flow structures similar to those studied in this work could still develop.   The detailed impact of variations in $\alpha$ will be evaluated in future work.

\section{Model Results}

Cases of both outward and inward transport were examined.  The outward transport of particles from the inner nebula was considered by looking at how materials formed at temperatures in excess of 1100 K were redistributed throughout the nebula.  This temperature was chosen so that this simulation would be analogous to looking at how amorphous precursors could have been annealed in the hot inner disk and then transported outward, potentially to where comets formed in our own solar nebula \citep{nuth00,gail01,bockelee02,cieslasci07}, or to the large distances from the central star that they observed around other young stars \citep{vanboekel04,apai05,watson07}.  A higher temperature may be necessary if crystalline grains were formed purely from condensation \citep[][and references therein]{wooden07}.  This would likely lead to slightly lower concentrations of materials in the outer disk as a smaller volume of the disk would be able to produce such high temperatures.  However, given the steep thermal profiles in the disk (Figure 1), the difference in volume would not be large, so the change in concentrations would likely not be significant.

For these outward transport calculations, a similar approach as \citet{kg04} is adopted where the source of the materials is defined to be constant throughout the simulation ($C=C_{0}$) where $T >$ 1100 K and where $z < H$.  This is similar to the approach used in \citet{kg04},  however, those authors considered the species as purely a vapor, so vertical settling or gas drag migration did not operate on their tracers.

Inward transport of dust is also considered where materials that originate in the outer disk are
tracked as they are redistributed throughout the protoplanetary disk.  In this case, $C=C_{0}$ at
the beginning of the simulation, but then the concentration evolves dynamically rather than being
constant throughout the simulation.  Such a situation would be akin to following the motions of short-lived radionuclides that are injected into the solar nebula from a nearby supernova
\citep{ouellette07},  materials with anomalous oxygen isotope ratios produced by CO self-shielding in the disk \citep{ly05}, or material that rains onto the nebula after potentially being launched from an X-wind \citep{shu96}.
It should be noted that while this consideration is labeled ``Inward Transport," outward transport of
this material also occurs. It is so named because the general interest  in such material or the natural application of these results has to do with
how these materials would be mixed into the meteorite parent bodies and planets that form in the inner solar system.  In the case of the ``Inward Transport'' calculations, the starting material originates in a radial span of $\sim$10 AU, centered at 
$\sim$25 AU, and at a vertical span of one scale-height ($H$), centered at 2$H$ at each radial location.

In the case of the 10$^{-6}$ $M_{\odot}$/yr disks, the models were run to a model time of 10$^{5}$ years, during which 0.1 $M_{\odot}$ of material would have been accreted onto the star.  As this corresponds to roughly the mass expected for a typical protoplanetary disk around a solar mass star, this represented a logical stopping point.  Similarly for the 10$^{-7}$ $M_{\odot}$/yr disks, models were run to a model time of 10$^{6}$ years for the same reason.  The 10$^{-8}$ $M_{\odot}$/yr were also modeled for 10$^{6}$ years (representing a period when 0.01 $M_{\odot}$ was accreted by the central star) as typical protoplanetary disk lifetimes are on this order \citep{haisch01}.  Also, in the case of our solar nebula, giant planet formation would likely have been at an advanced stage at that point \citep{hubickyj05}, meaning that the assumptions of axial symmetry and smooth surface density profiles likely would no longer hold for longer times.

\subsection{Outward Transport}

\subsubsection{$\dot M$=10$^{-6}$ $M_{\odot}$/yr}

Figure 4 shows the temporal evolution of the concentration of solids formed at $T>$1100 K for the case of $\dot M$=10$^{-6}$ $M_{\odot}$/yr and
$a$=5 $\mu$m. Each contour in the plot represents a change in concentration by a factor of 2 compared to neighboring contours, with bright contours representing high concentrations and lower concentrations becoming progressively darker (the lowest contour has a value of 0.001 of $C_{0}$ and the highest contour has a value of 0.512 of $C_{0}$).   With time, it is seen that outward transport
does occur, with the concentration at the midplane exceeding the value at higher altitudes as 
reported by \citet{cieslasci07}.  This is due to outward transport being favored along the
midplane as this is the location of the outward, meridonal flows in the disk.  In these flows, solids  diffusing outwards are moving with the flow of the gas, rather than against it as would be the case in the accretional (towards the star) flows at higher altitudes or the net inward flows considered in 1D models.  

Little difference is seen in these results and those where the outward transport of solids with $a$=0.5 mm in a disk with $\dot M$=10$^{-6}$ $M_{\odot}$/yr (Figure 5), despite these latter particles being two orders of magnitude larger in size (and six orders of magnitude larger in mass).  This is due to the high temperatures and high gas densities that are present in a young protoplanetary disk which can accrete at a rate of $\dot M$=10$^{-6}$ $M_{\odot}$/yr.  Both of these factors contribute to reducing the stopping time of the dust particles and making them very small compared to the orbital timescales of the nebula at all heights.   As a result, the two particle sizes have very similar dynamic responses due to their interactions with the gas, meaning the distributions of solids at the end of the simulations are nearly the same.  Thus, in both cases particles are carried well beyond 20 AU after 10$^{5}$ years of evolution.

\subsubsection{$\dot M$=10$^{-7}$ $M_{\odot}$/yr}

 Particles are not carried outwards as rapidly in the case of $\dot M$=10$^{-7}$ $M_{\odot}$/yr (Figures 6 and 7) as the outward flows are lower in magnitude at the midplane and because the 1100 K isotherm is located closer to the star.  However, even after just 10$^{5}$ years of evolution, materials from the hottest regions of the nebula still are transported beyond 20 AU, resulting in concentrations of a few percent in the region where comets like Wild 2 would have originated \citep{charnoz07}.  Here, though, because the disk is less massive and cooler, contrasts in the dynamics of the different sized particles begin to be more readily seen.  Firstly, the larger particles are not lofted as high as the smaller dust particles as shown at the 100,000 year mark, where outside of 15 AU small dust particles are present 4 AU from the disk midplane, whereas the greater settling velocities of $a$=0.5 mm particles keep them more concentrated around the midplane.  As the larger particles are kept at lower altitudes as compared to the smaller dust, they spend longer periods of time in the regions where outward transport occurs more efficiently.  As a result, the concentration of the $a$=0.5 mm particles around the midplane  in the outer disk slightly exceeds that of the $a$=5 $\mu$m particles.  This effect is most noticeable in this region because the lower gas densities and the lower temperatures produce more significant differences in the stopping times of the particles than would be found in the hotter, denser, inner nebula.

\subsubsection{$\dot M$=10$^{-8}$ $M_{\odot}$/yr}

Figures 8 and 9 show the results for the $\dot M$=10$^{-8}$ $M_{\odot}$/yr cases.  Again, the drop in accretion rate results in lower outward flux of high temperature materials due to the lower velocities at the disk midplane and due to the fact that high temperatures are limited to the very inner edge of the nebula.  This decrease in outward flux means that particles of either size make it to just beyond 10 AU in 100,000 years, rather than beyond 20 AU as in the higher mass accretion rate cases.  Again the lower temperatures in the disk and lower gas densities result in greater differences in the dynamics of the two sizes of particles.  In this case, even in the inner disk, the $a$=0.5 mm particles are not lofted to the same heights as the $a$=5 $\mu$m dust.  As before, this also results in higher concentrations of the larger particles in the outer parts of the disk (at 10 AU in the 100,000 year mark, for example).  This is a result that is counter to what would be found in the 1D models typically used in the past, as the greater inward velocities of the larger particles due to gas drag and lower diffusivities of the particles would hinder their outward transport as compared to their smaller counterparts.

\subsection{Inward Transport}

\subsubsection{$\dot M$=10$^{-6}$ $M_{\odot}$/yr}

Here the ``Inward Transport'' cases are described, where the dynamics of particles originating at radial distances of 20-30 AU and 1.5-2.5 scale heights above the disk midplane are calculated.
Figure 10 shows the results of  the case of $a$=5 $\mu$m and  $\dot M$=10$^{-6}$ $M_{\odot}$/yr.  The contours
in these figures are the same as those used in the previous section.  As expected, these
materials are redistributed throughout the disk, with inward transport occurring most rapidly along
the disk surface, where the inward flows of gas due to viscous evolution are the greatest.   Thus immediately after these grains are ``injected'' or ``created'' (depending on what kind of materials they represent) in the disk, there are strong vertical gradients in the concentration of these species.  These gradients persist throughout the simulation, even as diffusion works to smooth them out.  This is due to the fact that while dust  mixes down to the midplane due to gravity and diffusion, this process lags
behind the rapid inward motions of the grains at higher altitudes.

Gravitational settling occurs more rapidly for the 0.5 mm particles as shown in Figure 11.  While vertical gradients similar to the type seen in the $a$=5 $\mu$m case are present during the early part of the simulation, after $\sim$50,000 years, the 0.5 mm particles cluster around the midplane where their vertical distribution approaches the canonical steady state situation--with vertical diffusion balancing vertical settling. (The steady state is approached but not fully reached because the vertical adjustment is not instantaneous, and in the presence of radial transport, the conditions throughout a vertical column are constantly changing.) Thus, vertical gradients are also present, but with the highest concentrations of 0.5 mm particles being achieved at the disk midplane and decreasing with height, as opposed to the other way around for the smaller particles.

That the $a$=5 $\mu$m grains remain lofted at high altitudes, and thus are entrained in the rapid inward flows of the disk, for longer periods of time make them easier to deliver to the inner disk.  This is shown in detail in the different panels in Figure 12.  Panels A and B show the concentrations of the 5 $\mu$m and 0.5 mm solids in the inner disk at the end of the simulation respectively.  It can be seen that the contours extend to smaller heliocentric distances in the case of the smaller dust particles than the larger dust particles.  This is reinforced in the plot in panel C, which shows the concentrations of both sized particles along the disk midplane at the end of the simulation.  Inside of 10 AU, the concentration of the smaller particles exceeds that of the larger ones everywhere despite the more rapid settling velocities of the larger particles.  This also means that the small particles are accreted onto the star, and thus lost from the disk, more readily than the larger particles.  This is shown in panel D, where the fraction of the injected material remaining in the disk is plotted as a function of time.  The fraction of small dust decreases more rapidly than the larger dust particles.  While the fraction remaining in the disk at the end of the simulation is still large ($\sim$95\%), this reflects the fact that the materials of interest started so far from the inner edge of the disk ($>$20 AU).  Had this material instead been injected at 10 AU, the fraction lost, and discrepancy between the two different sized particles would likely have been larger. 

\subsubsection{$\dot M$=10$^{-7}$ $M_{\odot}$/yr}

Figure 13 shows how the small dust grains are transported over time in the $\dot M$=10$^{-7}$ $M_{\odot}$/yr disk.  The overall evolution of the solids again matches general expectations in that the solids begin to settle out of the surface layers, while being redistributed radially due to the large-scale flows associated with viscous evolution and the diffusion caused by the turbulence.  While the particles do get redistributed throughout the disk, with outward transport carrying them all the way to the outer boundary of the disk, their motions are largely inward, as is expected given the net movement of disk materials toward the star under viscous evolution.  Thus after 1 million years of evolution, the particles are concentrated to the highest levels inside of 10 AU.

Note again that vertical gradients persist immediately after injection onto the surface of the outer disk, though their magnitude at 100,000 years is not as great as in the case of $\dot M$=10$^{-6}$ $M_{\odot}$/yr.  The vertical diffusion timescale for these particles, say to make it down to the midplane from where they were injected, is (2$H$)$^{2}$/$\nu$=4$H^{2}$/$\alpha c H$=4($\alpha \Omega_{K}$)$^{-1}$, meaning it is independent of the disk structure or mass accretion rate.  The radial flows in a disk, however, decrease in magnitude as the mass accretion rates drop, meaning that the conditions in a column do not vary as rapidly, allowing vertical diffusion to smooth out gradients more effectively.

Figure 14 shows the results for the $\dot M$=10$^{-7}$ $M_{\odot}$ disk, but with the larger particles.  Again, the large-scale flows and turbulence redistribute the particles throughout the disk.  However, because of the lower densities and temperatures in this disk as opposed to the $\dot M$=10$^{-6}$ $M_{\odot}$/yr case, the larger particles settle towards the midplane much more rapidly, resulting in very different dynamical evolution when compared to the smaller particles. The more rapid settling allows larger particles to quickly cluster in the region of the nebula where outward transport is favored (around the midplane), slowing their delivery to the inner disk and resulting in greater mass fluxes to the outer disk.  This is more clearly seen in Figure 15, in which the concentrations of the two different species in the inner disk are shown in the top panels at the end of the simulation.  There are not significant differences in terms of the concentrations in the inner part of the disk (panels A and B), however, panel C shows that the concentration of the larger species at the midplane well beyond 40 AU exceeds the concentration of the smaller dust particles.  This is in part due to the greater settling rates of the larger particles, but also due to the fact that the outward flows in this region preserve the larger particles from migrating to the inner edge of the disk as rapidly as the small particles.  This is reinforced in panel D, which again shows that the smaller particles are lost from the disk more rapidly than the larger particles, with the differences being larger here than in the $\dot M$=10$^{-6}$ $M_{\odot}$/yr case.

\subsubsection{$\dot M$=10$^{-8}$ $M_{\odot}$/yr}

Figures 16 through 18 show the results for a simulation with $\dot M$=10$^{-8}$ $M_{\odot}$/yr.  Again, this disk has lower densities and temperatures than those disks considered above.  As a result, the large-scale flows have decreased in magnitude, leading to lower levels of radial transport.  Vertical gradients in the concentration of materials are minor, with the small particles becoming well mixed in the vertical direction and the large particles clustering around the midplane.  The outward flows around the midplane decrease but are still present; however the lower velocities mean that they are not as effective at slowing the inward movement of the larger particles due to diffusion or gas drag.  As a result, there is little difference between the delivery of the materials to the inner disk.  Panel C of Figure 18 shows that the concentration of the large particles at the disk midplane exceeds that of the small particles inside of $\sim$40 AU.  This is largely due to the greater settling efficiency of the larger particles and not due to more rapid inward transport of the small particles in this case.  That the smaller particles have greater abundances at the midplane beyond 40 AU is due to the fact that gas drag migration serves to prevent the larger particles from migrating outwards to the same levels as in the previous cases.  And while the greater settling efficiency of the large particles removes them from the rapid inward flows that occur along the disk surface, the difference in the fraction of material lost to the star due to the accretion process is minor for the different sized particles (Panel D), though again the larger particles are still retained more efficiently than the small particles.


\section{Summary and Discussion}

The results of the different models presented here demonstrate that the vertical variations in the rates and preferred directions of radial transport of solids in a protoplanetary disk are important to consider when talking about the early evolution of primitive materials.  These variations arise due to the fact that the gas densities, pressure, and their respective gradients, all of which are responsible for driving motions in a protoplanetary disk, are strong functions of location.  In terms of the viscous evolution of the disk, the vertical variations in gas density are responsible for driving a flow structure in which the gas moves toward the star rapidly at high altitudes in the disk, but moves outward, away from the star around the disk midplane.  The inward motions of the gas are responsible for the mass accretion onto the central star, representing its final stages of pre-main sequence growth, whereas the outward motions are responsible for the outward transport of angular momentum to compensate.  The relative fraction of material transported outward at a given location can range from $\sim$5 to 30\% of the mass that is transported inward at a given time \citep{kg04}.

Further, the gas drag induced velocities of solids in the solar nebula will also vary with height as the radial pressure gradient, which is responsible for the gas rotating at a non-Keplerian rate, also varies with height.  Unlike the large-scale flows caused by viscous evolution, however, the gas drag induced velocities will reach their largest, negative (towards the central star) velocities near the disk midplane.  At
higher altitudes, the inward velocities diminish, and in fact, will become positive as the radial pressure gradient switches signs, leading the gas to orbit at super-Keplerian rates at higher altitudes.  

These variations in the radial transport models are overlooked in traditional one-dimensional models of material transport in protoplanetary disks \citep[e.g.][]{stepval97,gail01,bockelee02,bitw03,cieslacuz06}.  Instead,  the motions of the particles due to the viscous evolution of the disk in these models are found by assuming that those particles are uniformly suspended throughout the height of the disk (thus ignoring vertical settling) and their net velocity is found by calculating the vertically averaged motions of the gas at that location ($V_{r}$=-3$\nu$/2$r$ under steady-state condtions).  For those 1D models that account for the radial motions due to gas drag \citep[e.g.][]{stepval97,bitw03,cieslacuz06}, these velocities are generally found by calculating the inward velocity at the disk midplane and applying that velocity to all particles, regardless of where they are located in the vertical column above the disk midplane.  While this may be appropriate for particles which are tightly clustered around the disk midplane due to vertical settling ($\sim$meter-sized bodies which do not experience significant vertical diffusion, for example), this would overestimate the inward gas drag velocities of particles lofted to higher altitudes.  Thus while 1D models are instructive for investigating the general dynamical evolution of materials in protoplanetary disks, consideration of the 2D effects can lead to important differences in the dynamical evolution of primitive materials.

Here, as in \citet{cieslasci07}, it was shown that the outward flows that are generated around the midplane in a viscously evolving disk aid in the outward transport of materials
by providing an environment in which materials can diffuse outwards without being hindered by the
accretional flows associated with protoplanetary disk evolution.  This provides a path by which high temperature materials (refractory inclusions and crystalline grains) collected from comet Wild 2 or seen in comet Halley and Tempel 1 could be delivered to the outer solar nebula.  Predicting the particular fraction of silicate grains in the outer disk  that would be crystalline in each of the models presented here is not straightforward, unfortunately.  This fraction depends on the amount of material that is delivered to the outer disk as well as the amount of unprocessed material that is present.  The abundance of this material will be determined by its dynamical interaction with the gas in the outer disk, the extent of the disk (the initial inventory of unprocessed materials) and the rate at which these unprocessed materials coagulate into larger bodies.  A first order approximation, however, would be that the total processed and unprocessed silicate grain abundance would add up to the solar ratio ($C_{crystal}+C_{amorphous} \sim$ 0.005).  This would suggest that the crystalline grain fraction at 10 AU at the end of each simulation would be $\sim$40\% for each of the $\dot M=10^{-6} M_{\odot}$/yr cases, $\sim$10\% for the $\dot M=10^{-7} M_{\odot}$/yr cases, and $<1\%$ for the $\dot M=10^{-8} M_{\odot}$/yr cases.  These numbers should only be taken as rough estimates, however, and future work should follow both the unprocessed and processed materials simultaneously.

These numbers are generally comparable to those found by \citet{wehrstedt08} in their two-dimensional models, though direct comparisons are not straightforward due to different model treatments.   For example, \citet{wehrstedt08} assumed that grains could become crystalline at temperatures as low as 800 K, allowing for a larger volume of grains to be processed at any given time.  More importantly, however, as discussed above, these authors focused on the dynamics of dust particles in the region around the midplane, rather than considering the dynamics of solid particles at very high altitudes.  These higher altitudes still contribute significantly to the inward mass flux in the disk.  Thus these models likely overestimate the outward flux of high temperature materials and underestimate the inward flux of unprocessed materials.  It should be noted that the differences due to the particular choice of viscosity model ($\alpha$-viscosity in this work, $\beta$-viscosity in \citet{wehrstedt08}) needs to also be quantified in future studies.

Not only does the rate of mass accretion in the disk determine the amount of outward transport that occurs in a disk, but the timing of this transport as well.
For disks with mass accretion rates in excess of $\dot M >$10$^{-7} M_{\odot}$/yr,  high-temperature materials can be transported beyond 20 AU on timescales of $\sim$10$^{5}$ years.  Such rapid outward transport is likely
necessary given that  the formation timescale of the giant planets via core accretion is 
approximately 1-2 Myr \citep{hubickyj05}.  Delivery of high temperature materials on longer timescales,
such as a few million years, would be problematic as the outward transport of
the grains would likely be blocked by the young giant planets, either because the cores would accrete
the grains themselves or because the planets would begin to open a gap in the disk through which particle transfer would likely be difficult.  

At lower mass accretion rates, with  $\dot M \sim$10$^{-8} M_{\odot}$/yr, typical of 10$^{6}$ year old T-Tauri stars \citep{hartmann98}, delivery of crystalline silicates or CAIs to the outer disk is difficult as high temperatures are generated only at the very inner edge of the disk.  The lower outward flow velocities along the disk midplane and the smaller volume of the disk in which high temperature materials can be formed or processed means that delivery of crystalline grains and refractory inclusions to beyond 20 AU will not occur as quickly as in the higher mass accretion rate models.  Thus it is likely that the high temperature material seen in comets were processed and delivered to the outer solar nebula during the early stages of disk evolution.

However, even at lower mass accretion rates, such as $\dot M \sim$10$^{-8} M_{\odot}$/yr, the flows at the disk midplane are likely still important in terms of understanding the dynamical evolution of sub-cm sized objects, as the outward flows velocities in the inner disk are $>$10 cm/s in the model presented here.  For the $a$=0.5 mm particles (which is roughly the size of chondrules) considered here, this means that the outward flows at the disk midplane exceed the inward drift velocities at the disk midplane that arise due to gas drag (Figure 3) out to about 20 AU.  In fact, given that inward drift velocity for small particles is roughly proportional to its stopping time \citep{cuzzweid06} the outward flows at the disk midplane would exceed the inward drift velocities of objects with radii of 5 mm (comparable to the larger CAIs found in CV chondrites) out to a few AU, offering a way of preserving these objects in the nebula for millions of years before the formation of chondrules and accretion of chondrites \citep{amelin02,kita05}.

The variations in flow structure also are important to consider when discussing the inward transport of materials from the outer parts of the protoplanetary disk.  Materials that would be injected into the solar nebula, rain down after being lofted upward in a protostellar jet, or are created due to photochemistry on the disk surface would initially be subjected to rapid inward flows in the very upper regions of the disk.  This would remain true until the particles are transported down towards the midplane where the inward flows diminish in magnitude or the flows reverse direction.  This removal from the upper part of the disk occurs more rapidly for larger particles due to their greater settling velocities, meaning that smaller particles can be transported to the inner disk on shorter timescales than larger ones.  In fact, because smaller particles can remain lofted at high altitudes for long periods of time, they are more likely to be accreted onto the central star in accretionary flows as compared to the larger bodies that settle around the disk midplane as demonstrated in the simulations presented above.  This remains true as long as the velocities due to viscous evolution of the disk exceed those of that arise due to gas drag.
This result is not limited to materials that begin at high altitudes.  Vertical diffusion would loft small particles to higher altitudes than larger ones, thus introducing them to the regions of rapid inward flow more readily.   This runs counter to the previous ideas that millimeter to centimeter sized objects would be more difficult than fine dust to retain in the solar nebula due to their inward drift velocities caused by gas drag.  Instead, it appears that aggretation into bodies of this size may actually have allowed them to be retained in the solar nebula more effectively.

It should be noted that millimeter-sized grains themselves do not necessarily serve as good analogs for the materials that would be injected into the solar nebula by a supernova \citep{ouellette07} or for
the heavy ($^{16}$O-poor) water ice that would be created by CO self-shielding in the outer nebula \citep{ly05}, as these
materials are likely contained within or adsorbed by grains much smaller and closer to the $a$=5 $\mu$m cases discussed above. However, 
it should be kept in mind that these smaller grains would likely have experienced collisions with other grains that led to sticking and growth.  \citet{weidenschilling97} showed that micron-sized grains would coagulate into millimeter-sized grains at 30 AU on timescales of less than 10,000 years.  Those coagulation calculations assumed a non-turbulent and non-evolving disk; growth at these small sizes may actually be accelerated by turbulence as the random motions of the dust particles would lead to more frequent collisions \citep{weidenschilling84,dd05,ccoag07,ormcuzzi07}.  Growth to very large sizes may be inhibited due to turbulence, as larger particles will develop relative velocities that become destructive in collisions, rather than accretional, but millimeter-sized grains appear to be below this threshold \citep[see discussion in][]{cuzzweid06}.  Thus, the results of the $a$=0.5 mm models are likely more reflective of how solids from the upper layers of the outer disk get transported to the inner disk.

The issue of retention is particularly important for the case of injection of short-lived radionuclides directly into the solar nebula \citep{ouellette07}.   If this material was injected, it would likely have been done in one or more Rayleigh-Taylor fingers \citep{boss07}, with materials being peppered along the surface of the nebula as assumed here.  \citet{ouellette07} demonstrated the ratios of $^{26}$Al/$^{27}$Al and $^{60}$Fe/$^{56}$Fe would be close to the inferred meteoritic values if the solar nebula were located 0.15 pc from a supernova.  These calculations, however, assumed that these materials became well mixed in the nebula.  However, as shown here, injected material that was present as small grains ($<$10 $\mu$m, as expected for the supernova grains) at the surface of the nebula would be lost to the Sun faster than larger solids, which would have settled around the midplane.  Thus to ensure that these injected materials become well mixed in the solids, they would quickly have to be accreted into the larger bodies that had already formed.  If they were not, the injected material would be lost at a faster rate than the native solids, thus leading to lower ratios of the abundance of short-lived radionuclides to stable isotopes.  This would be particularly true if solids were injected closer to the Sun than the $\sim$20 AU distance considered here as accretion onto the sun would happen on shorter timescales.  This would require the solar nebula to be closer to the supernova to achieve the meteoritic ratios.  Such considerations should be made in future studies.

An important point to note in each of these simulations is that the concentration at the end of each of the ``Inward Transport'' simulations shown here is not uniform throughout the disk, but rather, gradients exist.  This is particularly seen in panel C of Figures 12, 15, and 18, which show that the concentrations of the species at the disk midplanes vary with distance from the star.  Radial mixing timescales for cases such as these are typically defined as the diffusive timescale, $t_{mix}$=$r^{2}$/$\nu$,  which for the inward transport calculations presented here would be 6$\times$10$^{5}$, 1.5$\times$10$^{6}$, and 2.5$\times$10$^{6}$ years at 20 AU for each of the disks considered.  These numbers represent estimates for the amount of time it would take for the ``injected'' material tracked here to mix into the very inner part of the protoplanetary disk.  While the model simulations are not run to these times for the cases considered here, it can still be seen that delivery of significant amounts of material occurred on shorter timescales.  However, complete homogenization (little to no concentration gradients) of the injected material had not been achieved at the end of the simulations, nor is it expected to be achieved at any reasonable time afterward.  The evidence for this is that gradients exist at every point in the disk, regardless of the distance from where the materials originated.  Because of the variation in the direction and magnitudes of the large-scale flows in the disk, conditions are constantly changing, not allowing diffusion to work to smooth out the concentration gradients completely.

 The agreement between Pb-Pb ages and Al-Mg ages of CAIs and chondrules \citep{amelin02} suggests that $^{26}$Al was uniformly distributed relative to the stable isotope $^{27}$Al throughout the inner solar nebula where these materials formed, implying that if injected, $^{26}$Al was homogeneously distributed throughout the inner solar system where these objects formed.  
Such apparent homogenization in the injection model could be achieved if mixing had occurred in such a way that these objects formed near the concentration maximum, where the radial gradients in concentration are the lowest (but non-zero). 
Alternatively, near homogenization of injected materials could have been achieved if the materials were contained in small dust grains and they were injected at a late stage of nebular evolution.  Radial concentration gradients are smallest in the cases presented for the 5 $\mu$m grains in the disk accreting at 10$^{-8} M_{\odot}$/year, as the large-scale flows slowed to the point that diffusion became more effective at smoothing out gradients.  However, this low mass accretion rate also would not produce very high temperatures in the disk, meaning that CAIs are unlikely to form or their formation and redistribution was limited to only regions very close to the Sun.  Further, the small dust grains would likely have coagulated into larger objects, such as the 0.5 mm grains considered here, which show comparable or stronger radial gradients in all cases considered.   

Therefore, if $^{26}$Al was injected directly into the nebula, it would likely not have been perfectly homogenized, but rather gradients, or variations in its concentration, would have existed.  
These variations would lead to apparent age differences between chondritic components, the magnitude of which depended on the particular values of $^{26}$Al/$^{27}$Al that were recorded by those components.  For example, a 10\% variation in the $^{26}$Al/$^{27}$Al  ratio could easily develop over a scale of a few AU based on the results of the simulations presented here, and also is in line with the variations found by \citet{boss08} for materials injected into a marginally gravitationally unstable disk.  If two solids formed contemporaneously, one with $^{26}$Al/$^{27}$Al =5$\times$10$^{-5}$, the canonical value within CAIs \citep[see][for the most recent discussion on this issue]{jacobsen08}, and one with $^{26}$Al/$^{27}$Al =4.5$\times$10$^{-5}$, this would be interpreted as an age difference of $\sim$100,000 years, which is less than the uncertainty on the Pb-Pb ages.   If this were the case, and $^{26}$Al was injected directly into the solar nebula, the inherent uncertainty in Al-Mg ages would be at least $\sim$100,000 years.

Recently, measurements of $^{60}$Fe, whose presence in solar system materials \citep[e.g.][]{tachibana03} has been among the motivating factors for looking at the direct injection of short-lived nuclides into the young solar system from a supernova, have helped to shed light on how well homogenized potential supernova derived materials were in the solar nebula.
\citet{bizzarro07} originally  reported measurements of $^{60}$Ni deficits in meteorites and terrestrial samples  that indicated $^{60}$Fe was not present in planetesimals that accreted within the first 10$^{6}$ years of the formation of the solar system.  This finding supported the model of direct injection into the solar nebula by a supernova \citep{ouellette07}.  In this scenario, the 10$^{6}$ year interval represented the time it took for $^{60}$Fe to be injected and mixed in to the inner solar nebula.  More recently however, \citet{dauphas08} analyzed similar samples as \citet{bizzarro07}, but found that  $^{60}$Fe was homogenized in the terrestrial planet region of the solar nebula such that its concentration varied by less than 10\%.  Again, these results could be achieved in the injection model if the maximum in the concentration of $^{60}$Fe had migrated to the inner solar system at the time that planetesimals there began to form.  If this were not the case, or if the variation in $^{60}$Fe is much less than 10\% in planetary materials, the origin of short-lived radionuclides may be best understood as resulting from the injection into the natal molecular cloud prior to (or as the cause of) collapse \citep{boss08sn}.

The results presented here differ somewhat from those of \citet{boss08} who showed that the concentration of ``dye'' that was injected to a marginally gravitationally unstable disk would be mixed to the point that there would be less than 10\% variation in its concentration throughout the \emph{entire} disk after $\sim$1000 years.  This may represent a fundamental difference in how transport occurs in gravitationally unstable disks and viscous disks, or it may represent differences in the ways that calculations were performed.  For example, while the study described here tracked the evolution of solids, the dye in \citet{boss08} was more akin to vapor in the disk, meaning that the dynamics of vertical settling and gas drag were not considered.  While this may approximate the dynamical evolution of small dust particles which are well coupled to the gas, as described by \citet{hagboss03}, these processes are likely important as solids grow beyond a few microns in size.  Future work aimed at determining the differences in how solids evolve in marginally gravitationally unstable and viscous disks will help to identify which model better describes the early evolution of our solar nebula.

\emph{Acknowledgments}  I am grateful to Dr. Joseph Nuth and an anonymous reviewer who made comments and suggestions that resulted in a greatly improved manuscript.  This work was supported by NASA grant NNX08AY47G.

\bibliographystyle{natbib}

\begin{thebibliography}{}

\bibitem[{Adachi} {\em et~al.}(1976)]{adachi76}
{Adachi}, I., {Hayashi}, C., and {Nakazawa}, K. (1976).
\newblock {The gas drag effect on the elliptical motion of a solid body in the
  primordial solar nebula.}
\newblock {\em Progress of Theoretical Physics}, {\bf 56}, 1756--1771.

\bibitem[{Amelin} {\em et~al.}(2002)]{amelin02}
{Amelin}, Y., {Krot}, A.~N., {Hutcheon}, I.~D., and {Ulyanov}, A.~A. (2002).
\newblock {Lead Isotopic Ages of Chondrules and Calcium-Aluminum-Rich
  Inclusions}.
\newblock {\em Science}, {\bf 297}, 1678--1683.

\bibitem[{Apai} {\em et~al.}(2005)]{apai05}
{Apai}, D., {Pascucci}, I., {Bouwman}, J., {Natta}, A., {Henning}, T., and
  {Dullemond}, C.~P. (2005).
\newblock {The Onset of Planet Formation in Brown Dwarf Disks}.
\newblock {\em Science}, {\bf 310}, 834--836.

\bibitem[{Balbus} and {Hawley}(1991)]{bh91}
{Balbus}, S.~A. and {Hawley}, J.~F. (1991).
\newblock {A powerful local shear instability in weakly magnetized disks. I -
  Linear analysis. II - Nonlinear evolution}.
\newblock {\em Astrophysical Journal}, {\bf 376}, 214--233.

\bibitem[{Bizzarro} {\em et~al.}(2007)]{bizzarro07}
{Bizzarro}, M., {Ulfbeck}, D., {Trinquier}, A., {Thrane}, K., {Connelly},
  J.~N., and {Meyer}, B.~S. (2007).
\newblock {Evidence for a Late Supernova Injection of $^{60}$Fe into the
  Protoplanetary Disk}.
\newblock {\em Science}, {\bf 316}, 1178--1180.

\bibitem[{Bockel{\'e}e-Morvan} {\em et~al.}(2002)]{bockelee02}
{Bockel{\'e}e-Morvan}, D., {Gautier}, D., {Hersant}, F., {Hur{\'e}}, J.-M., and
  {Robert}, F. (2002).
\newblock {Turbulent radial mixing in the solar nebula as the source of
  crystalline silicates in comets.}
\newblock {\em Astronomy \& Astrophysics}, {\bf 384}, 1107--1118.

\bibitem[{Boley} {\em et~al.}(2006)]{boley06}
{Boley}, A.~C., {Mej{\'{\i}}a}, A.~C., {Durisen}, R.~H., {Cai}, K., {Pickett},
  M.~K., and {D'Alessio}, P. (2006).
\newblock {The Thermal Regulation of Gravitational Instabilities in
  Protoplanetary Disks. III. Simulations with Radiative Cooling and Realistic
  Opacities}.
\newblock {\em Astrophysical Journal}, {\bf 651}, 517--534.

\bibitem[{Boss}(2002)]{boss02}
{Boss}, A.~P. (2002).
\newblock {Evolution of the Solar Nebula. V. Disk Instabilities with Varied
  Thermodynamics}.
\newblock {\em Astrophysical Journal}, {\bf 576}, 462--472.

\bibitem[{Boss}(2007)]{boss07}
{Boss}, A.~P. (2007).
\newblock {Evolution of the Solar Nebula. VIII. Spatial and Temporal
  Heterogeneity of Short-lived Radioisotopes and Stable Oxygen Isotopes}.
\newblock {\em Astrophysical Journal}, {\bf 660}, 1707--1714.

\bibitem[{Boss}(2008)]{boss08}
{Boss}, A.~P. (2008).
\newblock {Mixing in the solar nebula: Implications for isotopic heterogeneity
  and large-scale transport of refractory grains}.
\newblock {\em Earth and Planetary Science Letters}, {\bf 268}, 102--109.

\bibitem[{Boss} {\em et~al.}(2008)]{boss08sn}
{Boss}, A.~P., {Ipatov}, S.~I., {Keiser}, S.~A., {Myhill}, E.~A., and
  {Vanhala}, H.~A.~T. (2008).
\newblock {Simultaneous Triggered Collapse of the Presolar Dense Cloud Core and
  Injection of Short-Lived Radioisotopes by a Supernova Shock Wave}.
\newblock {\em Astrophysical Journal}, {\bf 686}, L119--L122.

\bibitem[{Brauer} {\em et~al.}(2008)]{brauer08}
{Brauer}, F., {Dullemond}, C.~P., and {Henning}, T. (2008).
\newblock {Coagulation, fragmentation and radial motion of solid particles in
  protoplanetary disks}.
\newblock {\em Astronomy \& Astrophysics}, {\bf 480}, 859--877.

\bibitem[{Brownlee} {\em et~al.}(2006)]{brownlee06}
{Brownlee}, D., {Tsou}, P., {Al{\'e}on}, J., {Alexander}, C.~M.~O.~., {Araki},
  T., {Bajt}, S., {Baratta}, G.~A., {Bastien}, R., {Bland}, P., {Bleuet}, P.,
  {Borg}, J., {Bradley}, J.~P., {Brearley}, A., {Brenker}, F., {Brennan}, S.,
  {Bridges}, J.~C., {Browning}, N.~D., {Brucato}, J.~R., {Bullock}, E.,
  {Burchell}, M.~J., {Busemann}, H., {Butterworth}, A., {Chaussidon}, M.,
  {Cheuvront}, A., {Chi}, M., {Cintala}, M.~J., {Clark}, B.~C., {Clemett},
  S.~J., {Cody}, G., {Colangeli}, L., {Cooper}, G., {Cordier}, P., {Daghlian},
  C., {Dai}, Z., {D'Hendecourt}, L., {Djouadi}, Z., {Dominguez}, G., {Duxbury},
  T., {Dworkin}, J.~P., {Ebel}, D.~S., {Economou}, T.~E., {Fakra}, S.,
  {Fairey}, S.~A.~J., {Fallon}, S., {Ferrini}, G., {Ferroir}, T.,
  {Fleckenstein}, H., {Floss}, C., {Flynn}, G., {Franchi}, I.~A., {Fries}, M.,
  {Gainsforth}, Z., {Gallien}, J.-P., {Genge}, M., {Gilles}, M.~K., {Gillet},
  P., {Gilmour}, J., {Glavin}, D.~P., {Gounelle}, M., {Grady}, M.~M., {Graham},
  G.~A., {Grant}, P.~G., {Green}, S.~F., {Grossemy}, F., {Grossman}, L.,
  {Grossman}, J.~N., {Guan}, Y., {Hagiya}, K., {Harvey}, R., {Heck}, P.,
  {Herzog}, G.~F., {Hoppe}, P., {H{\"o}rz}, F., {Huth}, J., {Hutcheon}, I.~D.,
  {Ignatyev}, K., {Ishii}, H., {Ito}, M., {Jacob}, D., {Jacobsen}, C.,
  {Jacobsen}, S., {Jones}, S., {Joswiak}, D., {Jurewicz}, A., {Kearsley},
  A.~T., {Keller}, L.~P., {Khodja}, H., {Kilcoyne}, A.~L.~D., {Kissel}, J.,
  {Krot}, A., {Langenhorst}, F., {Lanzirotti}, A., {Le}, L., {Leshin}, L.~A.,
  {Leitner}, J., {Lemelle}, L., {Leroux}, H., {Liu}, M.-C., {Luening}, K.,
  {Lyon}, I., {MacPherson}, G., {Marcus}, M.~A., {Marhas}, K., {Marty}, B.,
  {Matrajt}, G., {McKeegan}, K., {Meibom}, A., {Mennella}, V., {Messenger}, K.,
  {Messenger}, S., {Mikouchi}, T., {Mostefaoui}, S., {Nakamura}, T., {Nakano},
  T., {Newville}, M., {Nittler}, L.~R., {Ohnishi}, I., {Ohsumi}, K.,
  {Okudaira}, K., {Papanastassiou}, D.~A., {Palma}, R., {Palumbo}, M.~E.,
  {Pepin}, R.~O., {Perkins}, D., {Perronnet}, M., {Pianetta}, P., {Rao}, W.,
  {Rietmeijer}, F.~J.~M., {Robert}, F., {Rost}, D., {Rotundi}, A., {Ryan}, R.,
  {Sandford}, S.~A., {Schwandt}, C.~S., {See}, T.~H., {Schlutter}, D.,
  {Sheffield-Parker}, J., {Simionovici}, A., {Simon}, S., {Sitnitsky}, I.,
  {Snead}, C.~J., {Spencer}, M.~K., {Stadermann}, F.~J., {Steele}, A.,
  {Stephan}, T., {Stroud}, R., {Susini}, J., {Sutton}, S.~R., {Suzuki}, Y.,
  {Taheri}, M., {Taylor}, S., {Teslich}, N., {Tomeoka}, K., {Tomioka}, N.,
  {Toppani}, A., {Trigo-Rodr{\'{\i}}guez}, J.~M., {Troadec}, D., {Tsuchiyama},
  A., {Tuzzolino}, A.~J., {Tyliszczak}, T., {Uesugi}, K., {Velbel}, M.,
  {Vellenga}, J., {Vicenzi}, E., {Vincze}, L., {Warren}, J., {Weber}, I.,
  {Weisberg}, M., {Westphal}, A.~J., {Wirick}, S., {Wooden}, D., {Wopenka}, B.,
  {Wozniakiewicz}, P., {Wright}, I., {Yabuta}, H., {Yano}, H., {Young}, E.~D.,
  {Zare}, R.~N., {Zega}, T., {Ziegler}, K., {Zimmerman}, L., {Zinner}, E., and
  {Zolensky}, M. (2006).
\newblock {Comet 81P/Wild 2 Under a Microscope}.
\newblock {\em Science}, {\bf 314}, 1711--1716.

\bibitem[{Calvet} {\em et~al.}(2005)]{calvet05}
{Calvet}, N., {Brice{\~n}o}, C., {Hern{\'a}ndez}, J., {Hoyer}, S., {Hartmann},
  L., {Sicilia-Aguilar}, A., {Megeath}, S.~T., and {D'Alessio}, P. (2005).
\newblock {Disk Evolution in the Orion OB1 Association}.
\newblock {\em Astronomical Journal}, {\bf 129}, 935--946.

\bibitem[{Cassen}(1994)]{cassen94}
{Cassen}, P. (1994).
\newblock {Utilitarian models of the solar nebula}.
\newblock {\em Icarus}, {\bf 112}, 405--429.

\bibitem[{Charnoz} and {Morbidelli}(2007)]{charnoz07}
{Charnoz}, S. and {Morbidelli}, A. (2007).
\newblock {Coupling dynamical and collisional evolution of small bodies. II.
  Forming the Kuiper belt, the Scattered Disk and the Oort Cloud}.
\newblock {\em Icarus}, {\bf 188}, 468--480.

\bibitem[{Chiang} and {Goldreich}(1997)]{chiang97}
{Chiang}, E.~I. and {Goldreich}, P. (1997).
\newblock {Spectral Energy Distributions of T Tauri Stars with Passive
  Circumstellar Disks}.
\newblock {\em Astrophysical Journal}, {\bf 490}, 368--376.

\bibitem[{Ciesla}(2007a)]{ccoag07}
{Ciesla}, F.~J. (2007a).
\newblock {Dust Coagulation and Settling in Layered Protoplanetary Disks}.
\newblock {\em Astrophysics Journal}, {\bf 654}, L159--L162.

\bibitem[{Ciesla}(2007b)]{cieslasci07}
{Ciesla}, F.~J. (2007b).
\newblock {Outward Transport of High-Temperature Materials Around the Midplane
  of the Solar Nebula}.
\newblock {\em Science}, {\bf 318}, 613--615.

\bibitem[{Ciesla} and {Cuzzi}(2006)]{cieslacuz06}
{Ciesla}, F.~J. and {Cuzzi}, J.~N. (2006).
\newblock {The evolution of the water distribution in a viscous protoplanetary
  disk}.
\newblock {\em Icarus}, {\bf 181}, 178--204.

\bibitem[{Cuzzi} and {Weidenschilling}(2006)]{cuzzweid06}
{Cuzzi}, J.~N. and {Weidenschilling}, S.~J. (2006).
\newblock {Nebula Evolution of Thermally Processed Solids: Reconciling Models
  and Meteorites}.
\newblock In {\em Meteorites and the Early
  Solar System II}, (D.~S. {Lauretta} and H.~Y.~J. {McSween}, editors) pages 353--381.

\bibitem[{Cuzzi} {\em et~al.}(2003)]{bitw03}
{Cuzzi}, J.~N., {Davis}, S.~S., and {Dobrovolskis}, A.~R. (2003).
\newblock {Blowing in the wind. II. Creation and redistribution of refractory
  inclusions in a turbulent protoplanetary nebula}.
\newblock {\em Icarus}, {\bf 166}, 385--402.

\bibitem[{Cuzzi} {\em et~al.}(2005)]{cuzzi05}
{Cuzzi}, J.~N., {Ciesla}, F.~J., {Petaev}, M.~I., {Krot}, A.~N., {Scott},
  E.~R.~D., and {Weidenschilling}, S.~J. (2005).
\newblock {Nebula Evolution of Thermally Processed Solids: Reconciling Models
  and Meteorites}.
\newblock In  {\em
  Chondrites and the Protoplanetary Disk}, volume 341 of {\em Astronomical
  Society of the Pacific Conference Series}, 
  (A.~N. {Krot}, E.~R.~D. {Scott}, and B.~{Reipurth}, editors)
  pp. 732--773.

\bibitem[{Cyr} {\em et~al.}(1998)]{cyr98}
{Cyr}, K.~E., {Sears}, W.~D., and {Lunine}, J.~I. (1998).
\newblock {Distribution and Evolution of Water Ice in the Solar Nebula:
  Implications for Solar System Body Formation}.
\newblock {\em Icarus}, {\bf 135}, 537--548.

\bibitem[{Dauphas} {\em et~al.}(2008)]{dauphas08}
{Dauphas}, N., {Cook}, D.~L., {Sacarabany}, A., {Frohlich}, C., {Davis}, A.~M.,
  {Wadhwa}, M., {Pourmand}, A., {Rauscher}, T., and {Gallino}, R. (2008).
\newblock {Iron-60 evidence for early injection and efficient mixing of stellar
  debris in the protosolar nebula}.
\newblock {\em Astrophysical Journal}, {\bf 686}, 560--569.

\bibitem[{Dullemond} and {Dominik}(2005)]{dd05}
{Dullemond}, C.~P. and {Dominik}, C. (2005).
\newblock {Dust coagulation in protoplanetary disks: A rapid depletion of small
  grains}.
\newblock {\em Astronomy \& Astrophysics}, {\bf 434}, 971--986.

\bibitem[{Dullemond} {\em et~al.}(2007)]{dullemond07}
{Dullemond}, C.~P., {Hollenbach}, D., {Kamp}, I., and {D'Alessio}, P. (2007).
\newblock {Models of the Structure and Evolution of Protoplanetary Disks}.
\newblock In  {\em
  Protostars and Planets V} (B.~{Reipurth}, D.~{Jewitt}, and K.~{Keil}, editors), pages 555--572.

\bibitem[{Fleming} and {Stone}(2003)]{flemstone03}
{Fleming}, T. and {Stone}, J.~M. (2003).
\newblock {Local Magnetohydrodynamic Models of Layered Accretion Disks}.
\newblock {\em Astrophysical Journal}, {\bf 585}, 908--920.

\bibitem[{Gail}(2001)]{gail01}
{Gail}, H.-P. (2001).
\newblock {Radial mixing in protoplanetary accretion disks. I. Stationary disc
  models with annealing and carbon combustion}.
\newblock {\em Astronomy \& Astrophysics}, {\bf 378}, 192--213.

\bibitem[{Gail}(2004)]{gail04}
{Gail}, H.-P. (2004).
\newblock {Radial mixing in protoplanetary accretion disks. IV. Metamorphosis
  of the silicate dust complex}.
\newblock {\em Astronomy \& Astrophysics}, {\bf 413}, 571--591.

\bibitem[{Gammie}(1996)]{gammie96}
{Gammie}, C.~F. (1996).
\newblock {Layered Accretion in T Tauri Disks}.
\newblock {\em Astrophysical Journal}, {\bf 457}, 355--362.

\bibitem[{Glassgold} {\em et~al.}(1997)]{glassgold97}
{Glassgold}, A.~E., {Najita}, J., and {Igea}, J. (1997).
\newblock {X-Ray Ionization of Protoplanetary Disks}.
\newblock {\em Astrophysical Journal}, {\bf 480}, 344--350.

\bibitem[{Haghighipour} and {Boss}(2003)]{hagboss03}
{Haghighipour}, N. and {Boss}, A.~P. (2003).
\newblock {On Gas Drag-Induced Rapid Migration of Solids in a Nonuniform Solar
  Nebula}.
\newblock {\em Astrophysical Journal}, {\bf 598}, 1301--1311.

\bibitem[{Haisch} {\em et~al.}(2001)]{haisch01}
{Haisch}, Jr., K.~E., {Lada}, E.~A., and {Lada}, C.~J. (2001).
\newblock {Disk Frequencies and Lifetimes in Young Clusters}.
\newblock {\em Astrophysical Journal}, {\bf 553}, L153--L156.

\bibitem[{Hallenbeck} {\em et~al.}(2000)]{hallenbeck00}
{Hallenbeck}, S.~L., {Nuth}, III, J.~A., and {Nelson}, R.~N. (2000).
\newblock {Evolving Optical Properties of Annealing Silicate Grains: From
  Amorphous Condensate to Crystalline Mineral}.
\newblock {\em Astrophysical Journal}, {\bf 535}, 247--255.

\bibitem[{Hartmann} {\em et~al.}(1998)]{hartmann98}
{Hartmann}, L., {Calvet}, N., {Gullbring}, E., and {D'Alessio}, P. (1998).
\newblock {Accretion and the Evolutiion of T Tauri Disks}.
\newblock {\em Astrophysical Journal}, {\bf 495}, 385--400.

\bibitem[{Hubickyj} {\em et~al.}(2005)]{hubickyj05}
{Hubickyj}, O., {Bodenheimer}, P., and {Lissauer}, J.~J. (2005).
\newblock {Accretion of the gaseous envelope of Jupiter around a 5 10
  Earth-mass core}.
\newblock {\em Icarus}, {\bf 179}, 415--431.

\bibitem[{Jacobsen} {\em et~al.}(2008)]{jacobsen08}
{Jacobsen}, B., {Yin}, Q.-Z., {Moynier}, F., {Amelin}, Y., {Krot}, A.~N.,
  {Nagashima}, K., {Hutcheon}, I.~D., and {Palme}, H. (2008).
\newblock {$^{26}$Al-$^{26}$Mg and $^{207}$Pb-$^{206}$Pb systematics of Allende
  CAIs: Canonical solar initial $^{26}$Al/$^{27}$Al ratio reinstated}.
\newblock {\em Earth and Planetary Science Letters}, {\bf 272}, 353--364.

\bibitem[{Keller} and {Gail}(2004)]{kg04}
{Keller}, C. and {Gail}, H.-P. (2004).
\newblock {Radial mixing in protoplanetary accretion disks. VI. Mixing by
  large-scale radial flows}.
\newblock {\em Astronomy \& Astrophysics}, {\bf 415}, 1177--1185.

\bibitem[{Kita} {\em et~al.}(2005)]{kita05}
{Kita}, N.~T., {Huss}, G.~R., {Tachibana}, S., {Amelin}, Y., {Nyquist}, L.~E.,
  and {Hutcheon}, I.~D. (2005).
\newblock {Constraints on the Origin of Chondrules and CAIs from Short-lived
  and Long-Lived Radionuclides}.
\newblock In {\em
  Chondrites and the Protoplanetary Disk}, volume 341 of {\em Astronomical
  Society of the Pacific Conference Series}, 
  (A.~N. {Krot}, E.~R.~D. {Scott}, and B.~{Reipurth}, editors) pp. 558--587.

\bibitem[{Kley} and {Lin}(1992)]{kleylin92}
{Kley}, W. and {Lin}, D.~N.~C. (1992).
\newblock {Two-dimensional viscous accretion disk models. I - On meridional
  circulations in radiative regions}.
\newblock {\em Astrophysical Journal}, {\bf 397}, 600--612.

\bibitem[{Lynden-Bell} and {Pringle}(1974)]{lbp74}
{Lynden-Bell}, D. and {Pringle}, J.~E. (1974).
\newblock {The evolution of viscous discs and the origin of the nebular
  variables.}
\newblock {\em Monthly Notices of the Royal Astronomical Society}, {\bf 168},
  603--637.

\bibitem[{Lyons} and {Young}(2005)]{ly05}
{Lyons}, J.~R. and {Young}, E.~D. (2005).
\newblock {CO self-shielding as the origin of oxygen isotope anomalies in the
  early solar nebula}.
\newblock {\em \nat}, {\bf 435}, 317--320.

\bibitem[McKeegan {\em et~al.}(2006)]{mckeegan06}
McKeegan, K.~D., Aleon, J., Bradley, J., Brownlee, D., Busemann, H.,
  Butterworth, A., Chaussidon, M., Fallon, S., Floss, C., Gilmour, J.,
  Gounelle, M., Graham, G., Guan, Y., Heck, P.~R., Hoppe, P., Hutcheon, I.~D.,
  Huth, J., Ishii, H., Ito, M., Jacobsen, S.~B., Kearsley, A., Leshin, L.~A.,
  Liu, M.-C., Lyon, I., Marhas, K., Marty, B., Matrajt, G., Meibom, A.,
  Messenger, S., Mostefaoui, S., Mukhopadhyay, S., Nakamura-Messenger, K.,
  Nittler, L., Palma, R., Pepin, R.~O., Papanastassiou, D.~A., Robert, F.,
  Schlutter, D., Snead, C.~J., Stadermann, F.~J., Stroud, R., Tsou, P.,
  Westphal, A., Young, E.~D., Ziegler, K., Zimmermann, L., and Zinner, E.
  (2006).
\newblock {Isotopic Compositions of Cometary Matter Returned by Stardust}.
\newblock {\em Science}, {\bf 314}, 1724--1728.

\bibitem[{Nuth} {\em et~al.}(2000)]{nuth00}
{Nuth}, J.~A., {Hill}, H.~G.~M., and {Kletetschka}, G. (2000).
\newblock {Determining the ages of comets from the fraction of crystalline
  dust}.
\newblock {\em \nat}, {\bf 406}, 275--276.

\bibitem[{Ormel} and {Cuzzi}(2007)]{ormcuzzi07}
{Ormel}, C.~W. and {Cuzzi}, J.~N. (2007).
\newblock {Closed-form expressions for particle relative velocities induced by
  turbulence}.
\newblock {\em Astronomy \& Astrophysics}, {\bf 466}, 413--420.

\bibitem[{Ouellette} {\em et~al.}(2007)]{ouellette07}
{Ouellette}, N., {Desch}, S.~J., and {Hester}, J.~J. (2007).
\newblock {Interaction of Supernova Ejecta with Nearby Protoplanetary Disks}.
\newblock {\em Astrophysical Journal}, {\bf 662}, 1268--1281.

\bibitem[{Press} {\em et~al.}(1992)]{nr}
{Press}, W.~H., {Teukolsky}, S.~A., {Vetterling}, W.~T., and {Flannery}, B.~P.
  (1992).
\newblock {\em {Numerical recipes in C. The art of scientific computing}}.
\newblock Cambridge: University Press, |c1992, 2nd ed.

\bibitem[{Richard} and {Zahn}(1999)]{richard99}
{Richard}, D. and {Zahn}, J.-P. (1999).
\newblock {Turbulence in differentially rotating flows. What can be learned
  from the Couette-Taylor experiment}.
\newblock {\em Astronomy \& Astrophysics}, {\bf 347}, 734--738.

\bibitem[{Rozyczka} {\em et~al.}(1994)]{rozyczka94}
{Rozyczka}, M., {Bodenheimer}, P., and {Bell}, K.~R. (1994).
\newblock {A Numerical Study of Viscous Flows in Axisymmetric alpha -Accretion
  Disks}.
\newblock {\em Astrophysical Journal}, {\bf 423}, 736--747.

\bibitem[{Ruden} and {Pollack}(1991)]{rp91}
{Ruden}, S.~P. and {Pollack}, J.~B. (1991).
\newblock {The dynamical evolution of the protosolar nebula}.
\newblock {\em Astrophysical Journal}, {\bf 375}, 740--760.

\bibitem[{Shakura} and {Sunyaev}(1973)]{ss73}
{Shakura}, N.~I. and {Sunyaev}, R.~A. (1973).
\newblock {Black holes in binary systems. Observational appearance.}
\newblock {\em Astronomy \& Astrophysics}, {\bf 24}, 337--355.

\bibitem[{Shu} {\em et~al.}(1996)]{shu96}
{Shu}, F.~H., {Shang}, H., and {Lee}, T. (1996).
\newblock {Toward an Astrophysical Theory of Chondrites}.
\newblock {\em Science}, {\bf 271}, 1545--1552.

\bibitem[{Stepinski}(1998)]{step98}
{Stepinski}, T.~F. (1998).
\newblock {The Solar Nebula as a Process-An Analytic Model}.
\newblock {\em Icarus}, {\bf 132}, 100--112.

\bibitem[{Stepinski} and {Valageas}(1997)]{stepval97}
{Stepinski}, T.~F. and {Valageas}, P. (1997).
\newblock {Global evolution of solid matter in turbulent protoplanetary disks.
  II. Development of icy planetesimals.}
\newblock {\em Astronomy \& Astrophysics}, {\bf 319}, 1007--1019.

\bibitem[{Stevenson} and {Lunine}(1988)]{stevenson88}
{Stevenson}, D.~J. and {Lunine}, J.~I. (1988).
\newblock {Rapid formation of Jupiter by diffuse redistribution of water vapor
  in the solar nebula}.
\newblock {\em Icarus}, {\bf 75}, 146--155.

\bibitem[{Supulver} and {Lin}(2000)]{supulverlin00}
{Supulver}, K.~D. and {Lin}, D.~N.~C. (2000).
\newblock {Formation of Icy Planetesimals in a Turbulent Solar Nebula}.
\newblock {\em Icarus}, {\bf 146}, 525--540.

\bibitem[{Tachibana} and {Huss}(2003)]{tachibana03}
{Tachibana}, S. and {Huss}, G.~R. (2003).
\newblock {The Initial Abundance of $^{60}$Fe in the Solar System}.
\newblock {\em Asrophyscal Journal}, {\bf 588}, L41--L44.

\bibitem[{Takeuchi} and {Lin}(2002)]{tl02}
{Takeuchi}, T. and {Lin}, D.~N.~C. (2002).
\newblock {Radial Flow of Dust Particles in Accretion Disks}.
\newblock {\em Astrophysical Journal}, {\bf 581}, 1344--1355.

\bibitem[{Tanaka} {\em et~al.}(2005)]{tanaka05}
{Tanaka}, H., {Himeno}, Y., and {Ida}, S. (2005).
\newblock {Dust Growth and Settling in Protoplanetary Disks and Disk Spectral
  Energy Distributions. I. Laminar Disks}.
\newblock {\em Astrophysical Journal}, {\bf 625}, 414--426.

\bibitem[{Tscharnuter} and {Gail}(2007)]{tschar07}
{Tscharnuter}, W.~M. and {Gail}, H.-P. (2007).
\newblock {2-D preplanetary accretion disks. I. Hydrodynamics, chemistry, and
  mixing processes}.
\newblock {\em Astronomy \& Astrophysics}, {\bf 463}, 369--392.

\bibitem[{Turner} and {Sano}(2008)]{turnersano08}
{Turner}, N.~J. and {Sano}, T. (2008).
\newblock {Dead Zone Accretion Flows in Protostellar Disks}.
\newblock {\em Astrophyscal Journal}, {\bf 679}, L131--L134.

\bibitem[{Urpin}(1984)]{urpin84}
{Urpin}, V.~A. (1984).
\newblock {Hydrodynamic Flows in Accretion Disks}.
\newblock {\em Soviet Astronomy}, {\bf 28}, 50--56.

\bibitem[{van Boekel} {\em et~al.}(2004)]{vanboekel04}
{van Boekel}, R., {Min}, M., {Leinert}, C., {Waters}, L.~B.~F.~M., {Richichi},
  A., {Chesneau}, O., {Dominik}, C., {Jaffe}, W., {Dutrey}, A., {Graser}, U.,
  {Henning}, T., {de Jong}, J., {K{\"o}hler}, R., {de Koter}, A., {Lopez}, B.,
  {Malbet}, F., {Morel}, S., {Paresce}, F., {Perrin}, G., {Preibisch}, T.,
  {Przygodda}, F., {Sch{\"o}ller}, M., and {Wittkowski}, M. (2004).
\newblock {The building blocks of planets within the `terrestrial' region of
  protoplanetary disks}.
\newblock {\em \nat}, {\bf 432}, 479--482.

\bibitem[{van Leer}(1977)]{vanleer77}
{van Leer}, B. (1977).
\newblock {Towards the ultimate conservative difference scheme: IV. A new
  approach to numerical convection}.
\newblock {\em Journal of Computational Physics}, {\bf 23}, 276--299.

\bibitem[{Watson} {\em et~al.}(2007)]{watson07}
{Watson}, D.~M., {Leisenring}, J.~M., {Furlan}, E., {Bohac}, C.~J., {Sargent},
  B., {Forrest}, W.~J., {Calvet}, N., {Hartmann}, L., {Nordhaus}, J.~T.,
  {Green}, J.~D., {Kim}, K.~H., {Sloan}, G.~C., {Chen}, C.~H., {Keller}, L.~D.,
  {dAlessio}, P., {Najita}, J., {Uchida}, K.~I., and {Houck}, J.~R. (2007).
\newblock {Crystalline silicates and dust processing in the protoplanetary
  disks of the Taurus young cluster}.
\newblock {\em astro-ph}, {\bf 0704.1518}.

\bibitem[{Wehrstedt} and {Gail}(2008)]{wehrstedt08}
{Wehrstedt}, M. and {Gail}, H.-P. (2008).
\newblock {Radial mixing in protoplanetary accretion disks VII. 2-dimensional
  transport of tracers}.
\newblock {\em astro-ph}, {\bf 0804.3377}.

\bibitem[{Weidenschilling}(1977)]{weidgd77}
{Weidenschilling}, S.~J. (1977).
\newblock {Aerodynamics of solid bodies in the solar nebula}.
\newblock {\em Monthly Notices of the Royal Astronomical Society}, {\bf 180},
  57--70.

\bibitem[{Weidenschilling}(1984)]{weidenschilling84}
{Weidenschilling}, S.~J. (1984).
\newblock {Evolution of grains in a turbulent solar nebula}.
\newblock {\em Icarus}, {\bf 60}, 553--567.

\bibitem[{Weidenschilling}(1997)]{weidenschilling97}
{Weidenschilling}, S.~J. (1997).
\newblock {The Origin of Comets in the Solar Nebula: A Unified Model}.
\newblock {\em Icarus}, {\bf 127}, 290--306.

\bibitem[{Weidenschilling}(2004)]{weidenschilling04}
{Weidenschilling}, S.~J. (2004).
\newblock {From icy grains to comets}.
\newblock In {\em
  Comets II} (M.~{Fetsou}, H.~U. {Keller}, and H.~A. {Weaver}, editors), pages 97--104. University of Arizona Press.

\bibitem[{Weisberg} {\em et~al.}(2006)]{weisberg06}
{Weisberg}, M.~K., {McCoy}, T.~J., and {Krot}, A.~N. (2006).
\newblock {Systematics and Evaluation of Meteorite Classification}.
\newblock In D.~S. {Lauretta} and H.~Y.~J. {McSween}, editors, {\em Meteorites
  and the Early Solary System II}, pages 19--52. University of Arizona Press.

\bibitem[{Wooden} {\em et~al.}(2007)]{wooden07}
{Wooden}, D., {Desch}, S., {Harker}, D., {Gail}, H.-P., and {Keller}, L.
  (2007).
\newblock {Comet Grains and Implications for Heating and Radial Mixing in the
  Protoplanetary Disk}.
\newblock In  {\em
  Protostars and Planets V} (B.~{Reipurth}, D.~{Jewitt}, and K.~{Keil}, editors), pages 815--833.

\bibitem[{Wooden}(2008)]{wooden08}
{Wooden}, D.~H. (2008).
\newblock {Cometary Refractory Grains: Interstellar and Nebular Sources}.
\newblock {\em Space Science Reviews}, {\bf 138}, 75--108.

\bibitem[{Youdin} and {Lithwick}(2007)]{yl07}
{Youdin}, A.~N. and {Lithwick}, Y.
(2007).
\newblock {Particle Stirring in Turbulent Gas Disks: Including Orbital Oscillations}.
\newblock {\em Icarus}, {\bf 192}, 588-604.

\bibitem[{Zolensky} {\em et~al.}(2006)]{zolensky06}
{Zolensky}, M.~E., {Zega}, T.~J., {Yano}, H., {Wirick}, S., {Westphal}, A.~J.,
  {Weisberg}, M.~K., {Weber}, I., {Warren}, J.~L., {Velbel}, M.~A.,
  {Tsuchiyama}, A., {Tsou}, P., {Toppani}, A., {Tomioka}, N., {Tomeoka}, K.,
  {Teslich}, N., {Taheri}, M., {Susini}, J., {Stroud}, R., {Stephan}, T.,
  {Stadermann}, F.~J., {Snead}, C.~J., {Simon}, S.~B., {Simionovici}, A.,
  {See}, T.~H., {Robert}, F., {Rietmeijer}, F.~J.~M., {Rao}, W., {Perronnet},
  M.~C., {Papanastassiou}, D.~A., {Okudaira}, K., {Ohsumi}, K., {Ohnishi}, I.,
  {Nakamura-Messenger}, K., {Nakamura}, T., {Mostefaoui}, S., {Mikouchi}, T.,
  {Meibom}, A., {Matrajt}, G., {Marcus}, M.~A., {Leroux}, H., {Lemelle}, L.,
  {Le}, L., {Lanzirotti}, A., {Langenhorst}, F., {Krot}, A.~N., {Keller},
  L.~P., {Kearsley}, A.~T., {Joswiak}, D., {Jacob}, D., {Ishii}, H., {Harvey},
  R., {Hagiya}, K., {Grossman}, L., {Grossman}, J.~N., {Graham}, G.~A.,
  {Gounelle}, M., {Gillet}, P., {Genge}, M.~J., {Flynn}, G., {Ferroir}, T.,
  {Fallon}, S., {Ebel}, D.~S., {Dai}, Z.~R., {Cordier}, P., {Clark}, B., {Chi},
  M., {Butterworth}, A.~L., {Brownlee}, D.~E., {Bridges}, J.~C., {Brennan}, S.,
  {Brearley}, A., {Bradley}, J.~P., {Bleuet}, P., {Bland}, P.~A., and
  {Bastien}, R. (2006).
\newblock {Mineralogy and Petrology of Comet 81P/Wild 2 Nucleus Samples}.
\newblock {\em Science}, {\bf 314}, 1735--1739.

\end{thebibliography}

\newpage
\begin{table}[!h]
\begin{center}
Model Parameters for Star-Disk System \\
\begin{tabular}{l  r}
\hline
\hline
Stellar Mass & 1 $M_{\odot}$ \\
Stellar Temperature & 4000 K \\
Stellar Radii & 3 $R_{\odot}$ \\
Grazing Angle, $\phi$ & 0.05 \\
Disk Opacity, $\kappa$ & 5 cm$^{2}$/g \\
Disk Inner Radius, $r_{in}$ & 0.05 AU \\
Disk Outer Radius & 84 AU \\
\hline
\end{tabular}
\end{center}
\end{table}

\newpage
\begin{table}[!h]
\begin{center}
Structural Properties of Disks Considered \\
\begin{tabular}{l c c c }
\hline
\hline
$\dot M$ ($M_{\odot}$/yr) & 10$^{-6}$ & 10$^{-7}$ & 10$^{-8}$ \\
$r_{1100 ~\mathrm{K}}$ (AU) &  3.4 & 1.3 & 0.5 \\
$H$ at 1 AU (AU) &  0.14 &  0.088 &  0.056 \\
$H$ at 20 AU (AU) & 3.0 & 2.0 & 1.5 \\
Time Computed (yrs) & 10$^{5}$ & 10$^{6}$ & 10$^{6}$ \\
\hline
\end{tabular}
\end{center}
\end{table}

\newpage
\begin{figure}
\includegraphics[angle=90,width=3.1in]{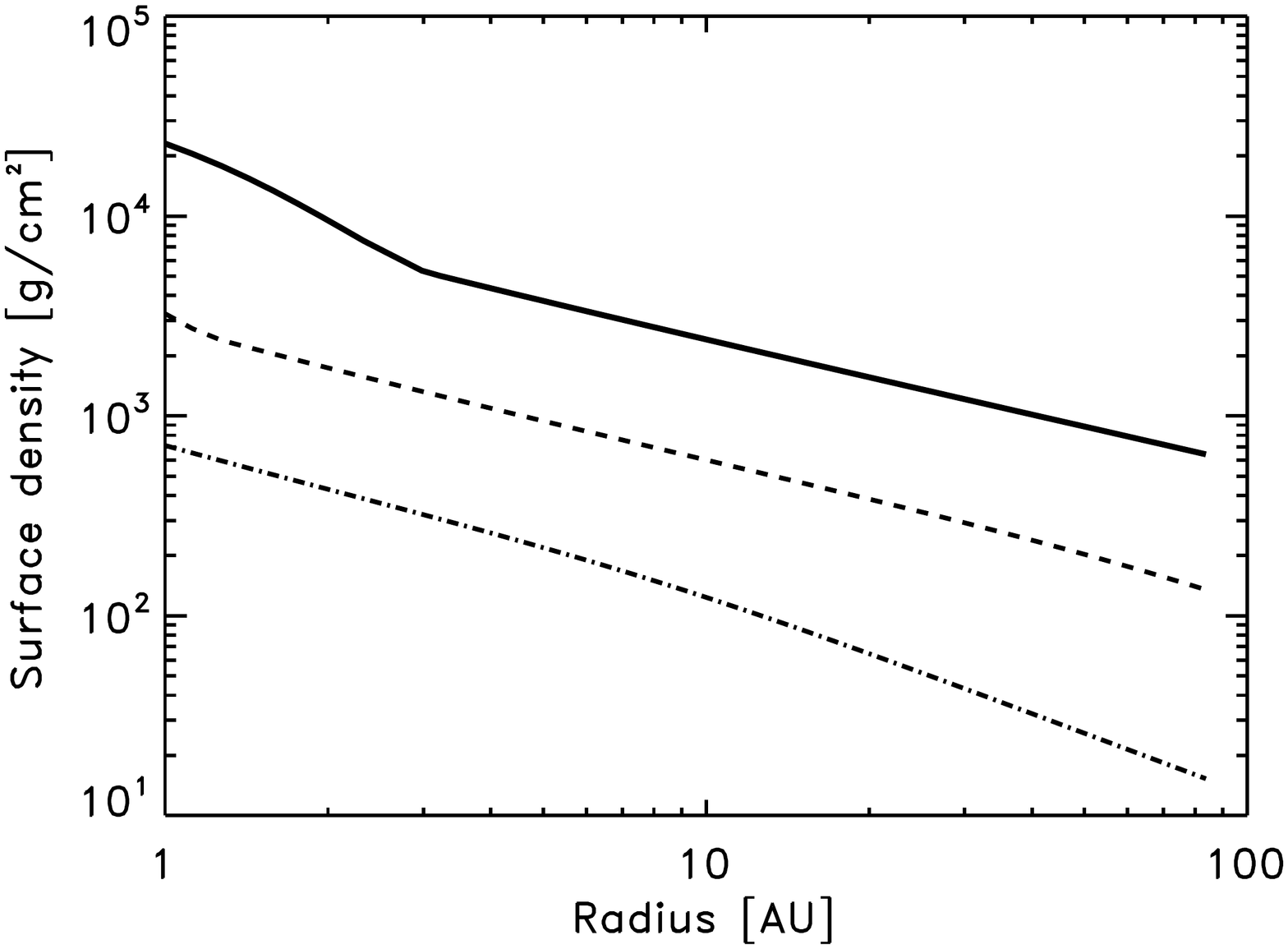}
\includegraphics[angle=90,width=3.1in]{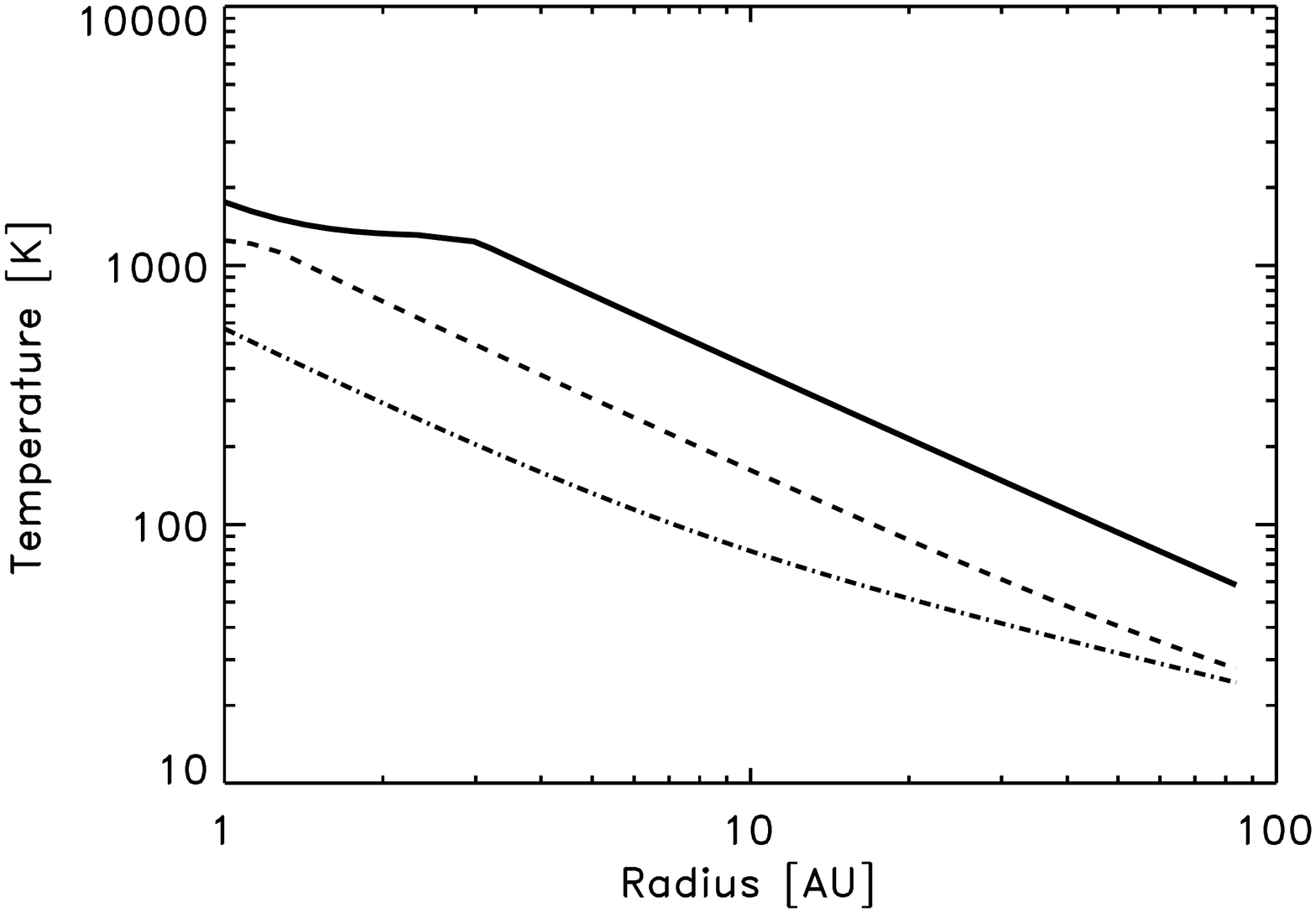}\\
\caption{The surface density and temperature profiles for the protoplanetary disks considered here with mass accretion rates of $\dot M$=10$^{-6}$, 10$^{-7}$, and 10$^{-8} M_{\odot}$/year.}
\end{figure}

\newpage
\begin{figure}
\includegraphics[angle=90,width=\textwidth]{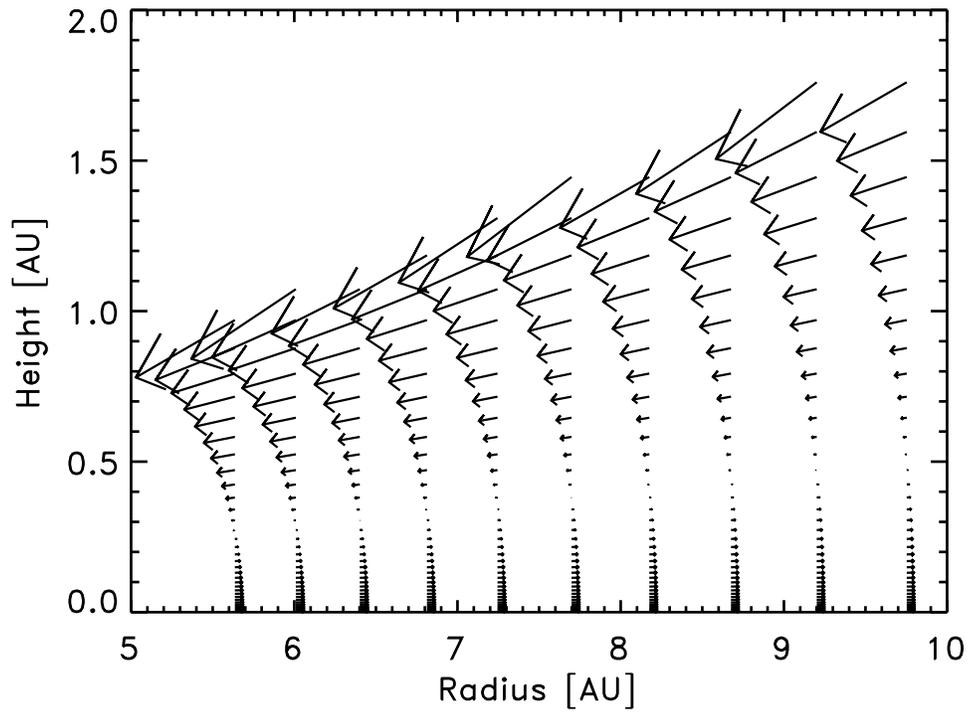}
\caption{The velocity profile for the gas in an viscously evolving protoplanetary disk.  Arrows indicate the direction of the flow, with the length of the arrow being proportional to the magnitude of the flow.  For
reference, the outward flow rates of the gas at the midplane for the different disks considered here are given in Figure 3.}
\end{figure}

\newpage
\begin{figure}
\includegraphics[angle=90,width=3.1in]{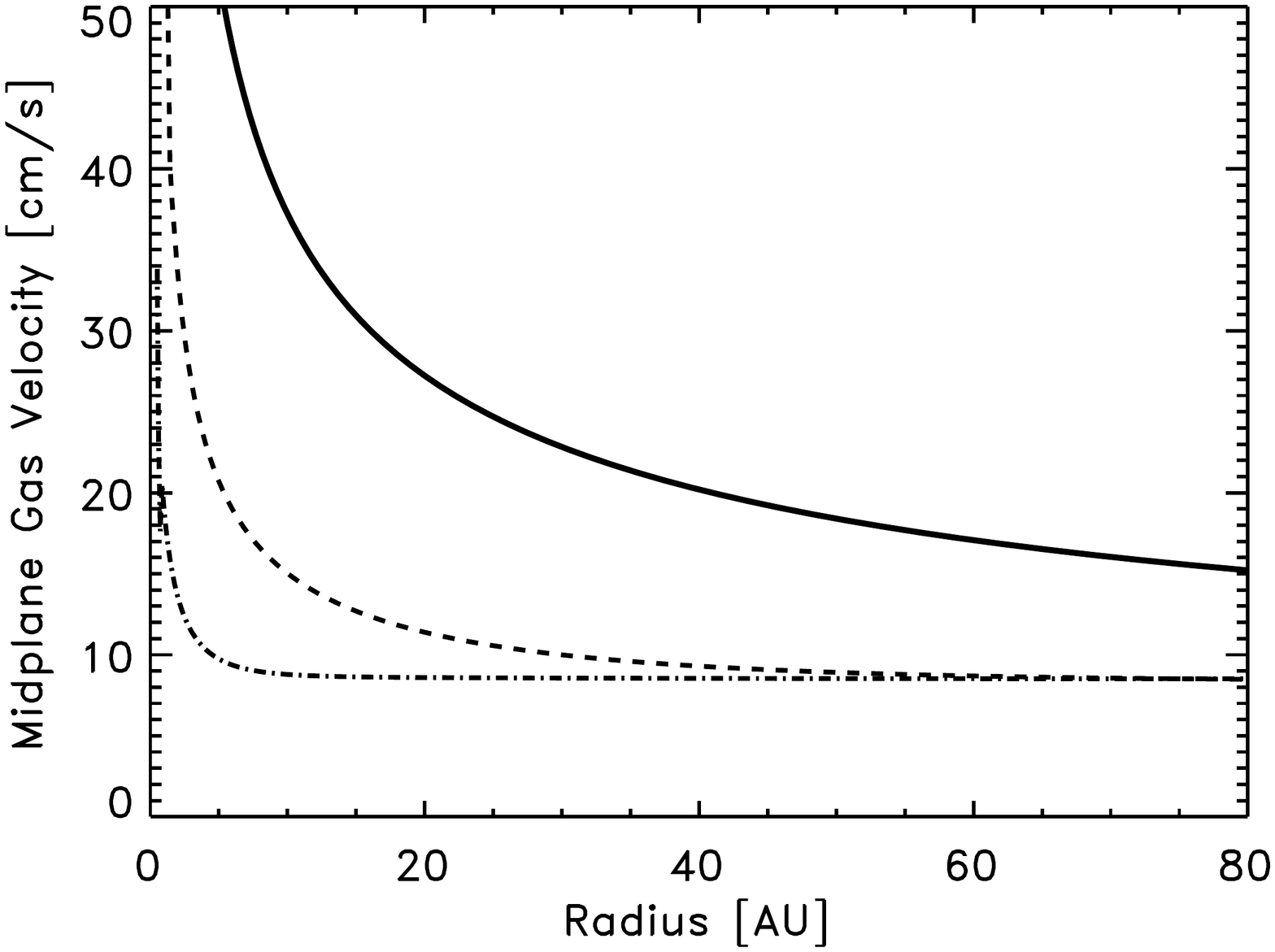}
\includegraphics[angle=90,width=3.1in]{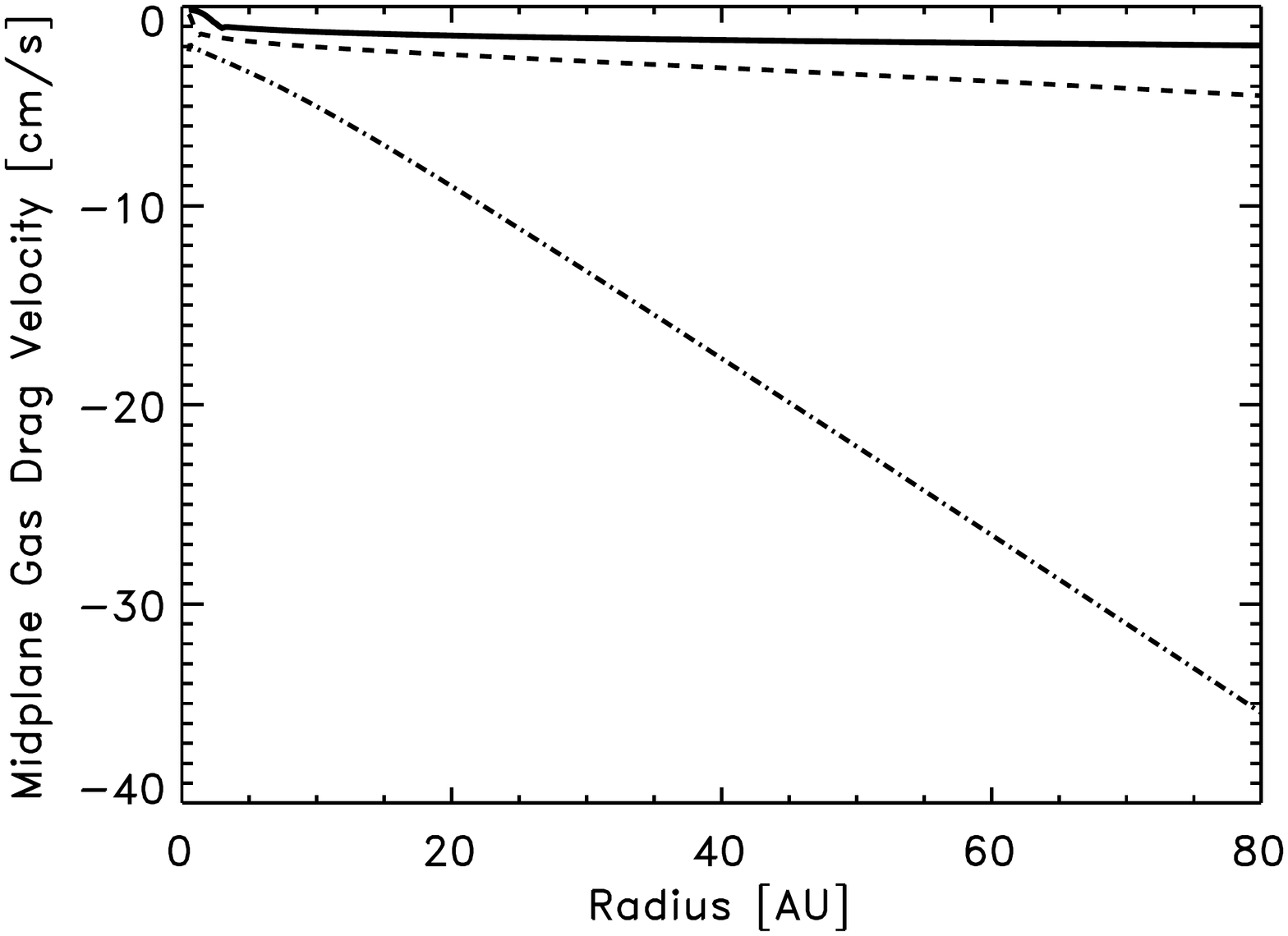}\\
\caption{The outward flow velocities of the gas (\emph{left}) and the inward velocities of the 0.5 mm particles due to gas drag (\emph{right}) both along the disk midplane.  Gas drag velocities for the 5 $\mu$m grains would be $\sim$100x smaller.  Outward flow velocities decrease with decreasing mass accretion rates, while gas drag velocities increase in magnitude.
As can be seen, only in the outer regions of the $\dot M$=10$^{-8} M_\odot$/year disk is the net velocity of the 0.5 mm particles inward at the disk midplane.  }
\end{figure}

\newpage
\begin{figure}
\includegraphics[angle=90,width=\textwidth]{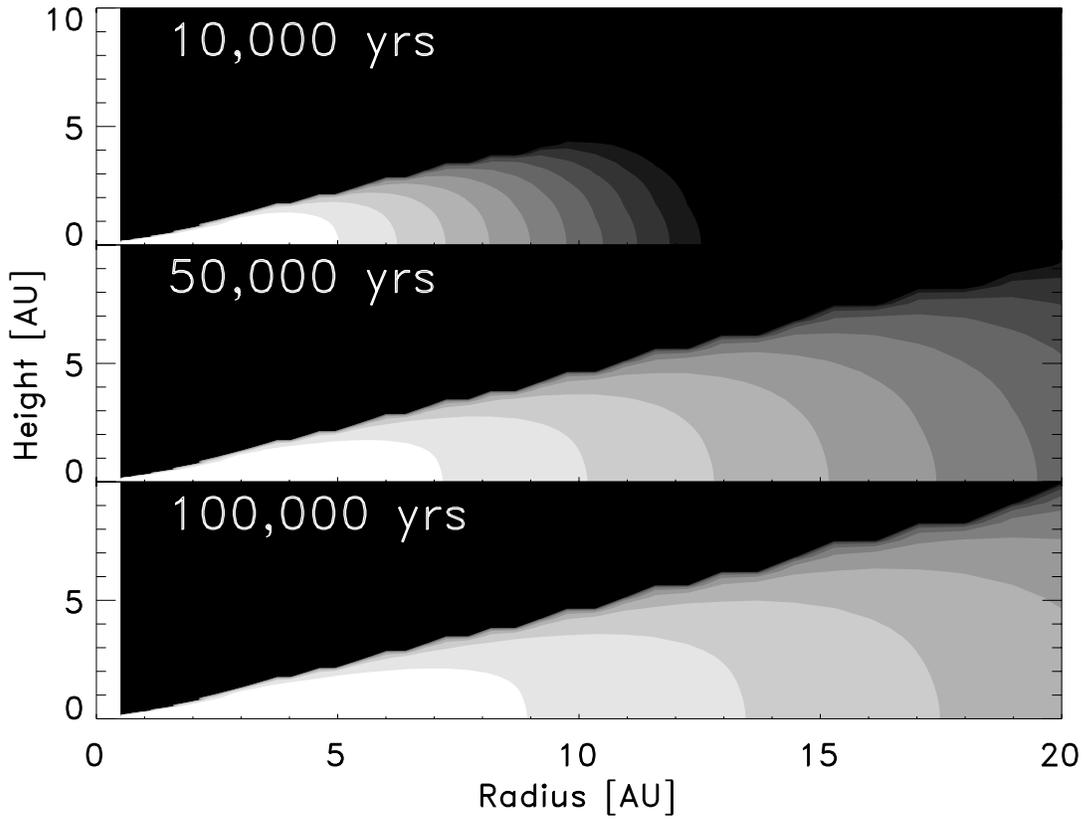}
\caption{Time evolution of the concentration of high temperature 5 $\mu$m grains in the protoplanetary disk with $\dot M$=10$^{-6} M_\odot$/year.  The concentration of materials was held constant at $C_{0}$=1 wherever T$>$1100 K and $z < H$.  Contours represent changes in concentration by a factor of 2, ranging from 0.001 to 0.512.}
\end{figure}

\newpage
\begin{figure}
\includegraphics[angle=90,width=\textwidth]{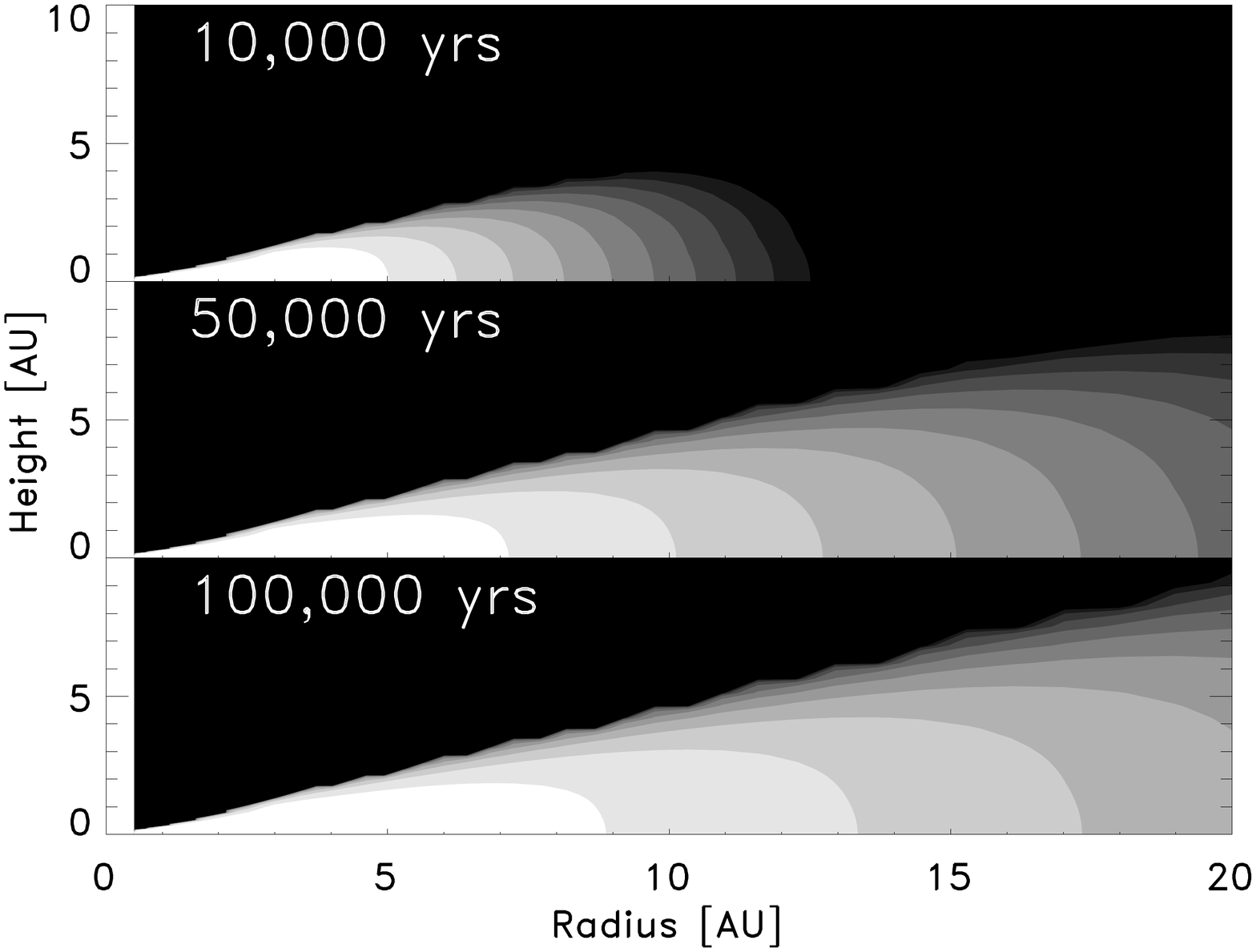}
\caption{Same as Figure 4, with $a$=0.5 mm.}
\end{figure}

\newpage
\begin{figure}
\includegraphics[angle=90,width=\textwidth]{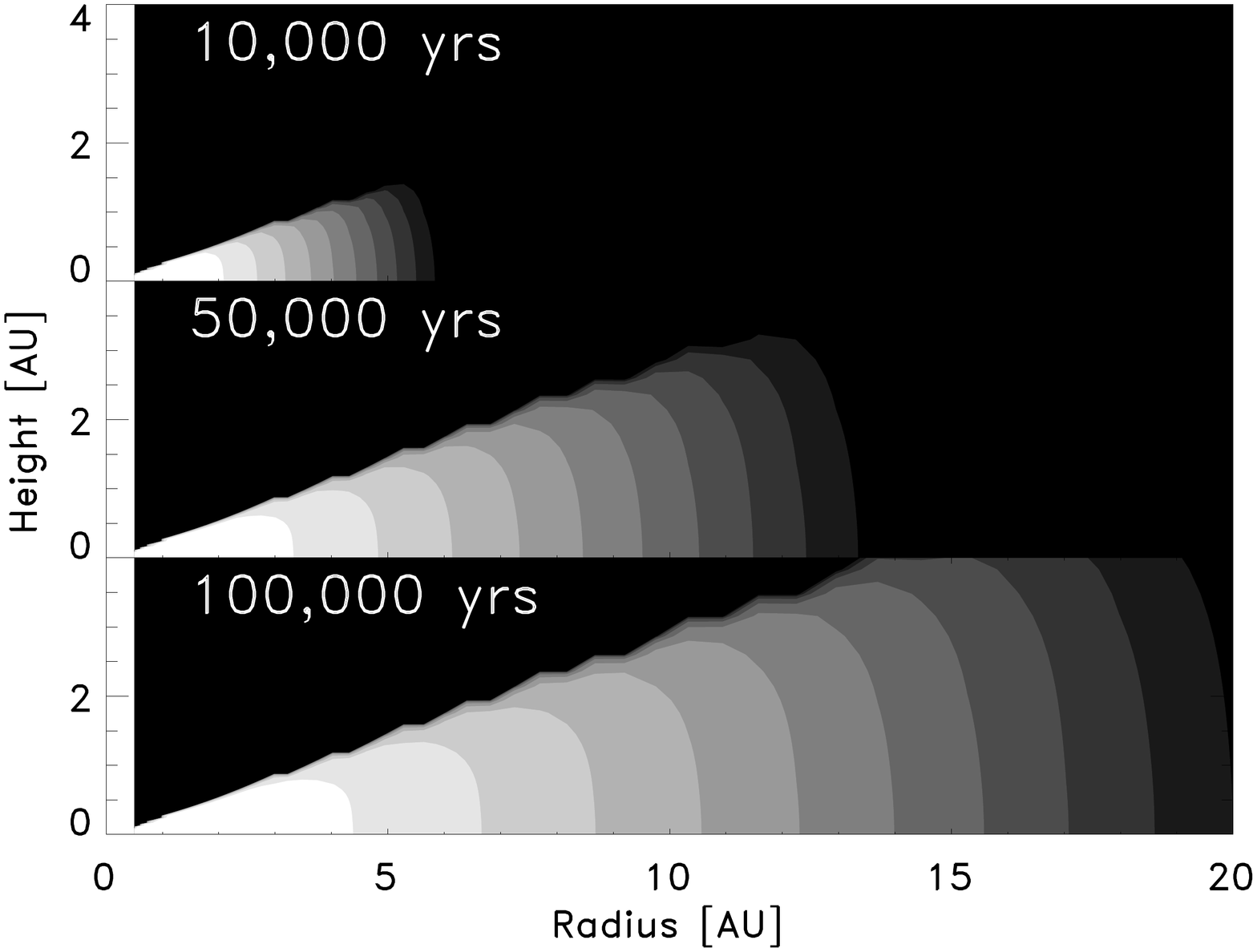}
\caption{Same as Figure 4, with $\dot M$=10$^{-7} M_\odot$/year.}
\end{figure}

\newpage
\begin{figure}
\includegraphics[angle=90,width=\textwidth]{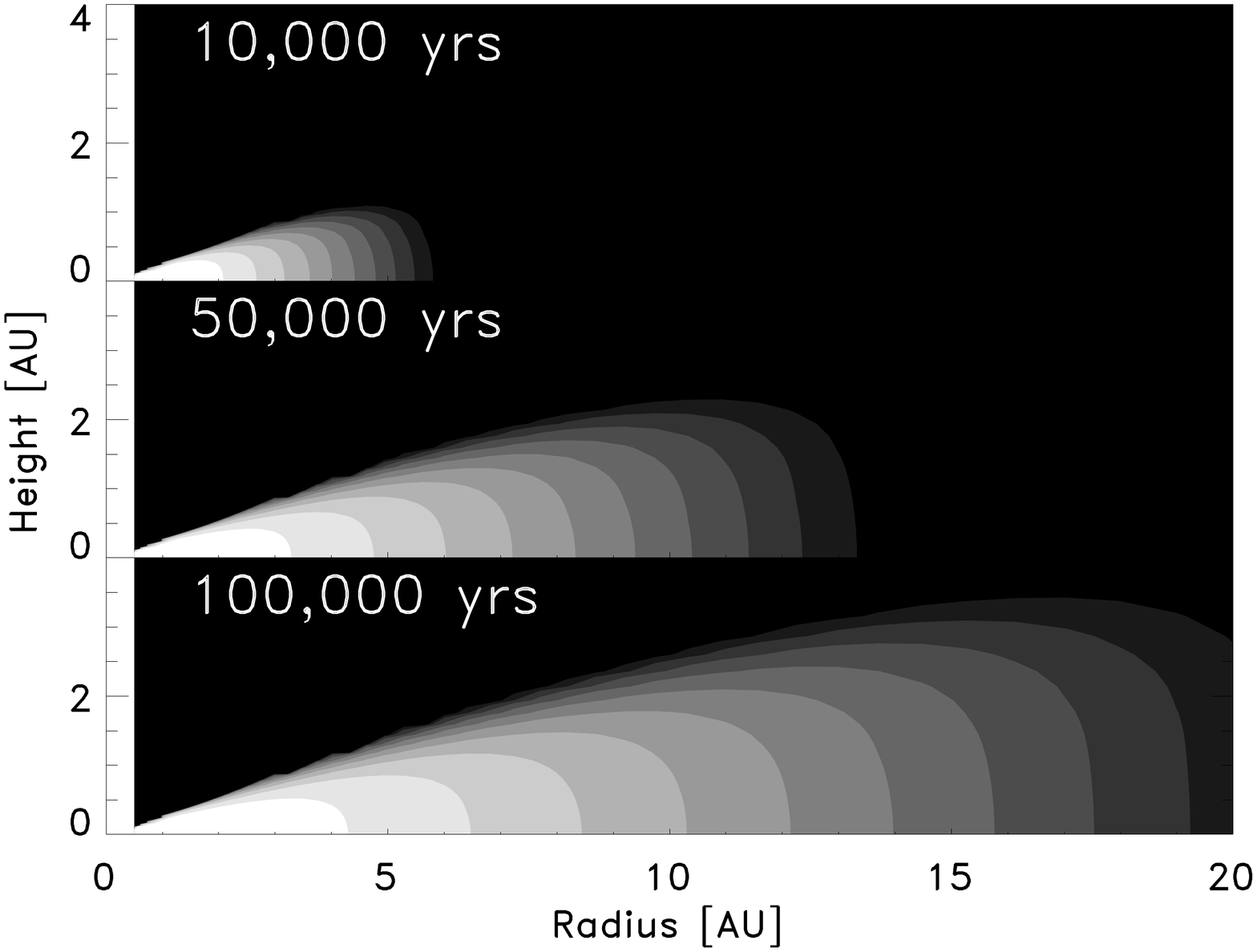}
\caption{Same as Figure 6, with $a$=0.5 mm.}
\end{figure}

\newpage
\begin{figure}
\includegraphics[angle=90,width=\textwidth]{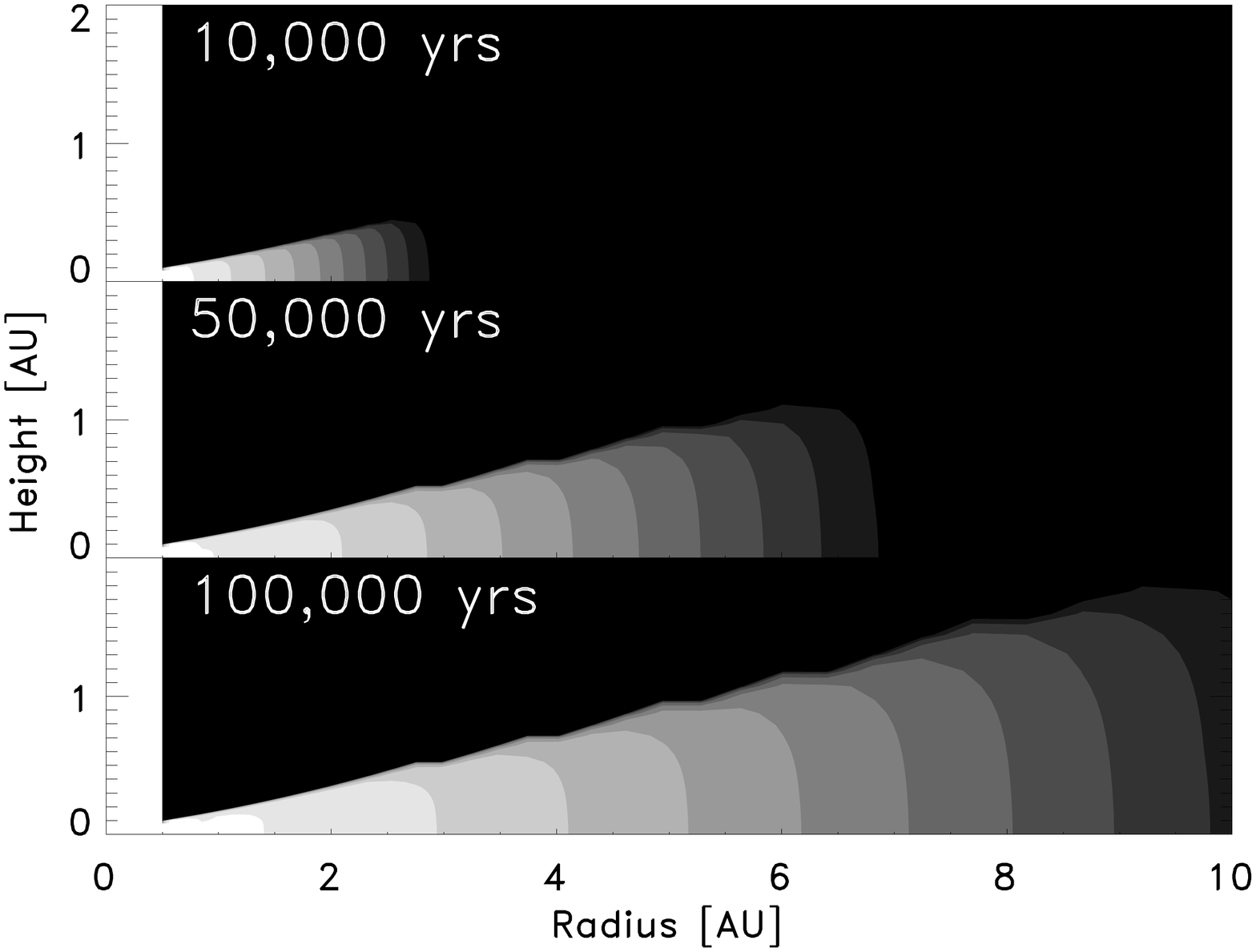}
\caption{Same as Figure 4, with $\dot M$=10$^{-8} M_\odot$/year.}
\end{figure}

\newpage
\begin{figure}
\includegraphics[angle=90,width=\textwidth]{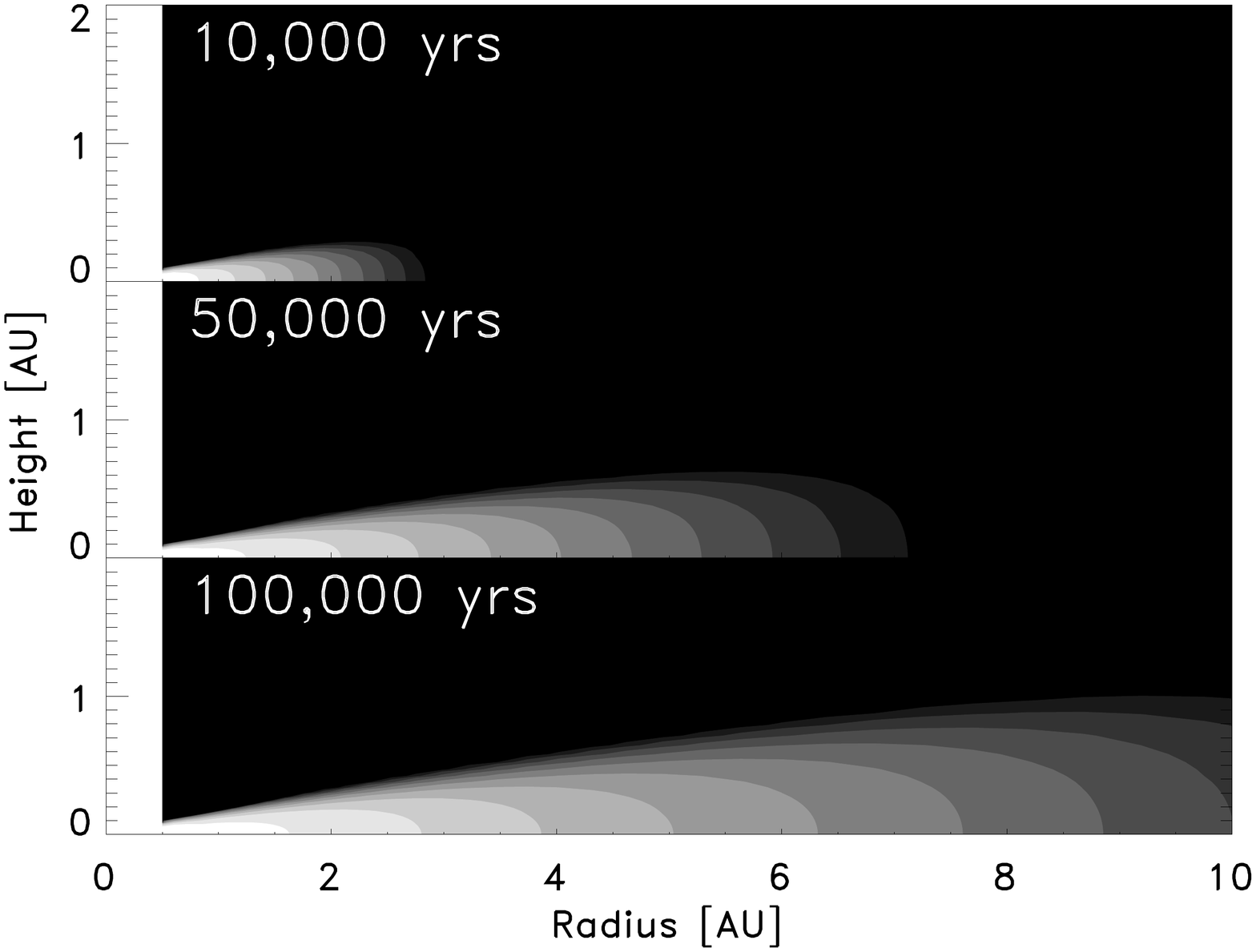}
\caption{Same as Figure 8, with $a$=0.5 mm.}
\end{figure}

\newpage
\begin{figure}
\includegraphics[angle=90,width=3.1in]{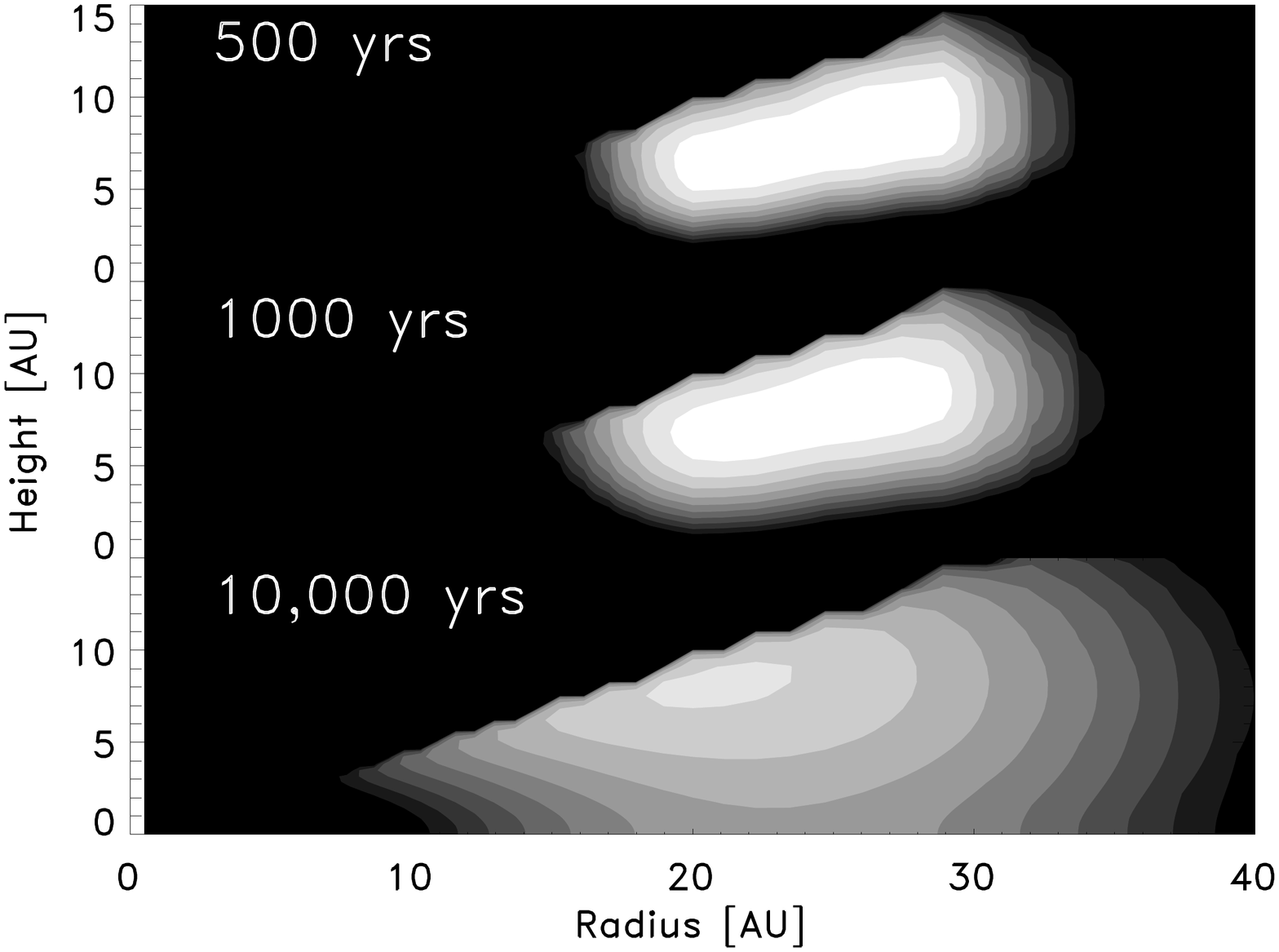}
\includegraphics[angle=90,width=3.1in]{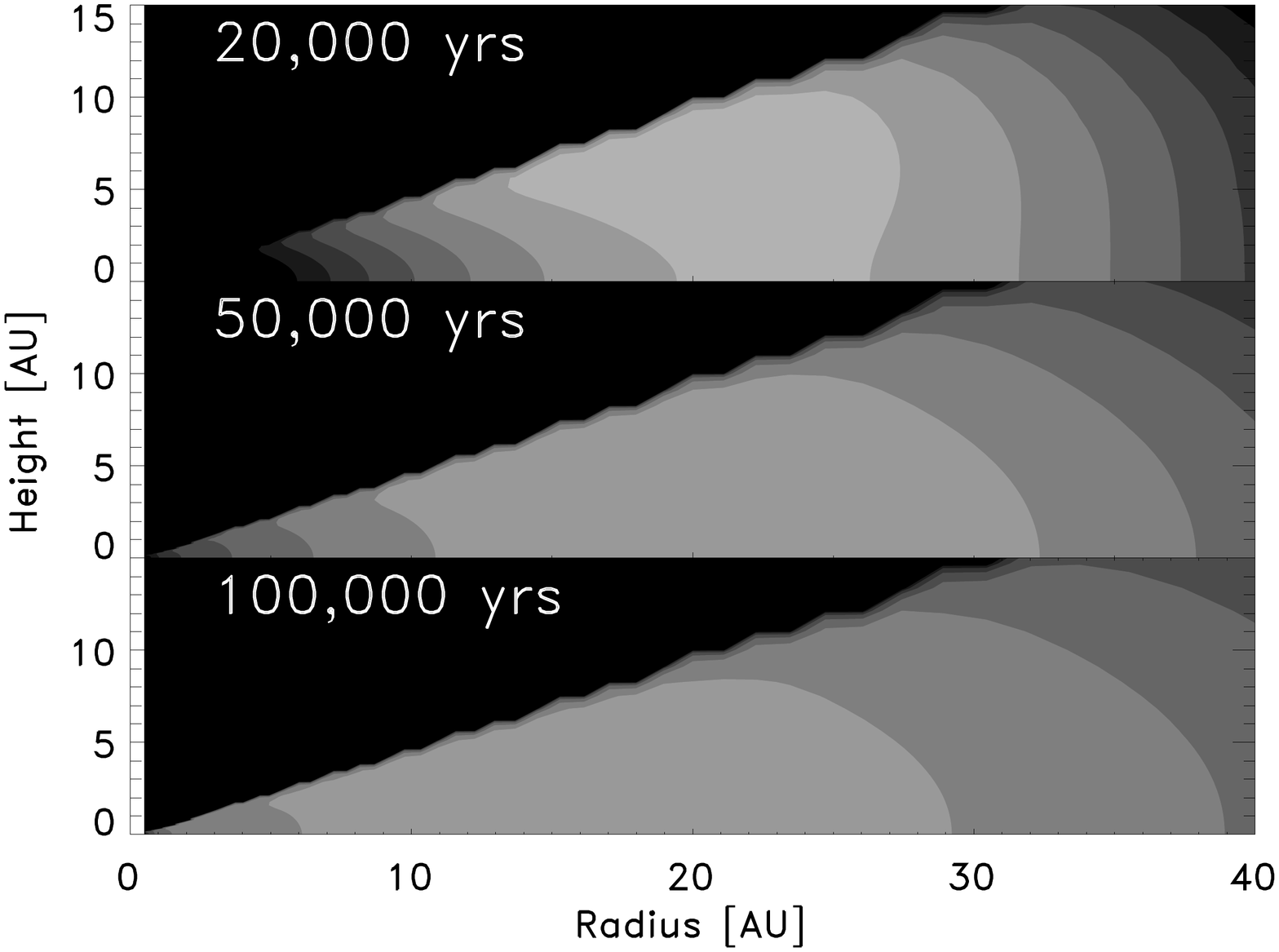}\\
\caption{Time evolution of the concentration of 5 $\mu$m grains in the protoplanetary disk with $\dot M$=10$^{-6} M_\odot$/year.  The concentration of materials was set at $C_{0}$=1 at t=0 between 20 and 30 AU and at heights between 1.5 to 2.5 $H$.  Contours are the same as those in Figure 4.}
\end{figure}

\newpage
\begin{figure}
\includegraphics[angle=90,width=3.1in]{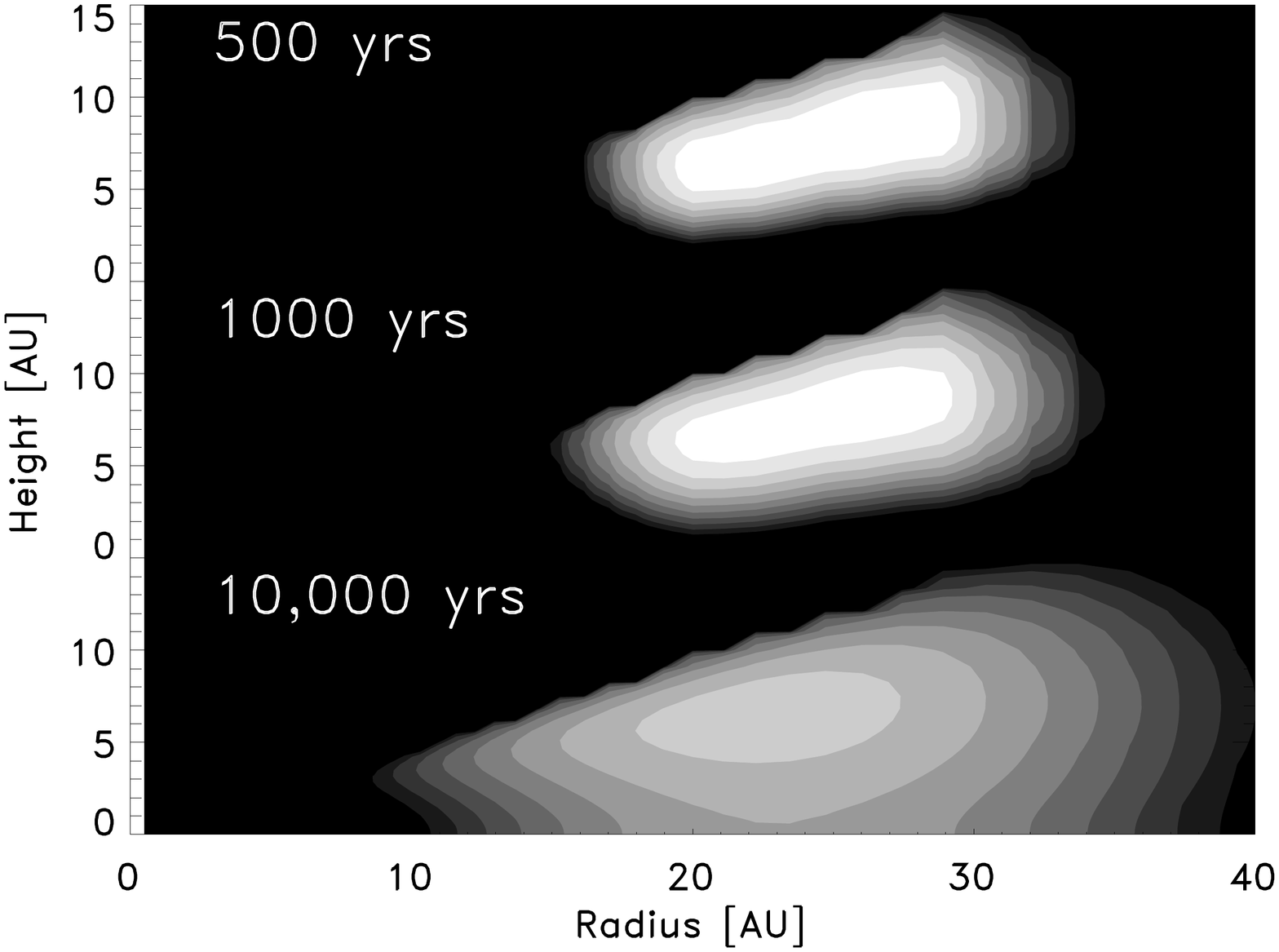}
\includegraphics[angle=90,width=3.1in]{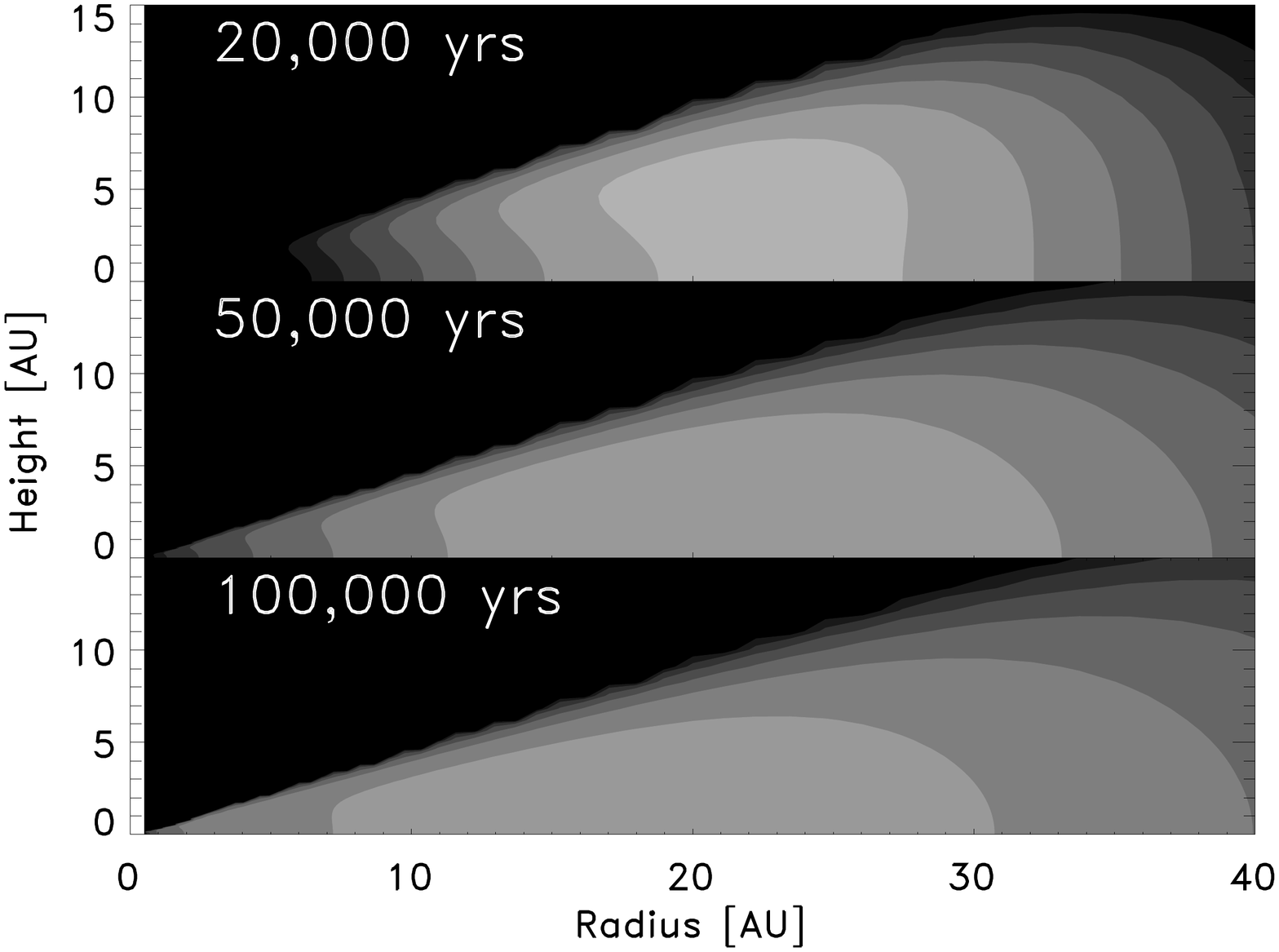}\\
\caption{Same as Figure 10, with $a$=0.5 mm.}
\end{figure}

\newpage
\begin{figure}
\includegraphics[angle=90,width=3.1in]{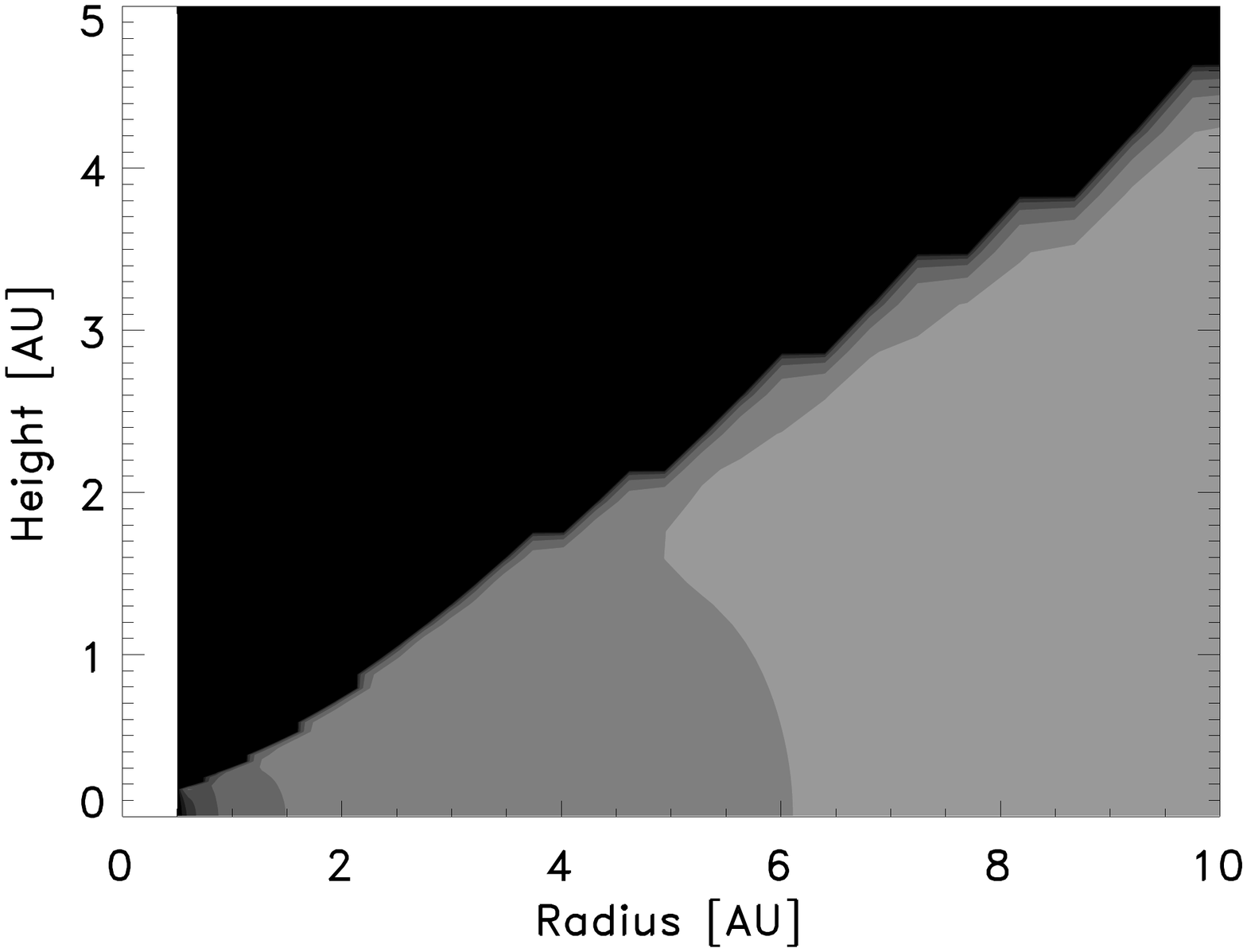}
\includegraphics[angle=90,width=3.1in]{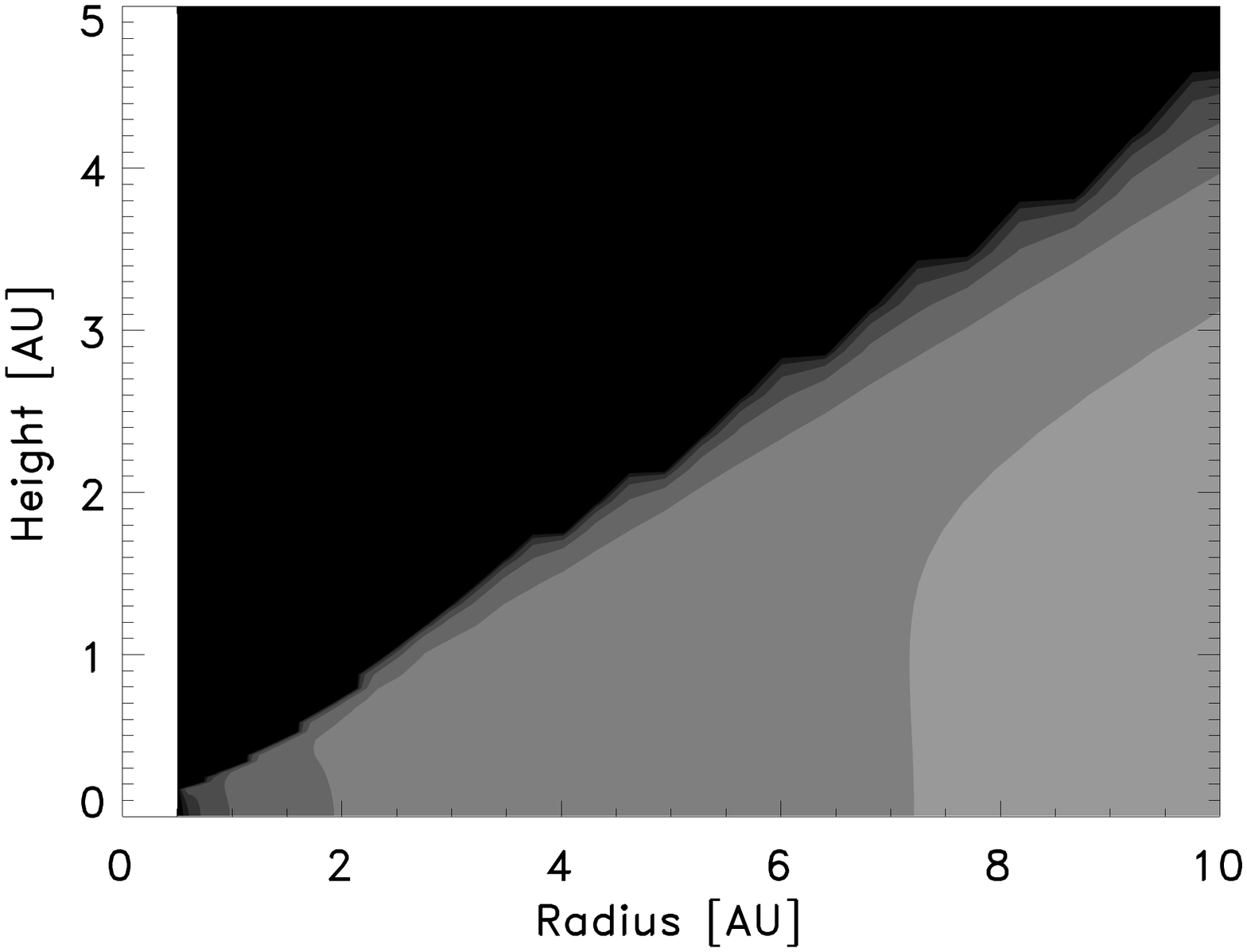}
\includegraphics[angle=90,width=3.1in]{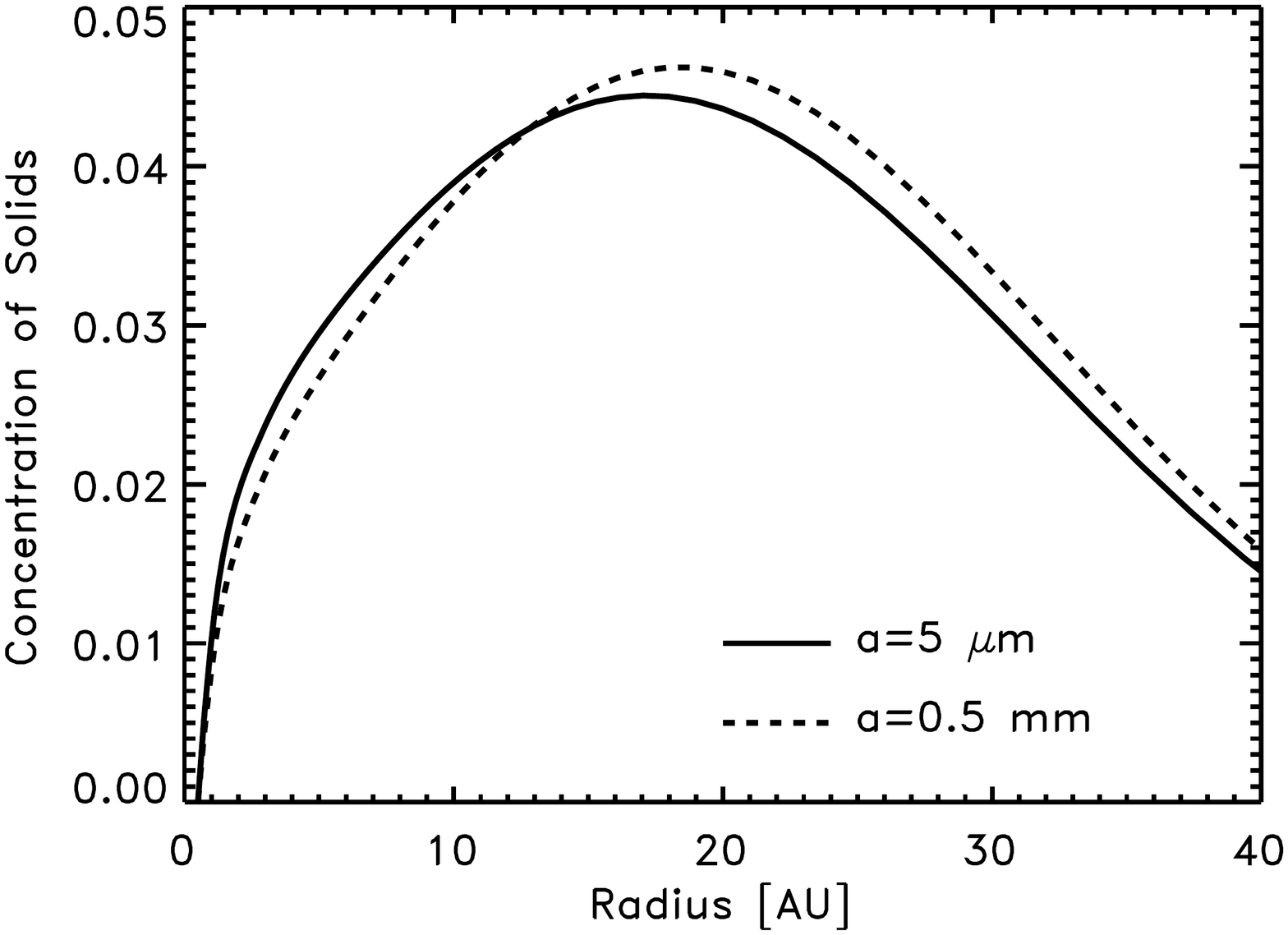}
\includegraphics[angle=90,width=3.1in]{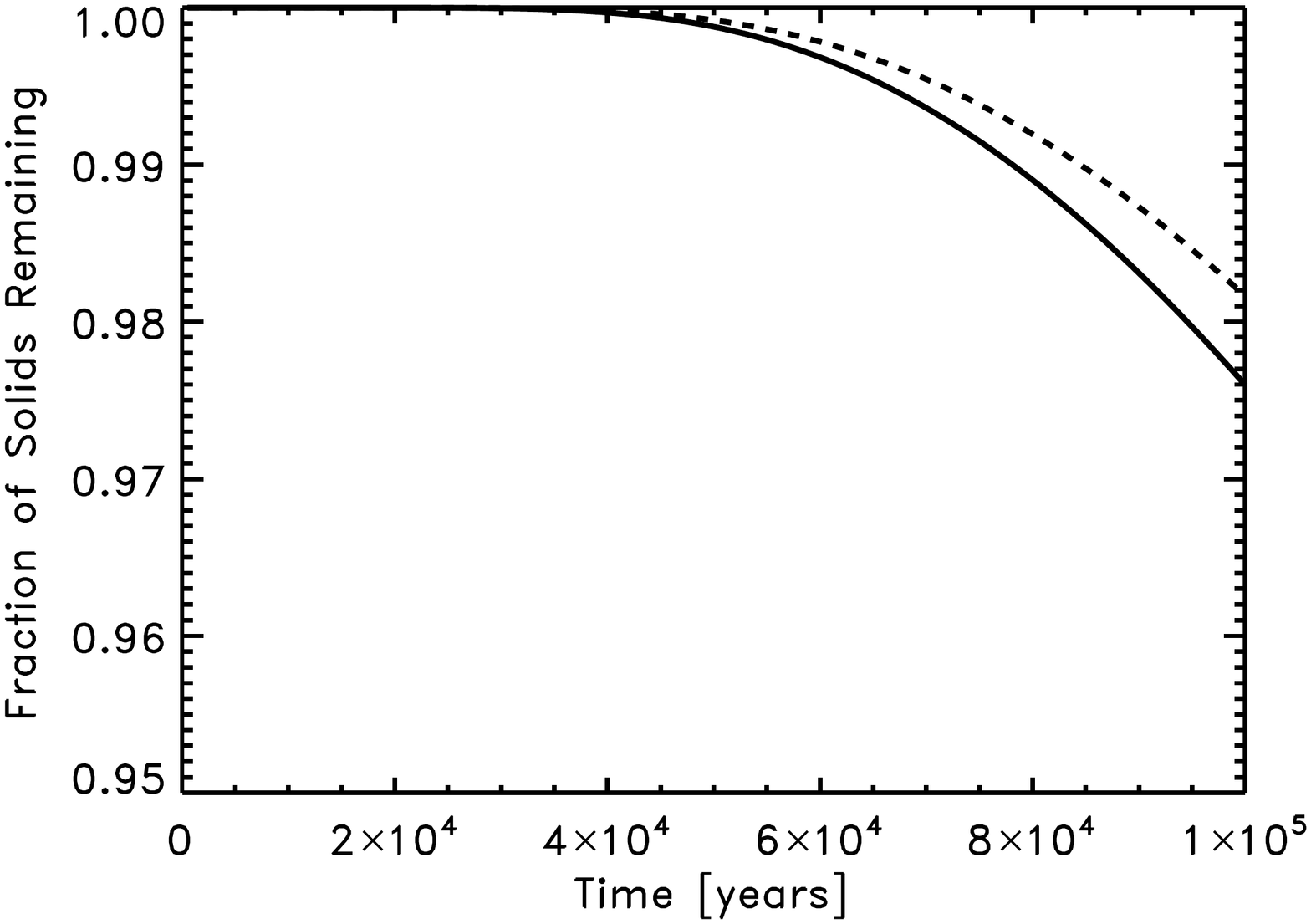}\\
\caption{A summary of the results of the inward transport models for the disk with $\dot M$=10$^{-6} M_\odot$/year.  Panels A and B show the distribution of the 5 $\mu$m and 0.5 mm grains at the end of the simulation respectively, with the contours having the same meaning as in Figure 4.  Panel C shows the concentration of the different sized particles particles at the disk midplane at the end of the simulation. Radial gradients in the concentrations are apparent.  Panel D shows the mass fraction of solids retained in the disk as a function of time.  That material that stays lofted at high altitudes and is pushed inward by the viscous flows of the disk begins to be accreted onto the central star after $\sim$10$^{4}$ years.  Because larger grains settle out of these upper regions more readily, they are retained more effectively than the smaller grains, an effect that runs counter to expectations based on previous one-dimensional models. }
\end{figure}

\newpage
\begin{figure}
\includegraphics[angle=90,width=3.1in]{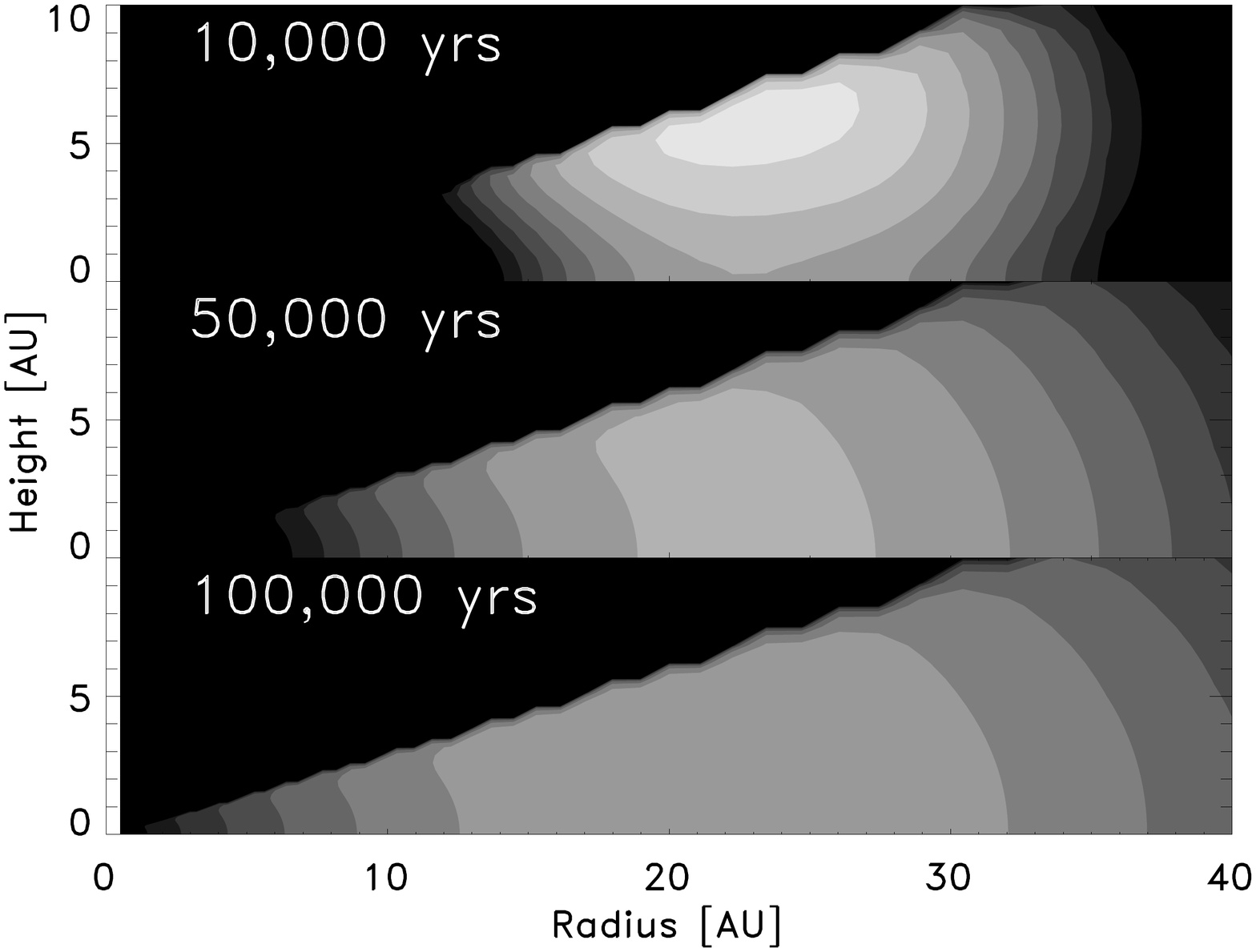}
\includegraphics[angle=90,width=3.1in]{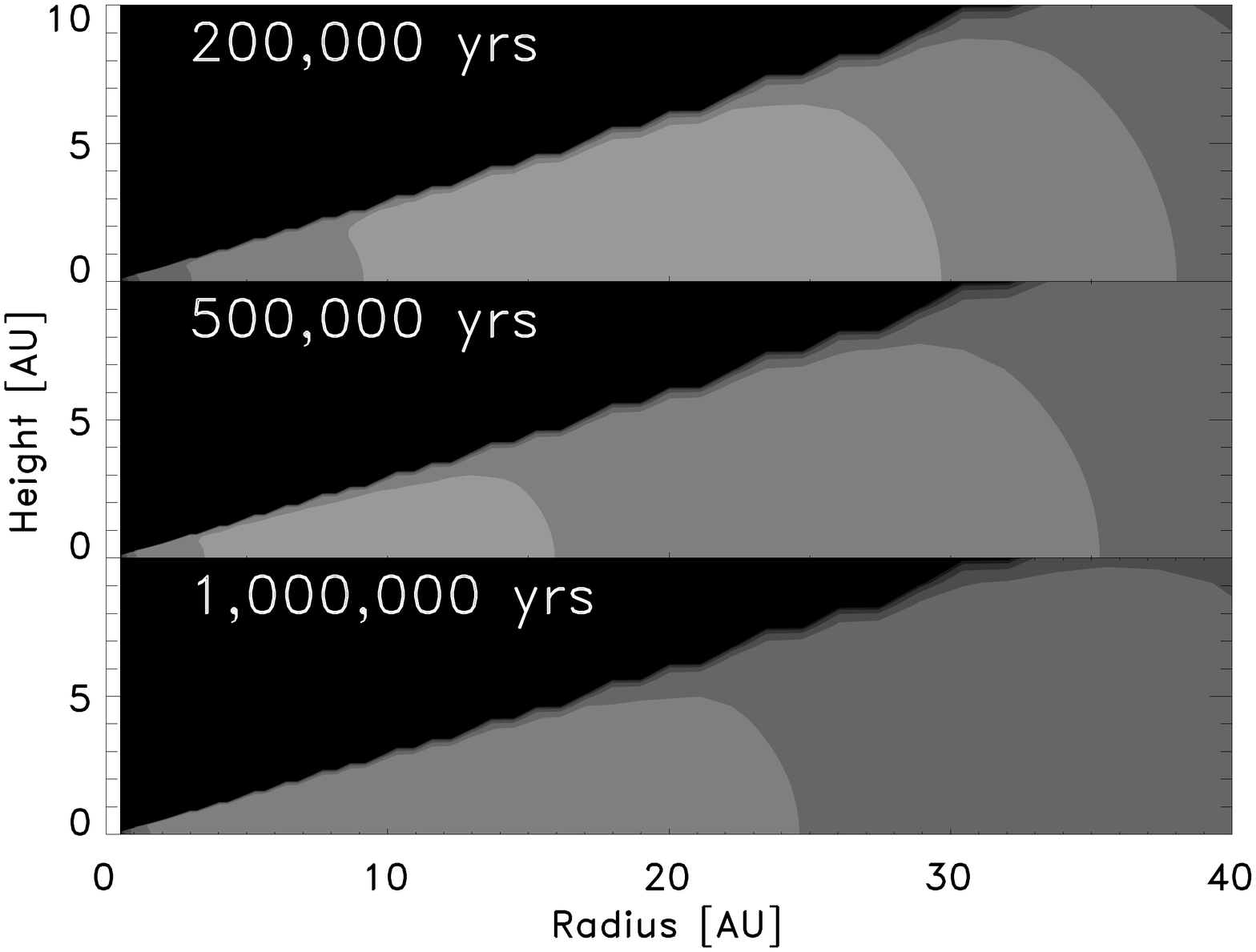}\\
\caption{Time evolution for the 5 $\mu$m grains for the $\dot M$=10$^{-7} M_\odot$/year disk.  The initial concentration was defined as in Figure 10.}
\end{figure}

\newpage
\begin{figure}
\includegraphics[angle=90,width=3.1in]{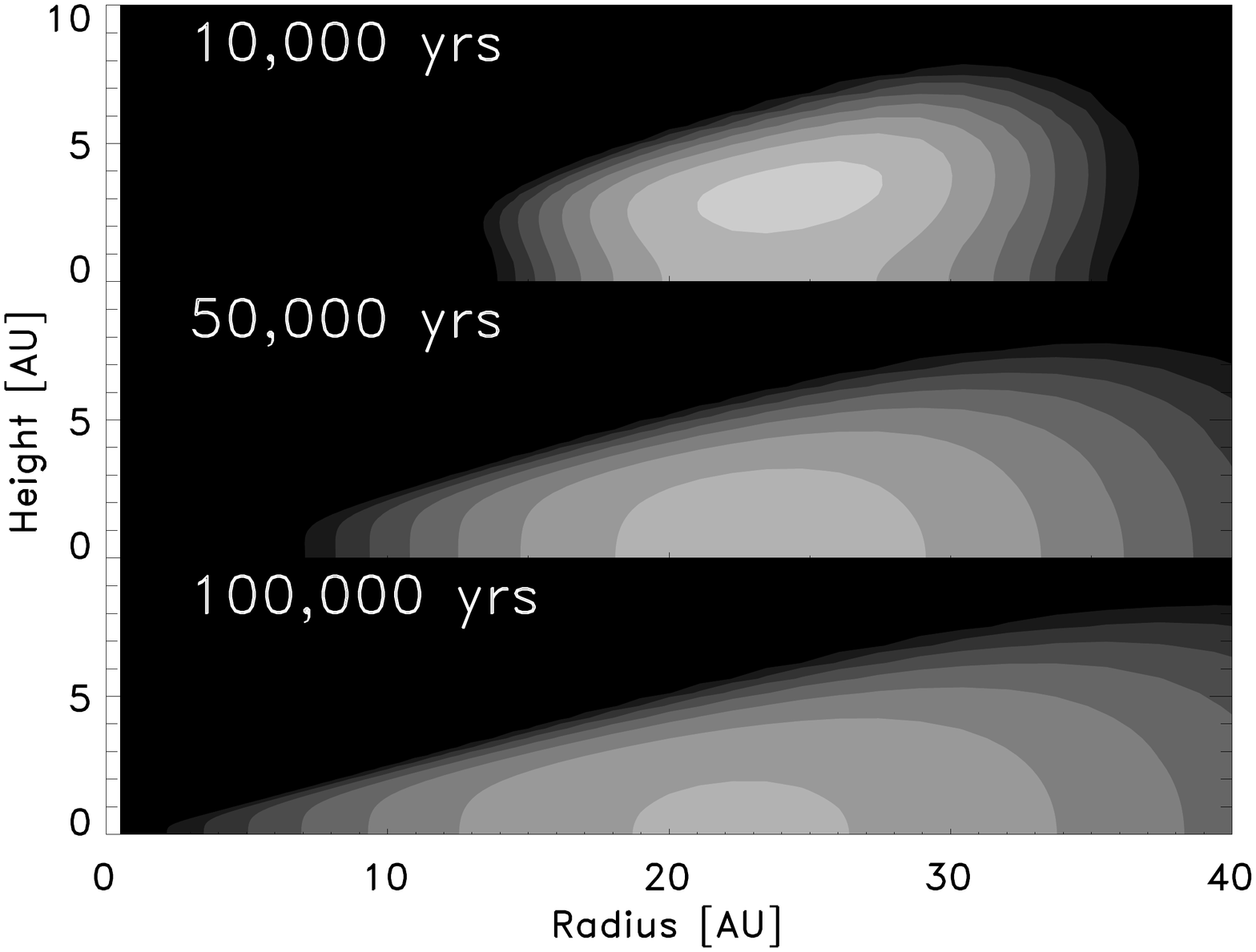}
\includegraphics[angle=90,width=3.1in]{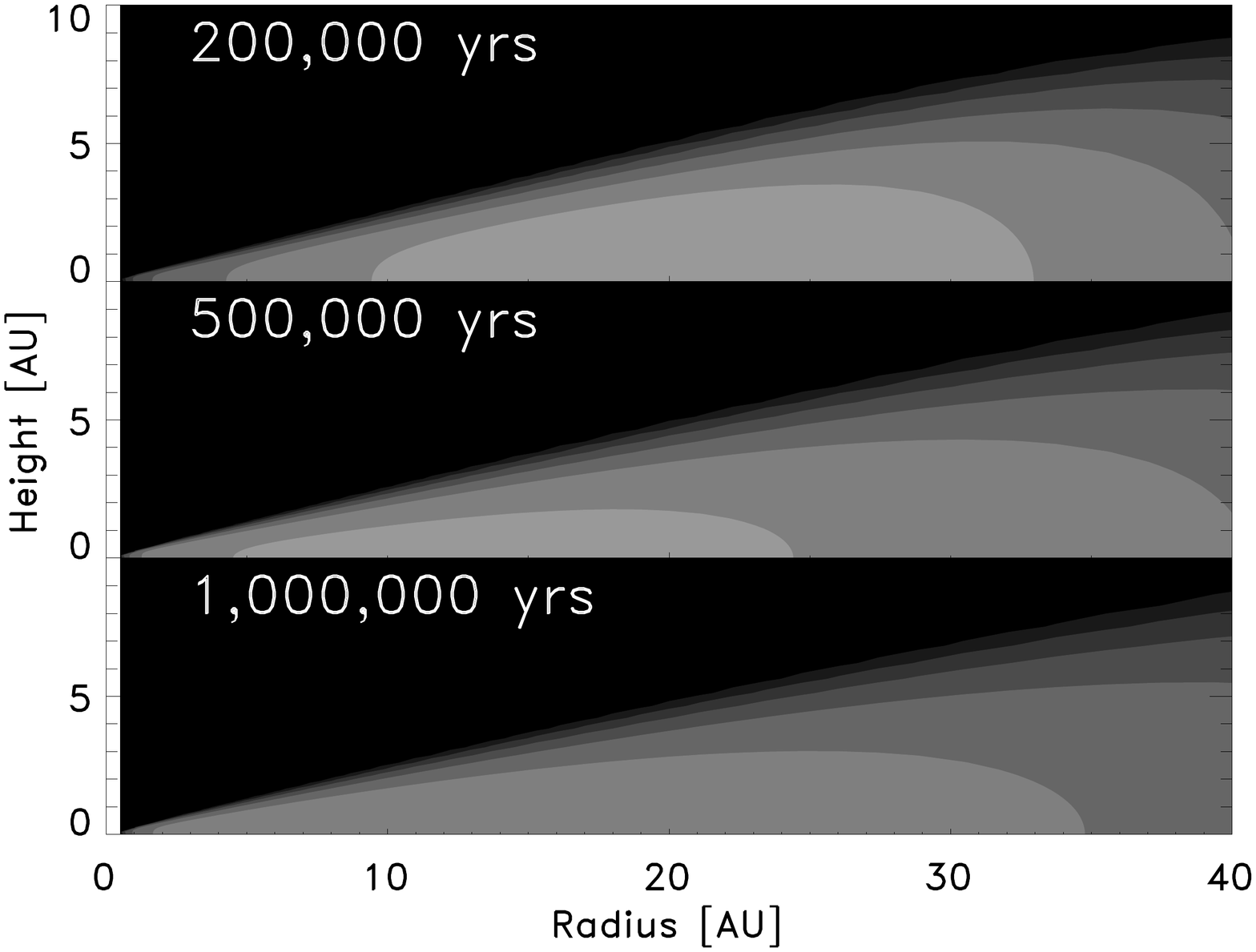}\\
\caption{Same as Figure 12, with $a$=0.5 mm.}
\end{figure}

\newpage
\begin{figure}
\includegraphics[angle=90,width=3.1in]{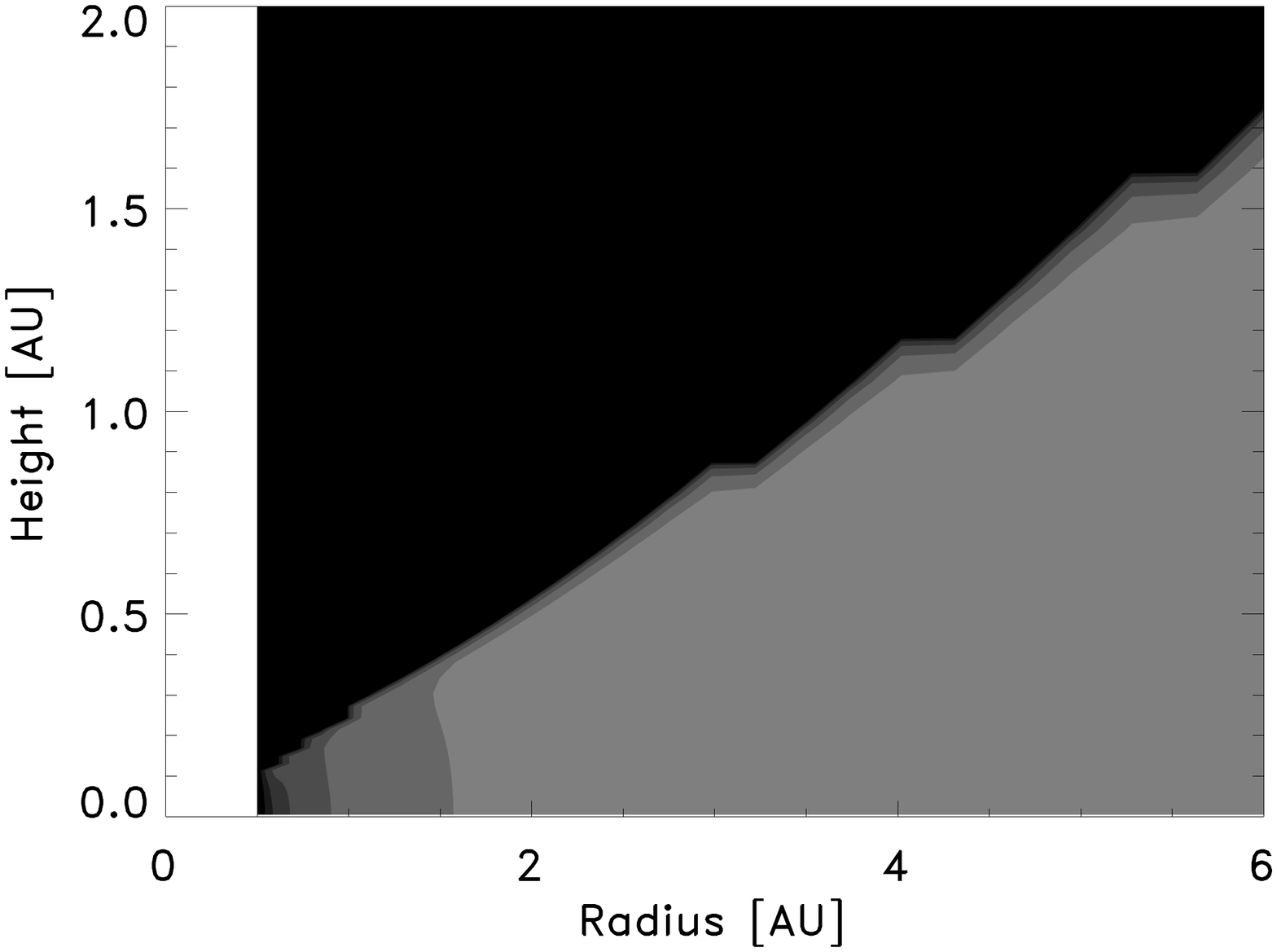}
\includegraphics[angle=90,width=3.1in]{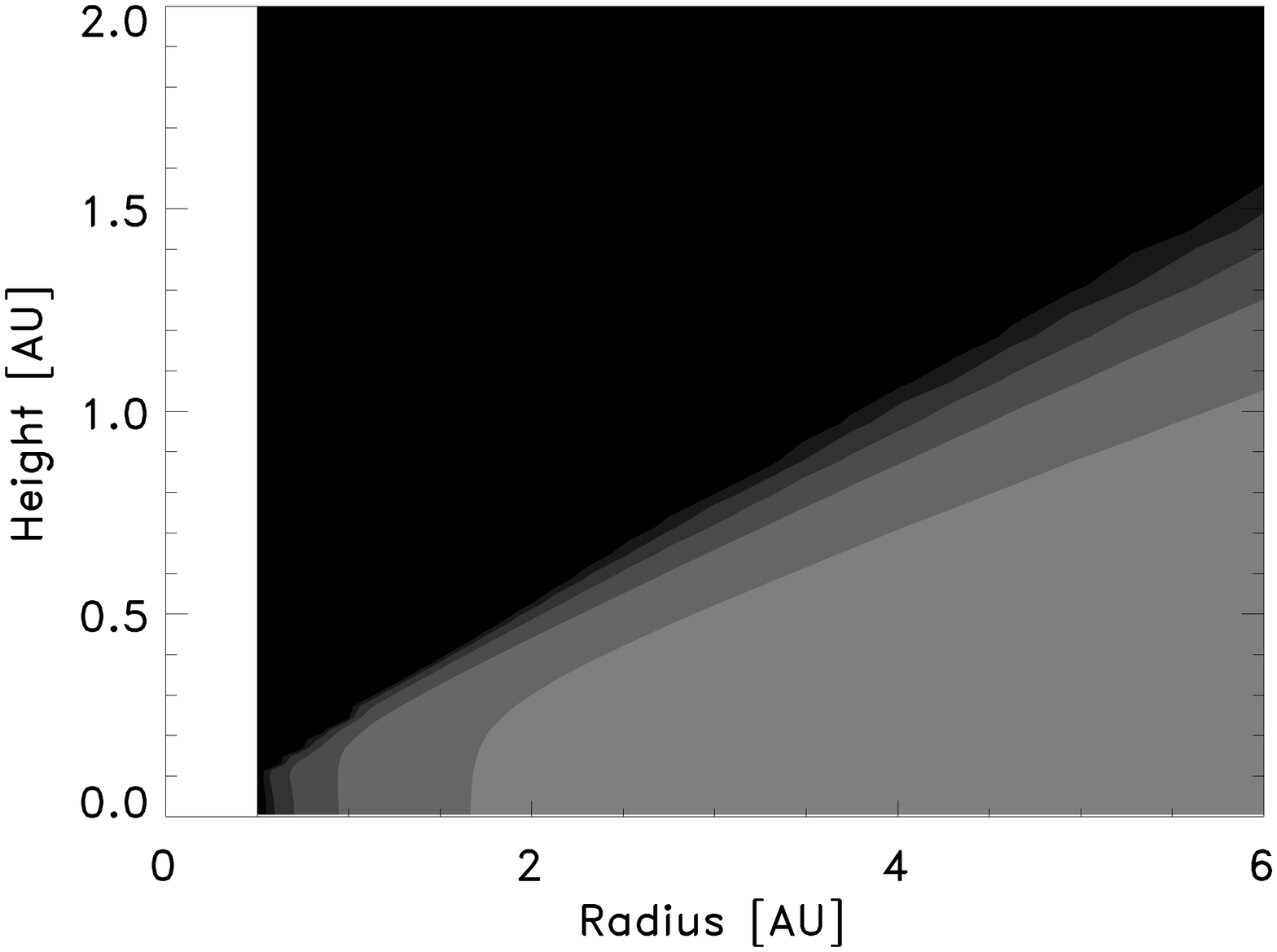}
\includegraphics[angle=90,width=3.1in]{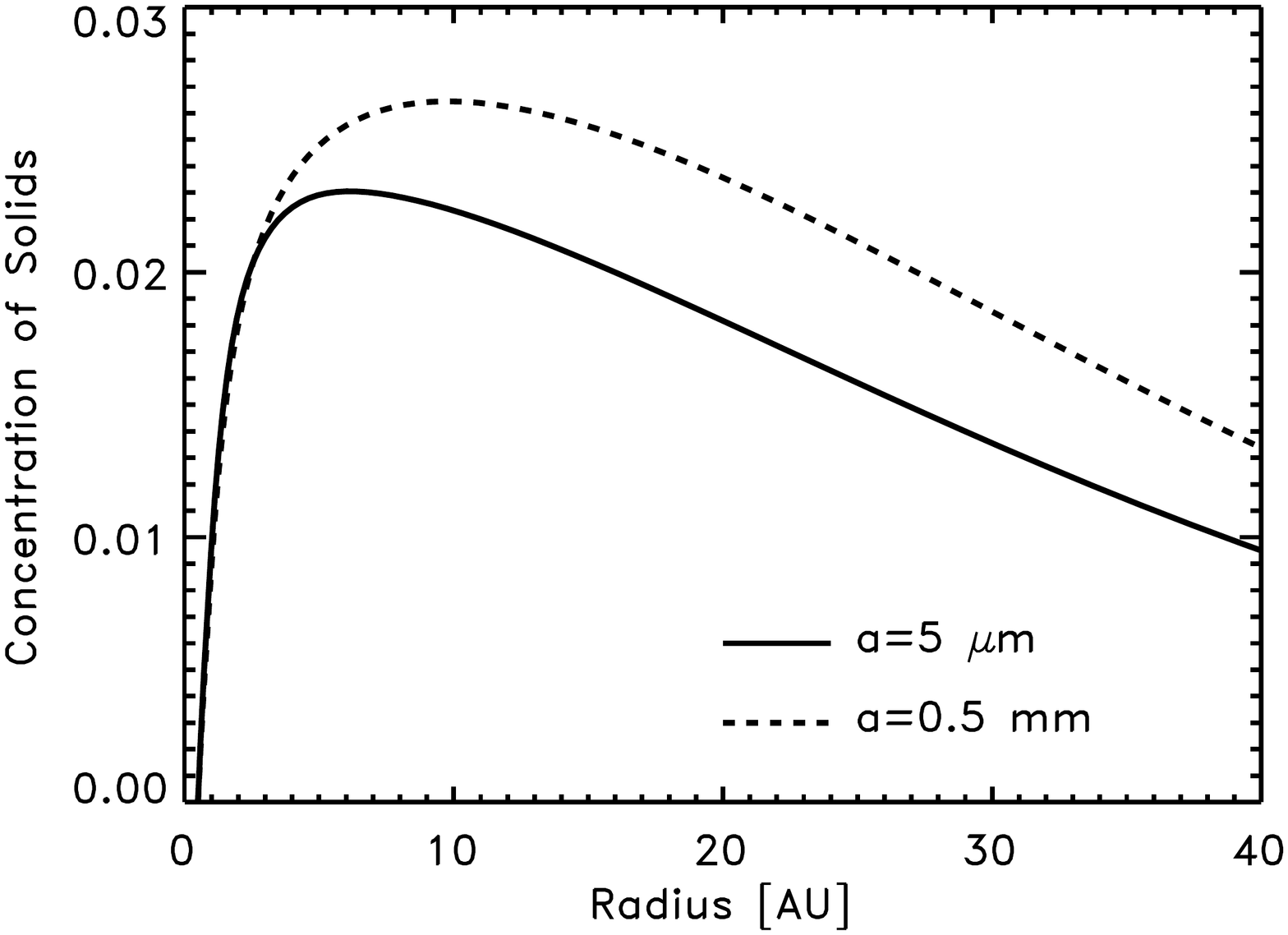}
\includegraphics[angle=90,width=3.1in]{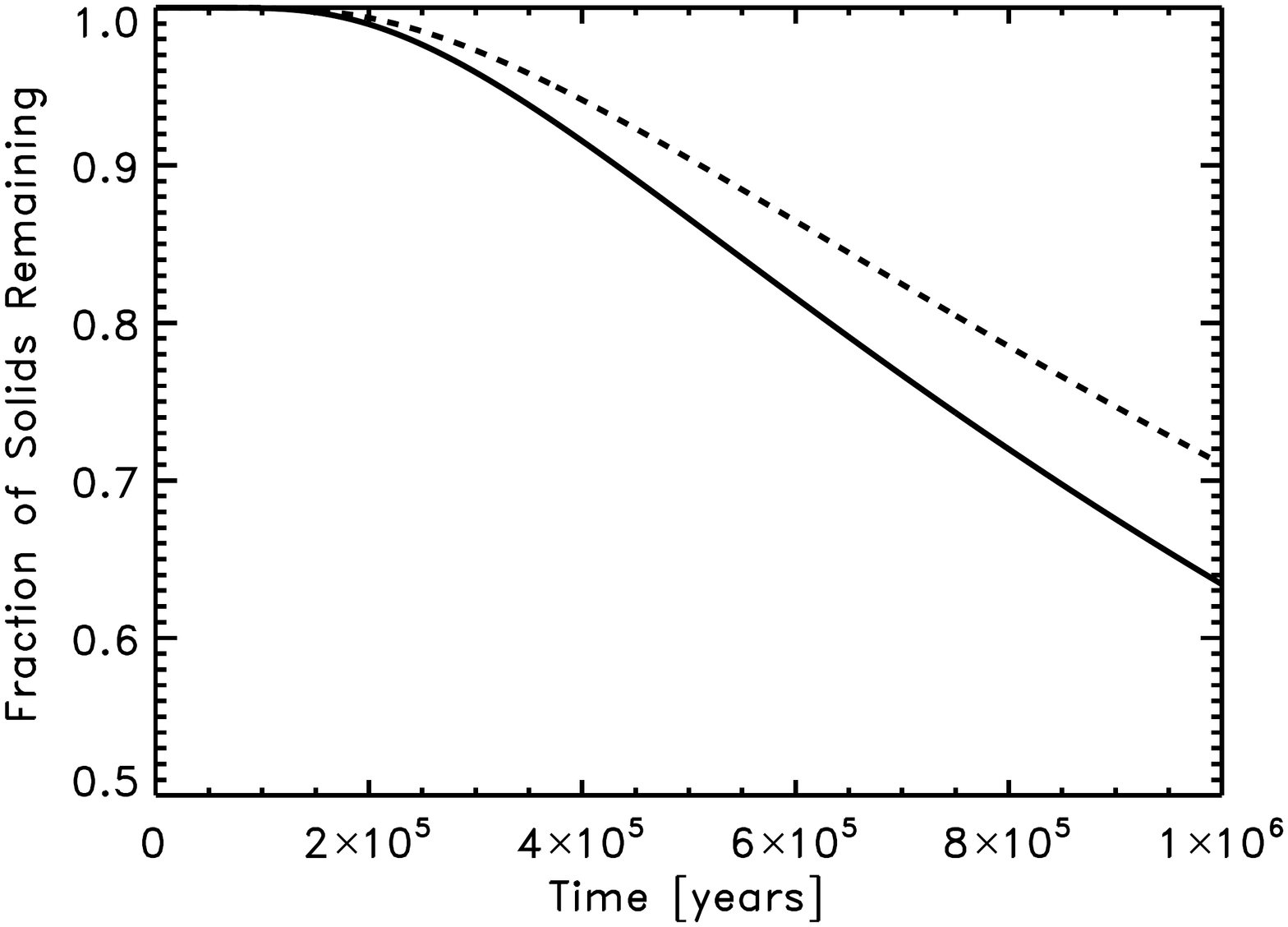}\\
\caption{Same as Figure 11, with $\dot M$=10$^{-7} M_\odot$/year.  Again, radial gradients remain in the disk everywhere and the smaller grains are lost from the disk faster than the large grains.}
\end{figure}

\newpage
\begin{figure}
\includegraphics[angle=90,width=3.1in]{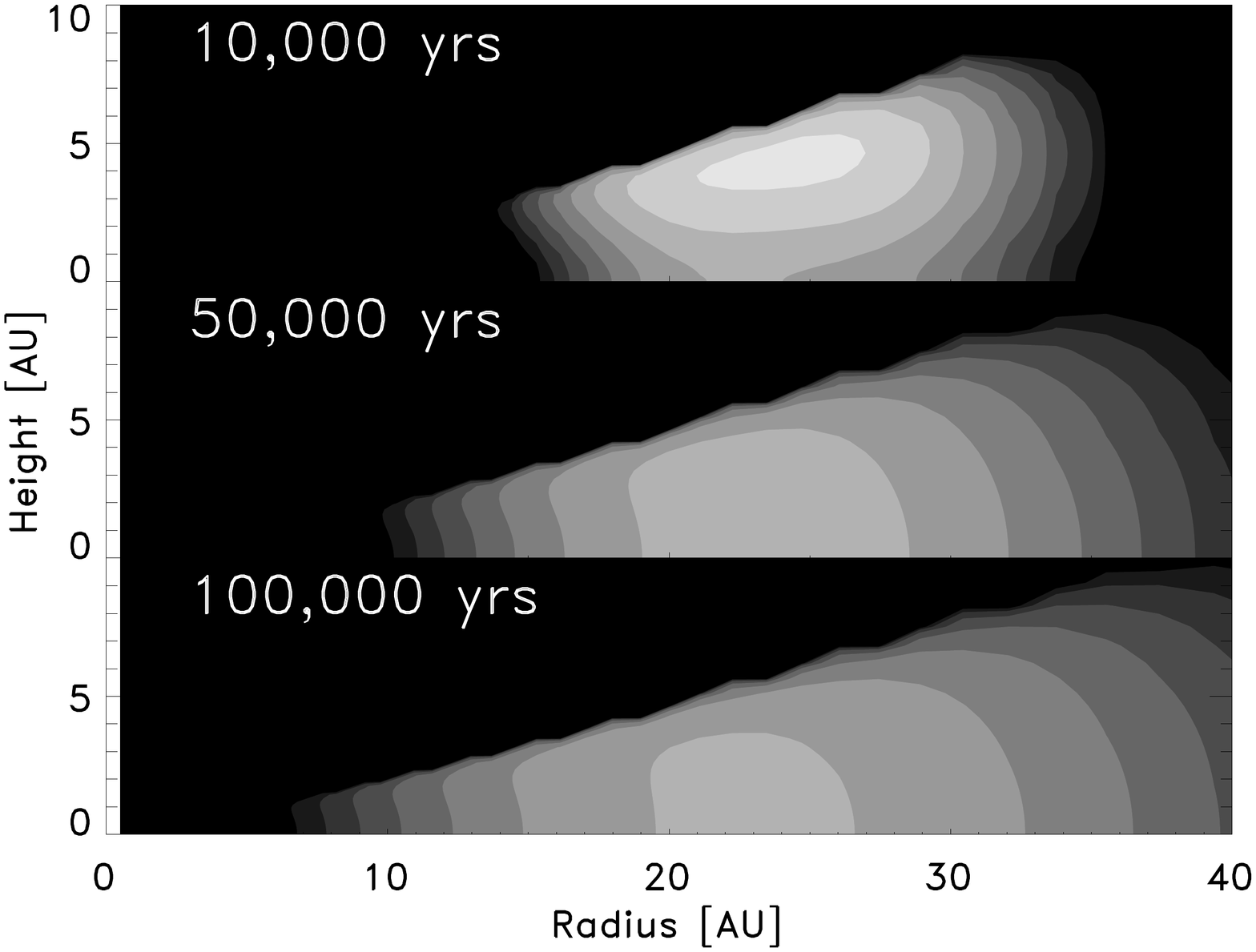}
\includegraphics[angle=90,width=3.1in]{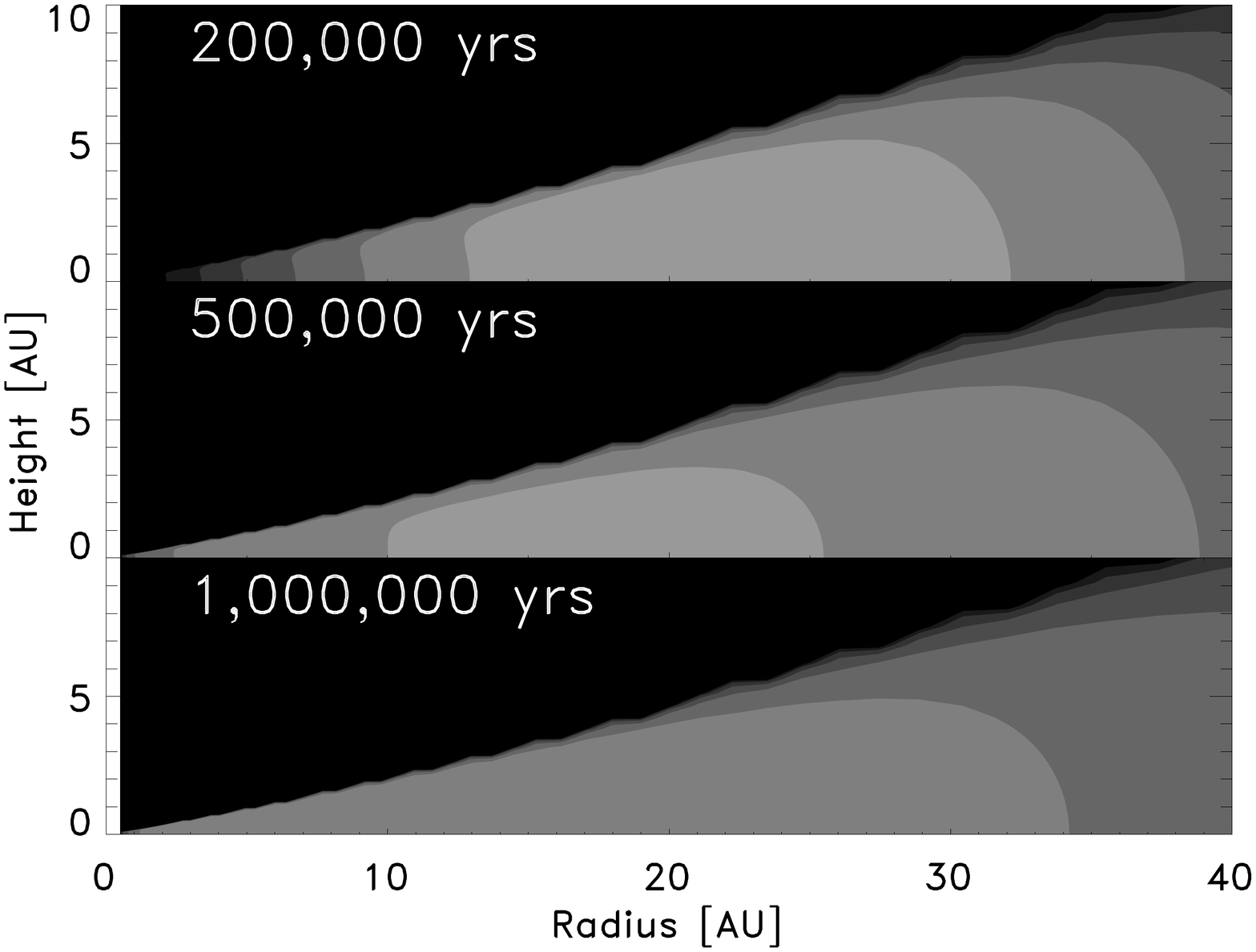}\\
\caption{Time evolution for the 5 $\mu$m grains for the $\dot M$=10$^{-8} M_\odot$/year disk.  The initial concentration was defined as in Figure 10.}
\end{figure}

\newpage
\begin{figure}
\includegraphics[angle=90,width=3.1in]{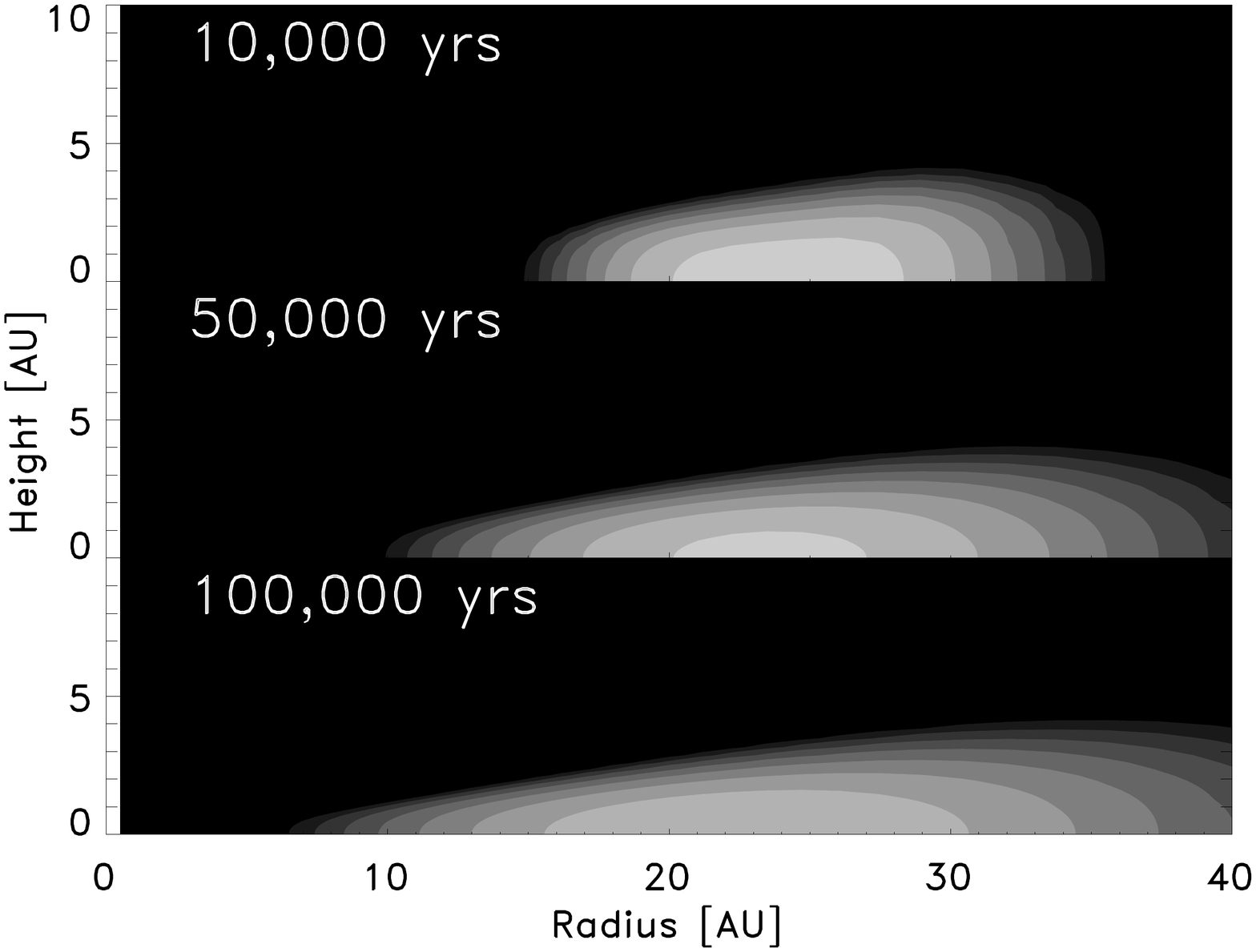}
\includegraphics[angle=90,width=3.1in]{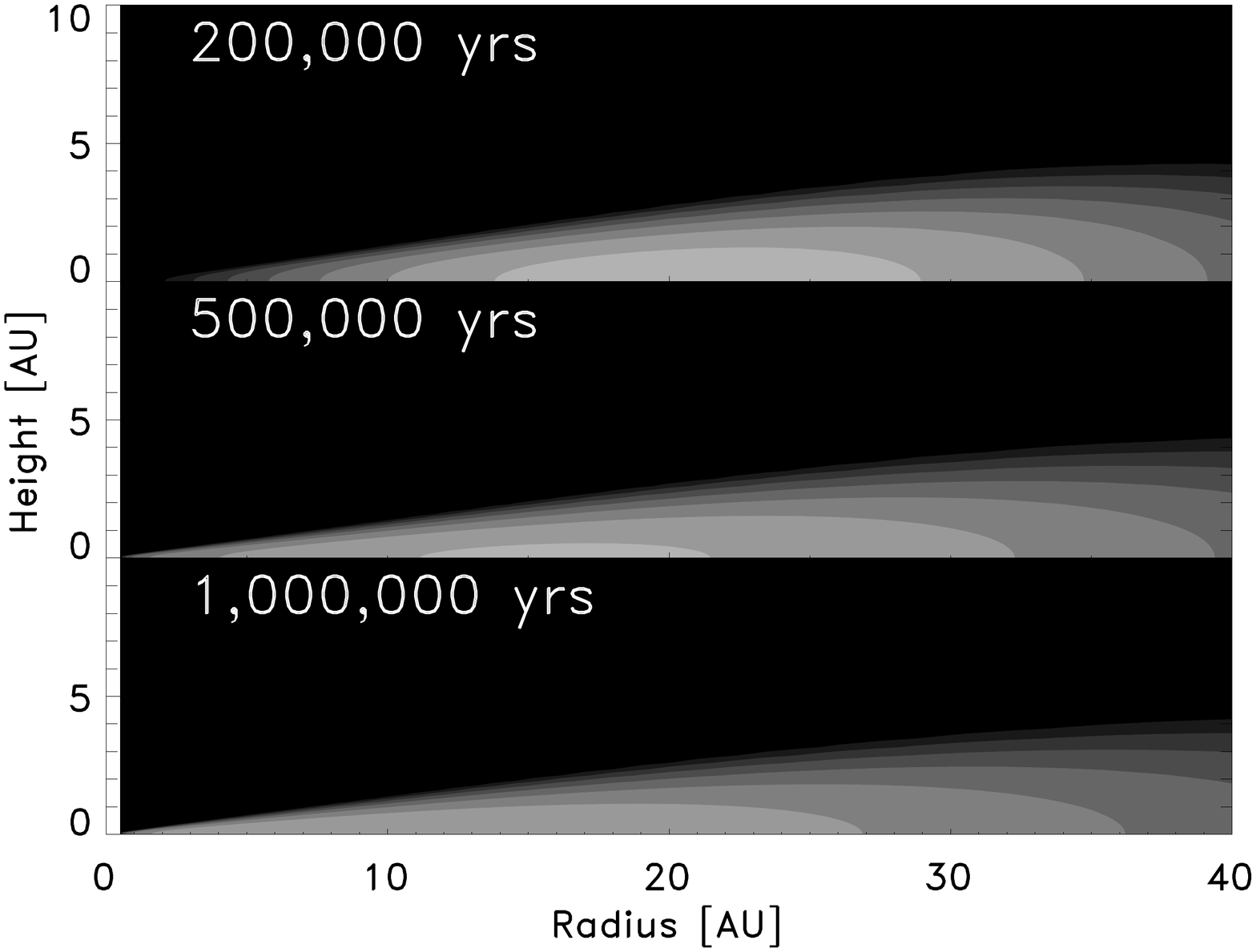}\\
\caption{Same as Figure 14, with $a$=0.5 mm}
\end{figure}

\newpage
\begin{figure}
\includegraphics[angle=90,width=3.1in]{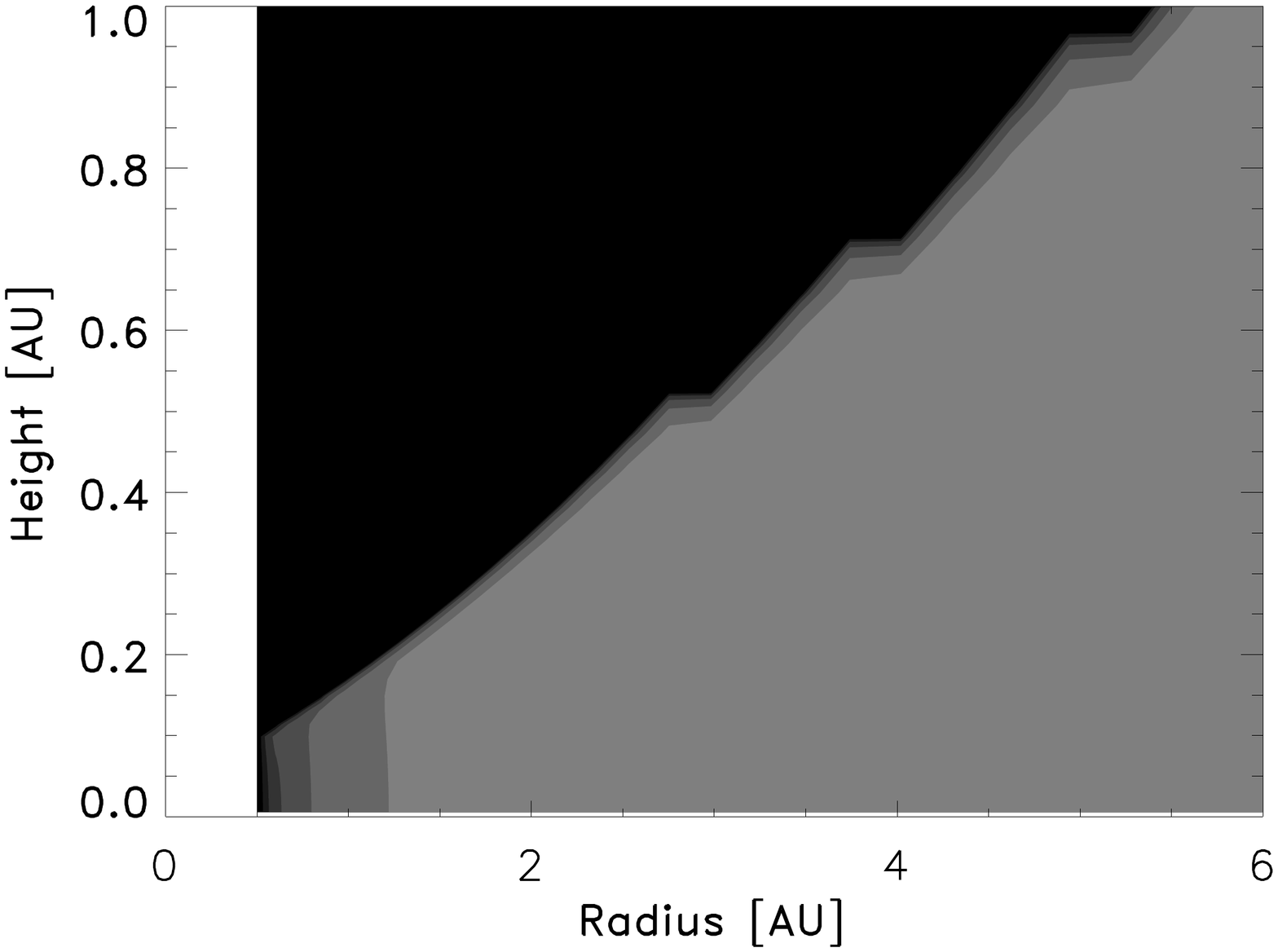}
\includegraphics[angle=90,width=3.1in]{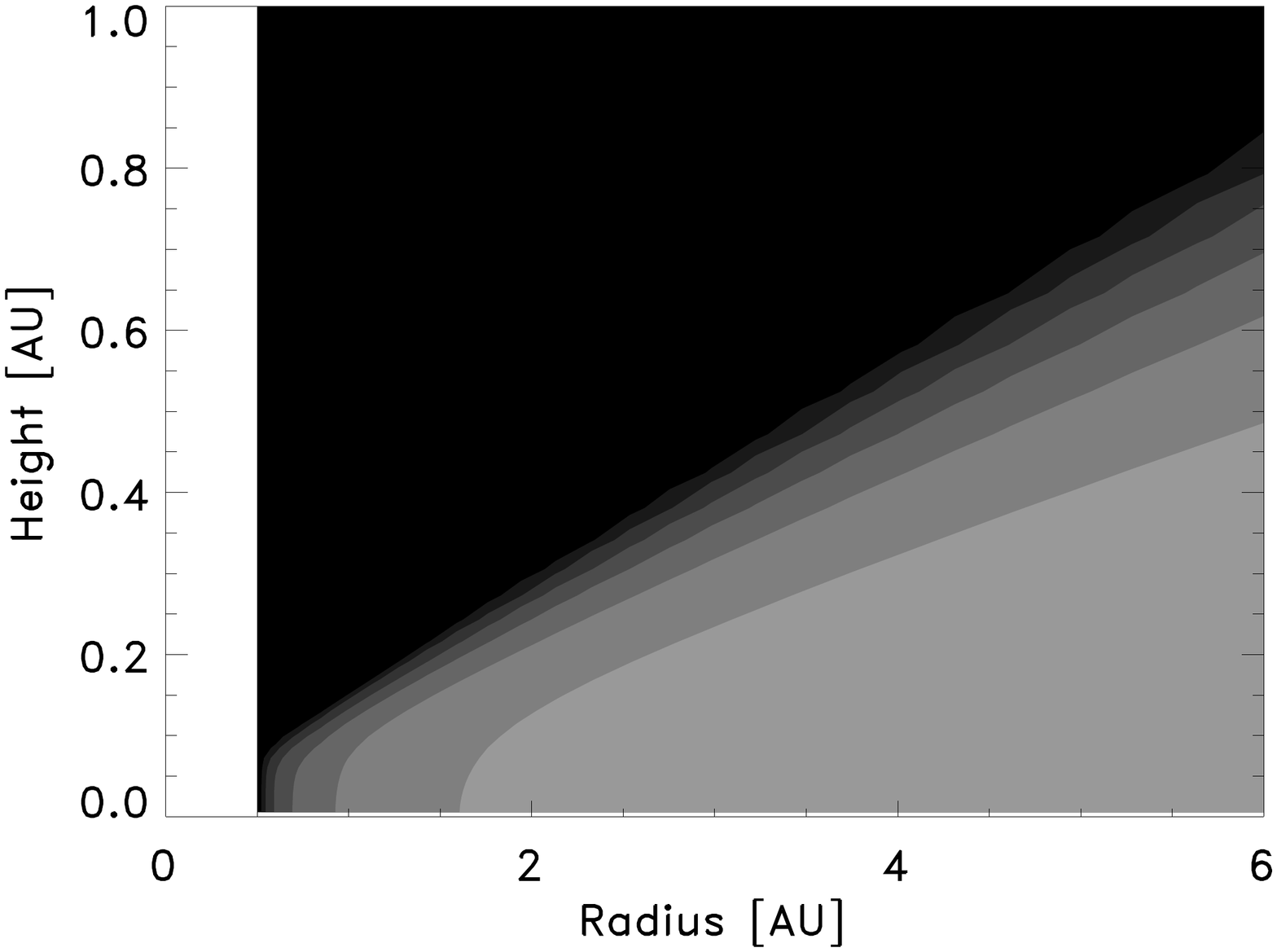}
\includegraphics[angle=90,width=3.1in]{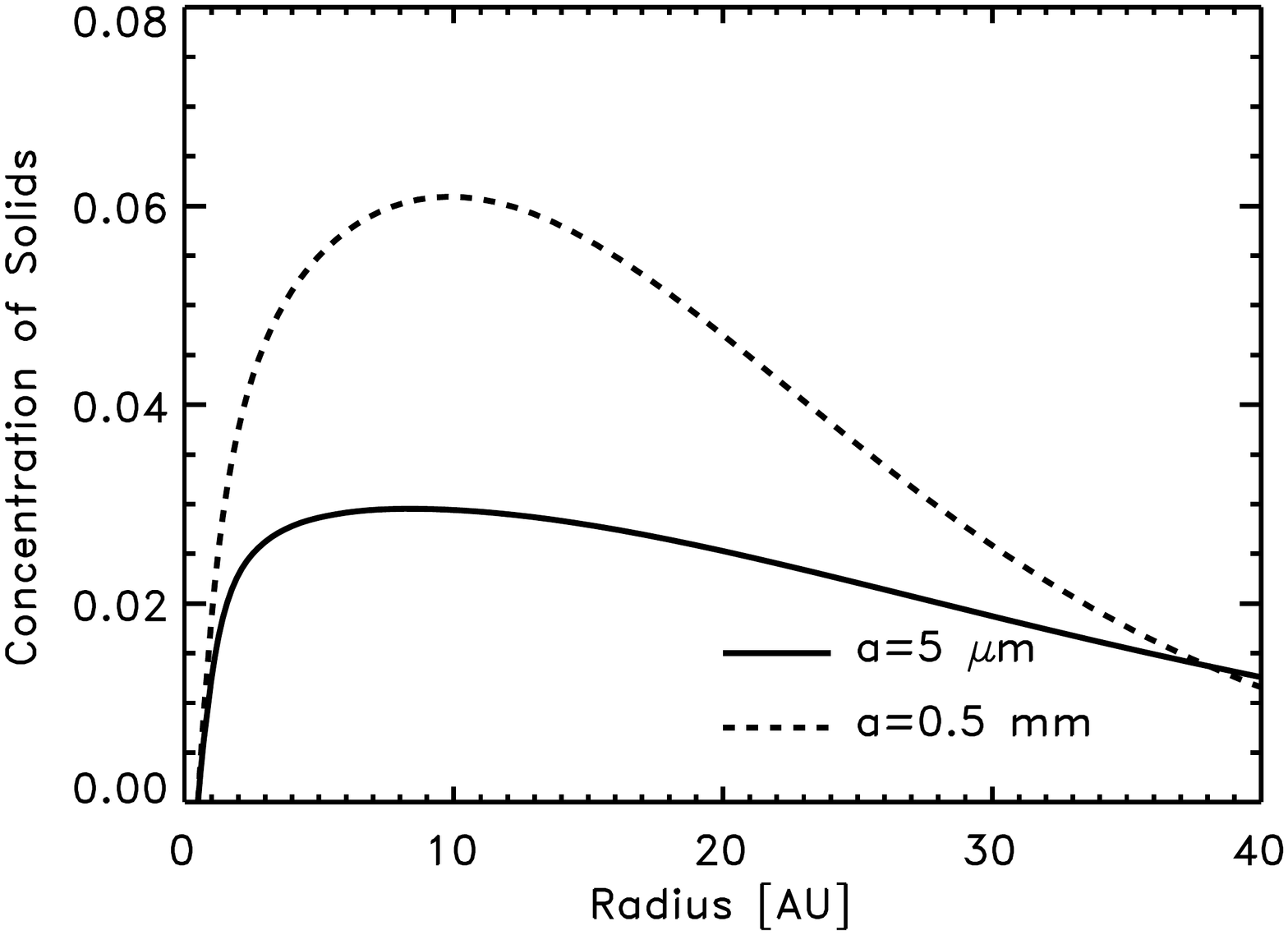}
\includegraphics[angle=90,width=3.1in]{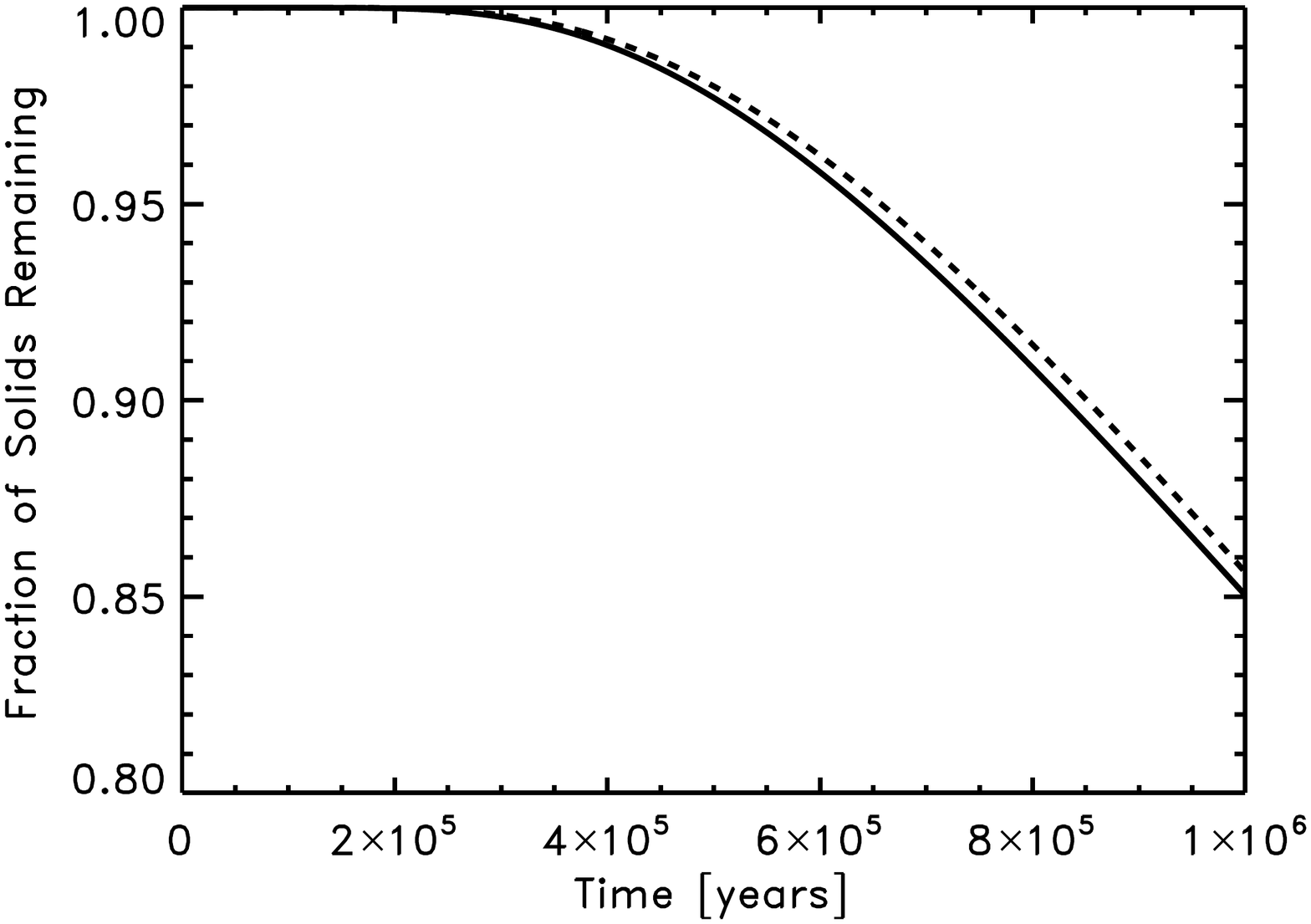}\\
\caption{Same as Figure 11, with $\dot M$=10$^{-8} M_\odot$/year.  The gradients in the distribution of the 5 $\mu$m grains are the smallest of the three simulations, as the large-scale flow velocities had diminished so that they were not significantly altering the gradients that diffusion was working to smooth.
This was not the case for the 0.5 mm grains, as gas drag velocities were high enough to constantly change the conditions on which diffusion was operating.}
\end{figure}

\end{document}